\documentclass[twocolumn]{aastex631}
\usepackage[varg]{txfonts}
\usepackage{graphicx}
\usepackage[printonlyused,withpage]{acronym}

\usepackage{ulem}
\usepackage{amsmath}
\usepackage{amssymb}
\usepackage{graphicx}
\usepackage{inputenc}
\usepackage{longtable}
\usepackage{float}
\usepackage{hyperref}
\accepted{31 March 2025 -- The Astrophysical Journal}
\shorttitle{Prediction of radio-quiet gamma-ray pulsar distances using the FP Relation}
\shortauthors{E.O. Angüner}
\graphicspath{{./}{figures/}}

\begin{document}

\title{Prediction of radio-quiet gamma-ray pulsar distances using the Fundamental Plane Relation}

\author[0000-0002-4712-4292]{Ekrem Oğuzhan Angüner}
\affiliation{TÜBİTAK Research Institute for Fundamental Sciences, 41470 Gebze, Kocaeli, Turkey}
\email{oguzhan.anguner@tubitak.gov.tr \\ oguzhananguner@gmail.com}

\begin{abstract}

The fundamental plane (FP) relation connects gamma-ray luminosity to intrinsic pulsar properties, offering the potential to estimate distances for radio-quiet (RQ) gamma-ray pulsars, where direct measurements are often unavailable. The Fermi Third Pulsar Catalog presents spectral data for 294 gamma-ray pulsars, including 72 RQ pulsars, of which only 9 have known distances. This study investigates the FP relation’s potential to predict distances for RQ gamma-ray pulsars using machine learning (ML) approaches. Ordinary least-squares regression was employed alongside ML methods, including random forest and support vector regression, to predict RQ gamma-ray pulsar distances. To ensure robustness, the analysis considered spectral cutoff significance and outlier data. Results confirm that the FP relation is valid only for pulsars exhibiting significant gamma-ray cutoffs, with FP exponents closely matching theoretical expectations. It was shown that gamma-ray emission origins significantly favor the curvature radiation regime over the synchrotron regime at the 4.2$\sigma$ level. The distances for 62 RQ pulsars, for which no traditional measurements exist, are predicted for the first time. The luminosity and spatial distributions of the considered RQ pulsar population match known pulsar distributions. Predictions for RQ pulsars with known distances are consistent with measured data. This study provides a proof of principle that the FP relation, combined with ML methods, is a promising tool for distance estimation in RQ gamma-ray pulsars, using gamma-ray emission directly for the first time in pulsar distance estimates. Future improvements can be achieved with larger data sets, precise spectral cutoff determinations, and corrections for key factors like the beaming factor.

\end{abstract}


\section{Introduction} \label{sec:intro}

Gamma-ray pulsars are a specific subclass of rapidly rotating neutron stars that emit pulsed gamma-ray radiation owing to accelerated particles in their highly magnetized environment. Before 2008, only seven gamma-ray pulsars, mostly young and detectable in radio wavelengths (radio-loud, RL), were identified~\citep{thompson2001}. Undoubtedly, a revolutionary era in gamma-ray pulsar astronomy began with the launch of the Fermi Gamma-ray Space Telescope on 2008 June 11 (see, e.g., \cite{caraveo2014}). The Fermi Large Area Telescope (LAT;~\cite{fermiLAT}) discovered numerous gamma-ray pulsars, revealing them as the dominant GeV sources in our Galaxy as reported in the First Fermi-LAT Catalog of Gamma-ray Pulsars (1PC;~\cite{fermi1PC}). One of the groundbreaking achievements of the LAT has been the discovery of radio-quiet\footnote{RQ pulsars remain undetected at radio frequencies down to a flux density threshold of 30 $\mu$Jy at 1400 MHz.} (RQ) gamma-ray pulsars~\citep{fermiRQ_det}, which was made possible only through advanced blind pulsation search techniques~\citep{kerr2011,Pletsch_2014}. Furthermore, detecting gamma-ray emissions from millisecond pulsars (MSPs;~\cite{fermiMSP_det}) and the discovery of the first RQ MSP~\citep{rq_msp_det} significantly extended the known gamma-ray pulsar population, remarkably improving the understanding of these extreme astrophysical objects.

The magnetospheric structures and emission mechanisms of pulsars can, in principle, be investigated through observed pulsed radiation. Significant theoretical progress has led to the development of two main gamma-ray pulsar emission model classes, distinguished by the location of high-energy emission origin. The first class includes polar cap (PC) models, proposing emission near the neutron star’s surface at low altitudes~\citep{sturrock1971, ruderman1975, daugherty1982, daugherty1996}. On the other hand, the second class, namely outer magnetosphere (OM) models, suggests higher-altitude emissions in the OM and beyond, such as outer gaps (OG;~\cite{cheng1986, romani1996, Hirotani_2008, Takata_2011}), two-pole caustic model or the slot gaps (SG;~\cite{arons1983, Muslimov_2004, harding_2008, Watters_2009, Harding_2011}), the striped wind model outside the light cylinder~\citep{petri2009,petri2011, petri2012,dirson2022}, and the widely accepted model based on particles accelerated at the equatorial current sheet beyond the light cylinder~\citep{Bai_2010, ecs_2010}. The OM models predict that gamma rays are mainly emitted through curvature radiation (CR) and synchrotron radiation (SR), exhibiting an exponential spectral cutoff around several GeV due to the radiation reaction limit. The PC models, by comparison, predict superexponential cutoffs at around a few GeV due to gamma-ray flux attenuation in the star's high magnetic field. It was shown in \cite{Harding_2021} that gamma-ray emission up to 100 GeV can originate from synchrocurvature (SC) radiation, with synchrotron self-Compton (SSC) emissions reaching $\sim$1~TeV energies. However, SSC emission below 100~GeV is obscured by the dominating SC radiation. At higher energies, inverse Compton (IC) emission arises from the most energetic particles in the current sheet scattering low-energy SR photons and can extend the spectral cutoff beyond 10~TeV (see, e.g.,~\cite{Hu_2024} for detection prospects of pulsar at higher energies). In addition, pulsar gamma-ray light curves published in the Second Fermi-LAT Catalog of Gamma-ray Pulsars (2PC; \cite{Abdo_2013_2PC}) typically show a double-peaked structure. In order to explain this feature, PC models require nearly aligned geometries, with magnetic inclination angles comparable to the angular extent of the PC, whereas OG models favor highly inclined rotators \citep{romani1996, caraveo2014} leading to a larger solid angle than the radio emission cones.

The Third Fermi-LAT Catalog of Gamma-ray Pulsars (3PC;~\cite{fermi3pc}) presents detailed spectral analyses of 294 gamma-ray pulsars from 12 yr of Fermi-LAT data, with energies extending beyond 50~GeV. This catalog is composed of three rotation-powered pulsar classes: 84 RL pulsars, 66 RQ pulsars, and 144 MSPs, of which 6 are RQ MSPs\footnote{These numbers are obtained from the most recent (v3) 3PC FITS file available at \url{https://fermi.gsfc.nasa.gov/ssc/data/access/lat/3rd_PSR_catalog/} using the \texttt{CHAR\_CODE} keyword.}. The unprecedented data set provided by 3PC allows in-depth studies of the known gamma-ray pulsar population, revealing various correlations and trends in the observed properties. Since gamma rays are key tracers for large-scale magnetic fields, analyzing Fermi spectra and light curves can help refine pulsar models by probing the magnetospheric configurations. The results based on the 1PC and 2PC data excluded gamma-ray emission from the polar gap, supporting its association with the OM and favoring OM models \citep{Kerr_2012, caraveo2014}. In contrast, radio emission is associated with the polar gap as in PC models, suggesting that gamma-ray and radio beams are codirected, while OM models propose distinct geometries. This difference in geometry provides a natural explanation for RQ pulsars, where the radio beam does not align with Earth’s line of sight, while the gamma-ray beam does~\citep{Perera_2013, Petrova_2016}. The observed double-peaked, energy-dependent, and often non-phase-aligned light curves~\citep{Thompson2004, Abdo_2013_2PC} further support OM models by highlighting differences between radio and gamma-ray emission regions. Moreover, the similar numbers of detections for RL and RQ pulsars in 3PC are consistent with OG model predictions (see, e.g.~\cite{yadigaroglu1995}), and provide further empirical validation for OM models. The preference for OM models over PC models, initially inferred from the 2PC, is now further validated and strengthened by the 3PC data set.

About half of the gamma-ray pulsars in 3PC are MSPs with periods shorter than 30~ms. These ancient neutron stars~\citep{Kiziltan_2010}, usually found in binary systems, have been recycled to high rotation rates through accretion from a companion star~\citep{annual_MSP, Lorimer_2008}. The rapid rotation of MSPs results in small light cylinders, where the intense curvature of the magnetic field plays a crucial role in beam broadening~\citep{Harding2022}. Radio emission originates near the magnetic poles along open field lines but extends to higher altitudes owing to the weaker magnetic fields of MSPs compared to young pulsars \citep{Bhattacharyya2022}. The relativistic effects induced by the rapid rotation of MSPs, combined with a larger fraction of open field lines due to the smaller light-cylinder radius, contribute to the broadening of the radio beam. In contrast, gamma-ray emission is produced in the OM at high altitudes, where relativistic effects further increase beam broadening~\citep{Harding_2005, Story_2007}. As a result, both radio and gamma-ray emissions exhibit broad beams, with gamma-ray beams being more extended, therefore improving the detectability of MSPs in high-energy observations. Similarly, the broad nature of radio beams leads to the rarity of RQ MSPs ~\citep{fermi3pc}. In fact, MSPs are among the most efficient gamma-ray sources, converting a large fraction of their spin-down luminosity into gamma rays. Fermi-LAT observations show that MSPs are distributed spherically, suggesting a local origin, and their numerous detections support the idea that they contribute significantly to the gamma-ray background of our Galaxy~\citep{psr_bkg}.

In many cases, measuring pulsar distances is a challenging problem, yet it remains of fundamental importance in the field. Accurate distance information is essential for estimating luminosity, proper motion, and mapping pulsar distributions within the Galaxy. Pulsars' remarkable accurate periodicity also makes them useful for detecting low-frequency gravitational waves (GWs). The pulsar timing array (PTA) network, with known distances, is the sole technique for detecting nHz GWs~\citep{pta_1979, xieGW2023}, originating from phenomena like black hole binaries, inflation~\citep{gw_inf}, and cosmic strings~\citep{GASPERINI1993317, cosmic_strings}. Recently, a pioneering investigation of the GWs using PTAs observed in the Fermi gamma-ray band demonstrated the great potential of this approach~\citep{FermiGWB}. Thus, determining gamma-ray pulsar distances, as well as increasing their number, is crucial for understanding pulsar astrophysics, refining emission models, revealing the Milky Way's structure, improving cosmic distance calibration, and enhancing GW detection capabilities. Several established techniques exist for measuring pulsar distances. Among them, parallax measurement offers a direct and reliable method using X-ray, optical, or radio interferometric images, as well as precise timing. However, it is only applicable to nearby pulsars and often requires corrections for Lutz$-$Kelker bias \citep{lutz1973,Verbiest_2012}. The most prevalent method involves calculating the dispersion measure, which quantifies free electrons along the line of sight to the pulsar, derived from the frequency-dependent delay in radio pulse arrival times. This technique requires an accurate model of the free electron distribution within the interstellar medium, with the YMW16~\citep{ymw16} and NE2001~\citep{Cordes2002} models frequently used for this purpose. An alternative method for estimating pulsar distances uses Doppler-shifted HI absorption or emission lines in radio spectra, converted to distances via Galactic rotation curves~\citep{RomanDuval_2009}, known as the kinematic method. Pulsar distances can be further constrained by referencing associated objects like supernova remnants, pulsar wind nebulae, star clusters, or HII regions, especially when the pulsar is situated within or in the vicinity of molecular clouds. For X-ray-emitting pulsars, distances can be estimated from X-ray-absorbing columns~\citep{He_2013} or correlations between X-ray luminosity and parameters such as spin-down power or photon index~\citep{possenti2002, Gotthelf_2003, mattana2009}.

The gamma-ray emission mechanism in pulsars, driven by particles accelerated at the equatorial current sheet beyond the light cylinder~\citep{Bai_2010, ecs_2010}, was extensively investigated in~\cite{fp2019}. This research revealed the existence of a fundamental plane (FP) relation, characterized by four critical parameters: total gamma-ray luminosity (L$_{\gamma}$), spin-down luminosity ($\dot{\text{E}}$), surface magnetic field (B$_{*}$), and spectral cutoff energy ($\epsilon_{\text{cut}}$). The 4-dimensional FP relation was subsequently validated in~\cite{fp2022} through the analysis of 190 gamma-ray pulsars from the 2PC and Fermi 4FGL data sets~\citep{fermi_4fgl}, indicating that gamma-ray emission predominantly arises from the CR regime. In its general form, the theoretical FP relation can be expressed as 
\begin{equation}
\label{fp_der}
L_\gamma \propto
\begin{cases}
  \epsilon_\mathrm{cut}^{4/3}\,\text{B}_{*}^{1/6}\,\dot{\text{E}}^{5/12} & \text{(CR regime)} \\
  \epsilon_\mathrm{cut}\,\dot{\text{E}} & \text{(SR regime)}
\end{cases}
\end{equation}
or equivalently
\begin{equation}
\label{fp_obs}
L_\gamma \propto
\begin{cases}
  \epsilon_\mathrm{cut}^{4/3}\,\text{P}^{-7/6}\,\dot{\text{P}}^{1/2} & \text{(CR regime)} \\
  \epsilon_\mathrm{cut}\,\text{P}^{-3}\,\dot{\text{P}} & \text{(SR regime)}
\end{cases}
\end{equation}
in terms of direct pulsar observables (see Appendix~B of \cite{fp2019}), where P is the pulsar period and $\dot{\text{P}}$ is the spin-down rate. The exponents of FP relation variables, ($\epsilon_\mathrm{\text{cut}}$, B$_{*}$, $\dot{\text{E}}$) given in Eq.~\ref{fp_der} (CR regime), were estimated as (1.39$\pm$0.17, 0.12$\pm$0.03, 0.35$\pm$0.05) in~\cite{fp2022}, compatible with theoretical expectations. 

Pulsars, in an ideal scenario, can be conceptualized as a class of astrophysical objects exhibiting a high degree of uniformity, sharing consistent characteristics, such as extreme densities, similar masses ($\sim$1.4~M$_{\odot}$), and compact radii ($\sim$10~km). This uniformity results in comparable emission mechanisms and observable parameters, forming the basis for universal relations such as the FP relation. Thus, to a first approximation, pulsars with identical ($\epsilon_\mathrm{cut}$, B$_{*}$, $\dot{\text{E}}$), or equivalently ($\epsilon_\mathrm{cut}$, P, $\dot{\text{P}}$), should dissipate equal energy into gamma rays. When a pulsar's energy flux (F$_{\text{E}}$) is measured above a certain threshold, i.e.,~100~GeV, observed differences in gamma-ray luminosity L$_{\gamma}$ = 4$\pi$f$_{\Omega}$F$_{E}$D$^{2}$ among pulsars arise, mainly due to differences in distance (D), assuming a beaming factor of f$_{\Omega}$ = 1. Consequently, one of the most important applications of the FP relation is its potential to predict the distances to RQ pulsars, which are otherwise generally not possible to measure using traditional methods mentioned above.

The aim of this study is to further investigate and use the FP relation to predict unknown RQ pulsar distances with various machine learning (ML) methods, together with providing proof of principle with robust statistical techniques. The paper is structured as follows: Section~\ref{data_analysis} presents the 3PC data and details the derivation of spectral cutoff energies. In Sect.~\ref{fp_revisited}, the FP relation is revisited and further investigated using the 3PC data. Section~\ref{sect_4} details the proof of principle of pulsar distance estimation employing the FP relation with various ML approaches. Section~\ref{pred_results} presents the first-ever distance predictions for 62 RQ gamma-ray pulsars in 3PC, comparing the results across different methods and providing a cross-check using RQ pulsars with known distances. Finally, Section~\ref{conc} provides a summary and conclusions.

\section{Data analysis}
\label{data_analysis}
\begin{figure}
\centering
\includegraphics[width=9.4cm]{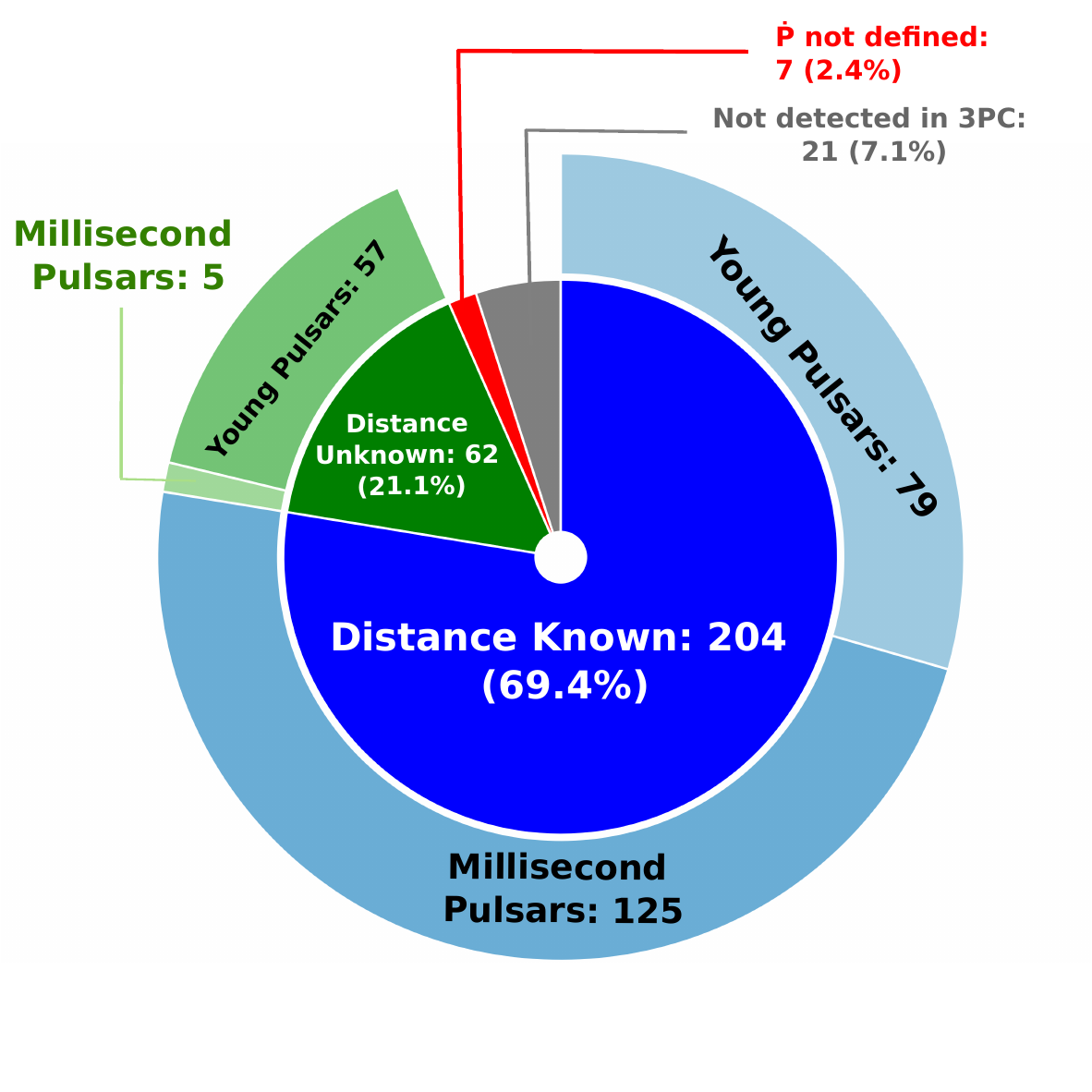}
\caption{The nested pie chart diagram of all pulsars included in the 3PC data. The inner circle shows the numbers and fraction of pulsars with known distances measured from various traditional techniques (blue shaded area), unknown distance (green shaded area), no significant detection in 3PC (gray shaded area), and missing $\dot{\text{P}}$ information (red shaded area). The outer circle categorizes pulsars with known and unknown distances into MSPs and young pulsars, represented by light-blue and light-green shaded regions, respectively.}
\label{3pc_chart}
\end{figure}

The 3PC includes comprehensive spectral information for a total of 294 gamma-ray pulsars identified in the LAT data. As illustrated in Fig.~\ref{3pc_chart}, 21 of these pulsars ($\sim$7.1$\%$) do not exhibit detectable gamma-ray emission (TS$<$25) and thus lack spectral information, while 7 of them ($\sim$2.4$\%$) have no $\dot{\text{P}}$ information available. Consequently, these 28 pulsars are excluded from the analysis presented in this paper (see Table~\ref{excluded_pulsar_list} for the list of excluded pulsars), as they cannot be used in the FP relation. From the remaining 266 pulsars detected significantly in 3PC, 204 ($\sim$69.4$\%$) have known distance estimates obtained through various traditional techniques, while the remaining 62 ($\sim$21.1$\%$) lack distance estimations owing to their RQ nature. Prominently, the majority of pulsars with known distances (125 of 204) are MSPs, and only a small fraction (5 of 62) of those with unknown distances are MSPs, which is expected owing to their small light cylinder radii increasing their visibility. It is important to note that this study exclusively uses pulsars with distances measured from traditional methods and provided in the 3PC data set. For instance, distance estimates for young pulsars\footnote{Throughout this paper, ``young pulsar'' refers to the pulsars that are not MSPs.} derived by~\cite{wang_2011} using the linear correlation between pulsar gamma-ray efficiencies ($\gamma$~=~L$_{\gamma}$/$\dot{\text{E}}$) and six pulsar parameters$-$period (P), characteristic age ($\tau$), magnetic field strength at the light cylinder (B$_{\text{LC}}$) and three generation order parameters ($\zeta_{1}$, $\zeta_{2}$, $\zeta_{3}$)$-$are excluded from the FP analysis\footnote{In contrast, these pulsars were included in the list of 190 pulsars analyzed in \cite{fp2022}. It is important to note that these three generation order parameters, $\zeta_{1}$, $\zeta_{2}$, $\zeta_{3}$, are derived assuming PC models~\citep{wang_2011}, which is disfavored by the 3PC data set.}. Instead, these pulsars are included in the list of pulsars with unknown distances, as their data could potentially bias the determination of the FP relation investigated in this study.

The study of FP relation requires the use of direct pulsar observables P and $\dot{\text{P}}$. Essentially, the spin-down luminosity, $\dot{\text{E}}$~=~4$\pi^{2}$I$_{0}\dot{\text{P}}$/P$^{3}$, and the magnetic field at the star surface, $\text{B}_{*}$~=~(1.5I$_{0}$c$^{3}$P$\dot{\text{P}}$)$^{1/2}$/(2$\pi$R$_{\text{NS}}^{3}$), are derived quantities and dependent on both P and $\dot{\text{P}}$. Evidently, proper motion of pulsars in space leads to an apparent change in the spin-down rate due to a kinematically induced Doppler shift, known as the Shklovskii effect~\citep{Shklovskii}, and should be corrected. The 3PC provides Shklovskii corrections applied to $\dot{\text{P}}$ values, as well as the corresponding corrected $\dot{\text{E}}$ values, wherever possible. For pulsars lacking this correction information, the observed $\dot{\text{P}}$ values are used instead. Additionally, the $\text{B}_{*}$ values given in 3PC do not account for this correction. Therefore, $\text{B}_{*}$ values used in the analysis presented in this study are corrected using the available Shklovskii-corrected $\dot{\text{P}}$ information provided in the 3PC data set.

\begin{table}[ht!]
\small
        \caption{List of pulsars excluded from the FP analysis presented in this study owing to their insufficient data in 3PC.}
        \centering
\begin{tabular}{cc}
        \hline\hline
        Pulsar Name (PSR) & Exclusion Reason \\
        \hline\hline
        J0117+5914    &  Not detected in 3PC (TS $<$ 25) \\
        J0139+5814    &  Not detected in 3PC (TS $<$ 25) \\
        J0154+1833    &  Not detected in 3PC (TS $<$ 25) \\
        J0636+5128    &  Not detected in 3PC (TS $<$ 25) \\
        J0729$-$1836  &  Not detected in 3PC (TS $<$ 25) \\
        J0737$-$3039A &  Not detected in 3PC (TS $<$ 25) \\
        J0834$-$4159  &   Not detected in 3PC (TS $<$ 25) \\
        J0955$-$3947  &  $\dot{\text{P}}$ is not defined \\
        J1224$-$6407  &  Not detected in 3PC (TS $<$ 25) \\
        J1259$-$8148  &  $\dot{\text{P}}$ is not defined \\
        J1306$-$6043  &  $\dot{\text{P}}$ is not defined \\
        J1327$-$0755  &  Not detected in 3PC (TS $<$ 25) \\
        J1513$-$5908  &  Not detected in 3PC (TS $<$ 25) \\
        J1623$-$6936  &  $\dot{\text{P}}$ is not defined \\
        J1646$-$4346  &  Not detected in 3PC (TS $<$ 25) \\
        J1748$-$2815  &  Not detected in 3PC (TS $<$ 25) \\
        J1757$-$6032  &  $\dot{\text{P}}$ is not defined \\
        J1816$-$0755  &  Not detected in 3PC (TS $<$ 25) \\
        J1832$-$0836  &  Not detected in 3PC (TS $<$ 25) \\
        J1835$-$1106  &  Not detected in 3PC (TS $<$ 25) \\
        J1841$-$0524  &  Not detected in 3PC (TS $<$ 25) \\
        J1856+0113    &  Not detected in 3PC (TS $<$ 25) \\
        J1858$-$5422  &  $\dot{\text{P}}$ is not defined \\
        J1909$-$3744  &  Not detected in 3PC (TS $<$ 25) \\
        J1928+1746    &  Not detected in 3PC (TS $<$ 25) \\
        J1946+3417    &  Not detected in 3PC (TS $<$ 25) \\
        J1954+3852    &  Not detected in 3PC (TS $<$ 25) \\
        J2029$-$4239  &  $\dot{\text{P}}$ is not defined\\
        \hline
\end{tabular}
\label{excluded_pulsar_list}
\end{table}

\subsection{Determination of Spectral Cutoff Energies}

The theoretical FP relation exhibits a strong dependence on the $\epsilon_\mathrm{cut}$, as evident from Equations~\ref{fp_der} and \ref{fp_obs}, due to the significant influence of the exponent of this variable on L$_\gamma$. Conceptually, the $\epsilon_\mathrm{cut}$ represents the maximum energy of gamma rays that can be effectively produced, thus intrinsically connected to the maximum energy of accelerated particles in the pulsar environment. Therefore, robust determination of the $\epsilon_\mathrm{cut}$ parameter is crucial for a reliable investigation of the FP relation. The analysis of the 3PC data has revealed that the majority of gamma-ray pulsars exhibit spectral peaks around E$_{\text{p}}$=$\sim$1.5~GeV~\citep{fermi3pc}. These characteristic peaked spectra are typically indicative of synchrotron emission. However, in certain pulsars, higher-energy emissions might originate from the IC mechanism. Consequently, the resultant observed spectra can be a superposition of these two distinct emission mechanisms, manifesting a spectral hardening at the high-energy part of the spectrum.

The spectral data for phase-averaged spectra of individual pulsars are provided in 3PC\footnote{The \url{https://fermi.gsfc.nasa.gov/ssc/data/access/lat/3rd_PSR_catalog/3PC_SEDPlotter.py} script provided by the Fermi Collaboration is used for extracting SED flux points and model data.}, both as spectral energy distribution (SED) data points and as best-fit parameters of the PLEC4 model defined as
\begin{equation}
\label{plec4}
\Phi_\mathrm{PLEC4}(E)=\text{N}_{0} \left(\frac{E}{E_{0}}\right)^{-\Gamma_{0} + \frac{d}{b}} \exp{\Bigg[ \frac{d}{b^{2}} \Bigg( 1 - \bigg(\frac{E}{E_{0}}\bigg)^{b} \Bigg)  \Bigg]}\,\mathrm{,}
\end{equation}
where N$_{0}$ is the flux density at the reference energy E$_{0}$, $\Gamma_{0}$ is the spectral index at E$_{0}$, d is the local curvature at E$_{0}$, and b is the index describing the sharpness of the exponential drop. Specifically, the d parameter controls the width of the peak seen in the spectrum, while the b parameter is responsible for the level of asymmetry. It is important to note that the d=0 case corresponds to the nested pure power-law (PL) spectral model with $\Gamma$=$\Gamma_{0}$. As was extensively discussed in \cite{fp2022}, the PLEC4 spectral model given in Eq.\ref{plec4} does not provide a spectral parameter that can be used for characterizing the high-energy part of the pulsar spectra, and therefore can not be used in the FP relation. Instead, the PLEC model, defined as 
\begin{equation}
\label{ecpl}
\Phi_\mathrm{PLEC}(E)=\Phi_{0}\mathrm{(E_{0})} \left(\frac{E}{E_{0}}\right)^{-\Gamma} \exp{\bigg[-\bigg(\lambda E \bigg)^{b}}\bigg]\mathrm{,}
\end{equation}
where $\lambda$=(1/$\epsilon_\mathrm{cut}$) is the inverse $\gamma$-ray cutoff energy, provides a better-suited parameter characterizing the highest-energy pulsar emission. Similarly, setting $\lambda$=0 reduces Eq.~\ref{ecpl} to a nested pure PL spectral model. Comparing Equations~\ref{plec4} and~\ref{ecpl}, one can interconnect the spectral parameters between the PLEC4 and PLEC models as 
\begin{equation}
\label{index_conn}
\Gamma = \Gamma_{0} - \frac{d}{b} 
\end{equation}
\begin{equation}
\label{ecut_conn}
\epsilon_\mathrm{cut,b} = \bigg(\frac{d}{b^{2}}\bigg)^{1/b} E_{0},
\end{equation}
while the peak energy E$_{\text{p}}$ values provided in 3PC can be obtained from the relation
\begin{equation}
\label{epeak_def}
E_{p} = E_{0} \bigg[\frac{d}{b} \left( 2 - \Gamma \right) \bigg]^{1/b}.
\end{equation}

In principle, the phase-averaged spectra observed in pulsars represent a superposition of multiple spectra originating from different magnetospheric subregions within the pulsar environment. Each of these subregion emits radiation characterized by either monoenergetic curvature or synchrotron emission, which can be effectively described by the PLEC model given in Eq.\ref{ecpl} with b=1. However, a more precise description of the phase-averaged spectra is often achieved using the subexponential cutoff model, characterized by b$<$1, as this model represents superposition of multiple exponential spectra with varying cutoff energies~\citep{Abdo_2010}. Although the spectral shapes predicted by the PLEC and PLEC4 models are basically identical, the preference for using the PLEC4 model in 3PC is due to its ability to reduce parameter degeneracies and provide smaller uncertainties compared to the PLEC model~\citep{fgl4_dr3, fermi3pc}. In Fermi 3PC, the PLEC4 spectral model parameters are provided with a fixed b=2/3 (b$_{23}$ model) for pulsars with low detection significance, while the b parameter is left free (b$_{\text{fr}}$ model) for bright gamma-ray pulsars. It has been noted that the b$_{23}$ model can potentially introduce biases, making results from the b$_{\text{fr}}$ models more robust and preferable whenever available~\citep{fermi3pc}. Essentially, in the pure PLEC model with a fixed parameter b=1, the $\epsilon_\mathrm{cut}$ parameter determines the energy level at which a spectrum begins to decline rapidly, thereby indicating the peak energy of the SED. However, deviations from b=1, particularly when b decreases as in the b$_{23}$ model, cause $\epsilon_\mathrm{cut}$ to lose its ability to accurately represent the peak energy. This effect has been extensively discussed and demonstrated through simulations of synthetic phase-averaged spectra in \cite{fp2022}. It was concluded that although the subexponential model function may offer a more accurate description of the SED, the cutoff energy defined by the pure PLEC model is often a more reliable indicator of peak energy and thus is more appropriate for use in the FP relation.

\begin{figure*}[ht!]
\centering
\includegraphics[width=18.0cm]{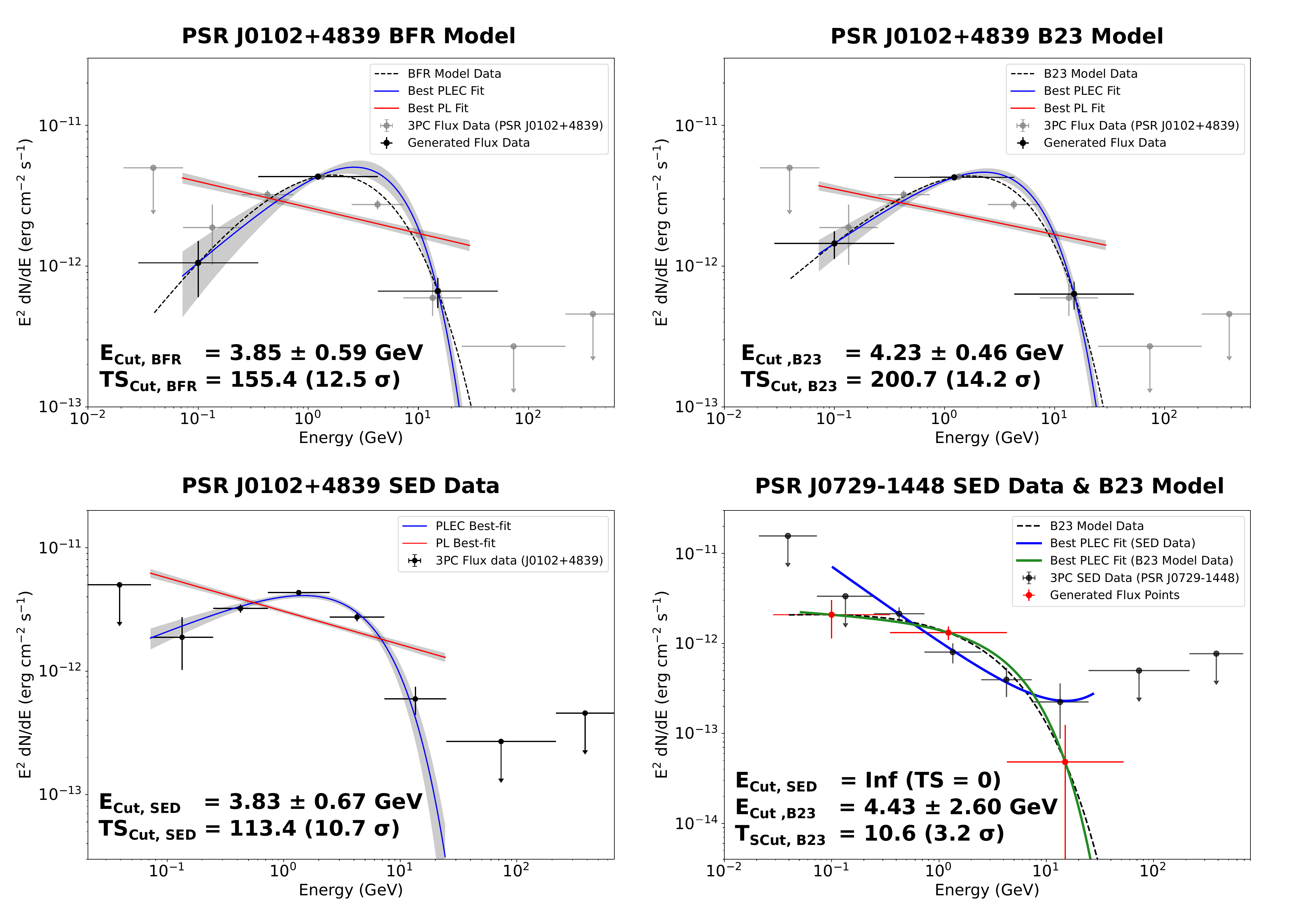}
\caption{Examples of determining spectral cutoff energies. Top panels: the synthetic SED data points of the pulsar PSR~J0102$+$4839 generated using PLEC4 b$_{\text{fr}}$ (left) and b$_{23}$ (right) models are shown in black, together with corresponding models shown with black dashed lines. The best-fit PLEC and pure PL models are shown with blue and red lines, respectively, along with their 1$\sigma$ error bands (gray shaded regions). Observational Fermi-LAT SED data of the pulsar are represented by gray points and upper limits. Bottom left panel: the best-fit PLEC (blue) and PL (red) models obtained from the fit to observational SED data (black points) of the pulsar PSR~J0102$+$4839. Bottom right panel: the observational SED data (black) and the PLEC4 b$_{23}$ synthetic SED data points (red) of the pulsar PSR~0729$-$1448. The best-fit PLEC models from the SED and synthetic data are indicated by blue and green lines, respectively, while the dashed black line represents the PLEC4 b$_{23}$ model provided in 3PC. Each figure provides the best-fit $\epsilon_\mathrm{cut}$ parameter and the corresponding test statistic values.}
\label{cutoffsignif}
\end{figure*}

In this study, the $\epsilon_{\text{cut}}$ parameter for each individual pulsar is determined using two independent methods based on the 3PC data for a consistency cross-check. The first method involves fitting the SED data of pulsars to a pure PLEC model, with the parameter b fixed at 1, which allows for a direct determination of the $\epsilon_\mathrm{cut}$ parameter from observational Fermi data; these values are referred to as $\epsilon_{\text{cut,SED}}$. In the second method, a similar approach to that used in \cite{fp2022} was employed. The previously discussed PLEC4 b$_{23}$ and b$_{\text{fr}}$ models for individual pulsars were reconstructed over the energy range of 0.1$-$15~GeV using the best-fit parameters provided in 3PC. To represent the spectral shapes of these models, three synthetic SED data points were generated at energies evenly spaced on a logarithmic scale along the respective PLEC4 curves (i.e., see the black data points in the top panels of Fig.~\ref{cutoffsignif}). The choice of three SED points was motivated to prevent overrepresenting the spectral curvature, as including more points could potentially exaggerate deviations from a simple PL shape. This approach ensures a balanced representation of the spectrum while minimizing artificial biases. Each synthetic data point includes 1$\sigma$ error bars, derived from the models' error bands, to account for the associated energy flux uncertainties. These synthetic SED data points were subsequently fitted to the pure PLEC model described in Eq.~\ref{ecpl} with b=1, to determine the $\epsilon_\mathrm{cut}$ parameter from both the b$_{23}$ model ($\epsilon_{\text{cut,b23}}$) and the b$_{\text{fr}}$ model ($\epsilon_{\text{cut,bfr}}$). It is important to point out that both of these methods allow the calculation of the statistical significance of the spectral cutoff feature (TS$_{\text{Cut}}$) through the application of likelihood ratio tests by comparing the PLEC and pure PL models as
\begin{equation}
\label{ts_lambda}
\mathrm{TS_{Cut}}=-2\ln\frac{\hat{L}(\lambda=0)}{\hat{L}(\lambda)},
\end{equation}
where $\hat{L}(\lambda_{\gamma})$ and $\hat{L}(\lambda_{\gamma}$=0) represent the maximum likelihoods over the full parameter space. The spectral fits presented throughout this study are performed using the {\tt Gammapy} Python package \citep{gammapy2023, gammapy_v11}. The cutoff values derived from the spectral analysis of SED data points, and from the b$_{23}$ and b$_{\text{fr}}$ PLEC4 models, along with the corresponding TS$_{\text{Cut}}$ values are provided in Appendix~\ref{A1}. Additionally, a hybrid cutoff value ($\epsilon_\mathrm{cut,HYB}$), which combines $\epsilon_{\text{cut,b23}}$ and $\epsilon_{\text{cut,bfr}}$, is defined. In instances where both the b$_{23}$ and b$_{\text{fr}}$ PLEC4 models are available, priority is given to the $\epsilon_{\text{cut,bfr}}$ values, following the recommendations outlined in 3PC. The derivation of the critical $\epsilon_\mathrm{cut}$ parameter of the FP relation using two distinct methods is motivated by two primary objectives. Firstly, employing independent methods allows cross-verification, ensuring that both approaches yield consistent results within the FP relation. Secondly, these methods allow the examination of the spectral curvature's influence on the FP relation and the determination of a reliable TS$_{\text{Cut}}$ threshold for its applications. Besides, it was discussed by \cite{fp2022} that the parameter $\epsilon_{10}$, defined as the energy value at which the SED power reaches 1/10 of its maximum, is closely correlated with $\epsilon_\mathrm{cut}$ and can therefore be substituted for $\epsilon_\mathrm{cut}$ in the FP relation. However, the $\epsilon_{10}$ parameter alone does not provide information about the statistical significance of the spectral cutoff. Indeed, to assess the cutoff significance, one of the methods mentioned above must be employed. Consequently, $\epsilon_{10}$ values are not used in this study.

Figure~\ref{cutoffsignif} presents an example spectral analysis of the pulsar PSR~J0102+4839. The top panels demonstrate the analysis of the pulsar’s synthetic SED points generated from b$_{\text{fr}}$ (left) and b$_{23}$ (right) PLEC4 models. For comparison, the 3PC SED data analysis for the same pulsar is shown in the bottom left panel. Similar to the PSR~J0102+4839 example given in Fig.~\ref{cutoffsignif}, the majority of pulsars listed in 3PC exhibit significant spectral cutoff features with TS$_{\text{Cut}}\geq$9.0 (equivalent to a 3$\sigma$ level) in both their SED data and PLEC4 models. However, as is demonstrated for the PSR~J0729$-$1448 case shown in the bottom right panel, there are several pulsars where spectral hardening is evident in the SED data. For such pulsars, the optimized best-fit PLEC model (blue line) results in negative $\epsilon_\mathrm{cut}$ values, indicating an upward-curved spectrum. This feature can be interpreted as indicative of an additional spectral component at high energies, potentially arising from emerging very high energy (VHE; E$>$100~GeV) emission that originated from IC scattering mechanisms. Due to the inability to derive a reliable spectral cutoff energy that can be used in the FP relation from such pulsars without prior knowledge of the precise second high-energy component, the corresponding values for TS$_{\text{Cut}}$ and $\epsilon_\mathrm{cut}$ are assigned as zero and infinity, respectively. In contrast to SED data, b$_{23}$ (and/or b$_{\text{fr}}$) PLEC4 models provided in 3PC can exhibit significant cutoff structures, as demonstrated by the example of PSR~J0729$-$1448 (b$_{23}$ model, green line). Table~\ref{cutoff_table} provides a summary of the number of 3PC pulsars for which $\epsilon_\mathrm{cut}$ values have been derived using the methods described above. As detailed in the table, 3PC SED data are available for all 266 pulsars. Among these, 210 pulsars exhibit significant spectral cutoffs, while 17 display an upward curvature in their spectra. The PLEC b$_{23}$ model parameters are provided for 255 pulsars, with 243 showing significant cutoffs, whereas b$_{\text{fr}}$ model parameters are available for 116 pulsars, all of which exhibit significant cutoffs. It is important to note that the numbers of pulsars classified as $\epsilon_{\text{cut,HYB}}$ are equivalent to those classified as $\epsilon_{\text{cut,b23}}$, with spectral cutoff values being substituted with $\epsilon_{\text{cut,bfr}}$ where available. 

\begin{table}
\footnotesize
\caption{Overview of the number of spectral cutoff values determined in this study. The first column indicates the method used: 3PC SED data ($\epsilon_{\text{cut,SED}}$), or synthetic SED data from PLEC4 b$_{\text{fr}}$ ($\epsilon_{\text{cut,bfr}}$) and b$_{23}$ ($\epsilon_{\text{cut,b23}}$) models. $\epsilon_{\text{cut,HYB}}$ combines  $\epsilon_{\text{cut,bfr}}$ and $\epsilon_{\text{cut,b23}}$, replacing $\epsilon_{\text{cut,bfr}}$ with $\epsilon_{\text{cut,b23}}$ where available. The TS$_{\text{Cut}}\geq$9.0 and TS$_{\text{Cut}}<$9.0 columns give the total number of significant ($\geq$3$\sigma$) and insignificant ($<$3$\sigma$) cutoffs, respectively, while the TS$_{\text{Cut}}$=0 column gives the number of spectra with upward-curved PLEC models. Numbers in parentheses indicate spectral cutoffs for pulsars with known distances.}
\centering
\renewcommand{\arraystretch}{1.3}
\begin{tabular}{cccccc}
\hline\hline
Cutoff Method & Total Number & TS$_{\text{Cut}}\geq$ 9.0 & TS$_{\text{Cut}}<$ 9.0 & TS$_{\text{Cut}}$ = 0 \\
\hline
$\epsilon_{\text{cut,SED}}$ & 266 (204) & 210 (150) & 39 (54) & 17 (0) \\
$\epsilon_{\text{cut,bfr}}$ & 116 (80)  & 116 (80)         & 0       & 0  \\
$\epsilon_{\text{cut,b23}}$ & 255 (193) & 243 (181) & 12 (12) & 0  \\
$\epsilon_{\text{cut,HYB}}$ & 255 (193) & 243 (181) & 12 (12) & 0  \\
\hline
\end{tabular}
\label{cutoff_table}
\end{table}

\begin{figure*}
\centering
\includegraphics[width=18.0cm]{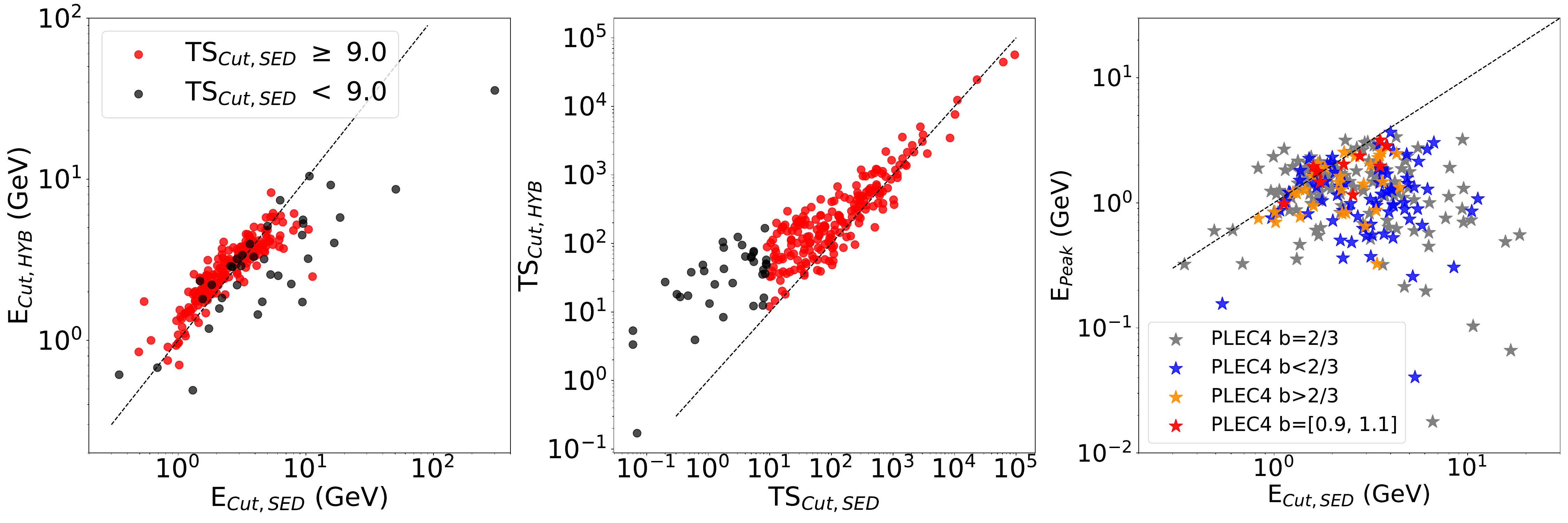}
\caption{The scatter plot of $\epsilon_{\text{cut}}$ values obtained from 3PC SED data and synthetic PLEC4 model SED data is shown in the left panel, while the middle panel shows the scatter of corresponding TS$_{\text{Cut}}$ values. The red and black circles show the significant ($\geq$3$\sigma$) and insignificant ($<$3$\sigma$) spectral cutoff features. The right panel shows the comparison of $\epsilon_{\text{cut,SED}}$ values obtained from spectral analysis of pulsars and corresponding E$_{\text{Peak}}$ values provided in 3PC. The E$_{\text{Peak}}$ values calculated for a fixed b value of 2/3 (b$_{23}$ models) are indicated with gray stars, while the blue and orange stars mark the E$_{\text{Peak}}$ values obtained from the b$_{\text{fr}}$ case for b$<$2/3 and b$>$2/3, respectively. In addition, b values scattered around b=1.0$\pm$0.1 are shown with red stars. The dashed lines indicate the reference y=x line in all plots.} 
\label{ec_ts_comparison}
\end{figure*}

The left panel of Fig.~\ref{ec_ts_comparison} presents a comparison between the derived $\epsilon_{\text{cut,SED}}$ and $\epsilon_{\text{cut,HYB}}$, while the middle panel shows the comparison of the corresponding TS$_{\text{Cut}}$ values. The compatibility between $\epsilon_{\text{cut,SED}}$ and $\epsilon_\mathrm{cut,HYB}$ suggests that cutoff values obtained from these two independent methods are exchangeable and applicable in the FP relation. It is important to point out that the majority of the outliers (indicated by black circles) originate from insignificant spectral cutoff features with TS$_{\text{Cut}}<$9.0. The PLEC4 models used in this study to derive $\epsilon_{\text{cut,HYB}}$ are provided in the current Fermi catalogs. For pulsars detected in the future, $\epsilon_{\text{cut,SED}}$ values can be determined by fitting the pulsar spectra to PLEC models (with $\beta$ fixed to 1), and can directly be used in the FP relation. This approach is justified by the exchangeability feature between $\epsilon_{\text{cut,HYB}}$ and $\epsilon_{\text{cut,SED}}$ demonstrated in Fig. 3 (left panel), provided that the spectral cutoffs are statistically significant. Evidently, a noticeable pattern emerges when comparing corresponding TS$_{\text{Cut}}$ values. At high TS$_{\text{Cut}}$ values, the $\epsilon_\mathrm{cut,HYB}$ values closely align with the $\epsilon_{\text{cut,SED}}$ values, indicating a strong correlation, while the divergence at low TS$_{\text{Cut}}$ values points to a systematic discrepancy, possibly due to the PLEC4 b$_{23}$ modeling of the faint pulsars (see \cite{fgl4_dr3} for the modeling details) provided in 3PC. The right panel of Fig.~\ref{ec_ts_comparison} compares derived $\epsilon_{\text{cut,SED}}$ and E$_{\text{Peak}}$ provided in the 3PC data, revealing a wide spread and lack of clear correlation. Specifically, the E$_{\text{Peak}}$ values corresponding to small b exhibit significant deviations from the $\epsilon_{\text{cut,SED}}$ values (see, e.g., blue circles for b$<$2/3). For comparison, the E$_{\text{Peak}}$ values corresponding to b=1, which scatter around the y=x line, are shown in red. These results further support the discussions in \cite{fp2022} that the E$_{\text{Peak}}$ parameter cannot effectively probe the maximum energy of accelerated particles, and therefore it is not appropriate to use in the FP relation.

\section{Investigation of the Fundamental Plane Relation using 3PC Data}
\label{fp_revisited}

The luminosity functions given in Equations \ref{fp_der} and \ref{fp_obs} for the CR regime can be written in loglinear form by taking the logarithm of both sides as
\begin{equation}
\label{fp_log_der}
\text{Log}(L_\gamma) = \zeta + \beta\text{Log}(\epsilon_\mathrm{cut}) + \xi\text{Log}(\text{B}_{*}) + \mu\text{Log}(\dot{\text{E}}),
\end{equation}
and 
\begin{equation}
\label{fp_log_obs}
\text{Log}(L_\gamma) = \alpha + \beta \text{Log}(\epsilon_\mathrm{cut}) + \gamma\text{Log}(\text{P}) + \delta\text{Log}(\dot{\text{P}}) 
\end{equation}
respectively, which allows application of linear regression methods to derive the exponent constants of the FP relation. Note that the exponent of the $\epsilon_\mathrm{cut}$ parameter ($\beta$) is identical for both expressions. The ordinary least-squares (OLS) method, a fundamental approach in linear regression that minimizes the sum of squared differences between observed and predicted values, is employed to estimate the exponents of the FP relation using the {\tt statsmodels}\footnote{\url{https://www.statsmodels.org/stable/index.html}} Python package. 

\begin{table*}
\caption{The FP relation exponential constants obtained from OLS regression. Note that all the pulsars included in these analyses have known distances. The first column indicates the $\epsilon_{\text{cut}}$ value used in the FP relation, while the second column shows the TS$_{\text{Cut}}$ condition applied to the analysis data set. The third column gives the remaining number of pulsars in the FP analysis satisfying the cut condition. The best-fit FP relation exponent values for Eq.~\ref{fp_log_obs} constant term ($\alpha$), $\epsilon_\mathrm{cut}$ ($\beta$), P ($\gamma$), $\dot{\text{P}}$ ($\delta$) and for Eq.~\ref{fp_log_der}, constant term ($\zeta$), B$_{*}$ ($\xi$) and $\dot{\text{E}}$ ($\mu$) are provided in respective columns together with their 1$\sigma$ errors. Note that $\beta$ values are identical for FP relations given in Equations~\ref{fp_log_der} and \ref{fp_log_obs}. The rows given in bold correspond to the TS$_{\text{Cut}}$ threshold shown in Fig.~\ref{exp_profiles} (see the top panel) indicated with dark-gray shaded regions, showing a strong deviation trend from expected theoretical values. The expected theoretical FP exponent values, $\beta$=4/3~(1.33), $\gamma$=-7/6~(-1.17), $\delta$=1/2~(0.5), $\xi$=1/6~(0.17), and $\mu$=5/12~(0.42), are provided for comparison.}
\centering
\footnotesize
\renewcommand{\arraystretch}{1.5}

\begin{tabular}{ccc|c|ccc|ccc|c}
\hline\hline
$\epsilon_{\text{cut}}$ & Cut & Number of &  $\beta$ & $\alpha$ & $\gamma$ & $\delta$ & $\zeta$ & $\xi$ & $\mu$ & Adj.~R$^{2}$\\
Method & Condition & Pulsars & ($\epsilon_\mathrm{cut}$) & (Const.) & (P) & ($\dot{\text{P}}$) & (Const.) & (B$_{*}$) & $(\dot{\text{E}}$) \\
\hline\hline

$\epsilon_{\text{cut,HYB}}$ & TS$_{\text{Cut}}\geq$9.0 & 181 & 1.31$\pm$0.19 & 39.48$\pm$0.51 & $-$1.22$\pm$0.17 & 0.49$\pm$0.04 & 17.26$\pm$1.63 & 0.12$\pm$0.03 & 0.43$\pm$0.05 & 0.730\\

$\textbf{$\epsilon_{\text{cut,HYB}}$}$ & $\textbf{TS$_{\text{Cut}}\geq$4.0}$ & $\textbf{184}$ & $\textbf{1.34$\pm$0.19}$ & $\textbf{39.42$\pm$0.50}$ & $\textbf{$-$1.21$\pm$0.17}$ & $\textbf{0.48$\pm$0.04}$ & $\textbf{17.35$\pm$1.63}$ & $\textbf{0.12$\pm$0.03}$ & $\textbf{0.42$\pm$0.05}$ & $\textbf{0.728}$\\

$\epsilon_{\text{cut,HYB}}$ & TS$_{\text{Cut}}\geq$1.0 & 190 & 1.05$\pm$0.17 & 40.04$\pm$0.49 & $-$1.44$\pm$0.17 & 0.54$\pm$0.04 & 15.33$\pm$1.59 & 0.09$\pm$0.03 & 0.49$\pm$0.05 & 0.714\\

$\epsilon_{\text{cut,HYB}}$ & TS$_{\text{Cut}}>$0 & 193 & 0.89$\pm$0.15 & 40.24$\pm$0.48 & $-$1.51$\pm$0.16 & 0.55$\pm$0.04 & 14.72$\pm$1.56 & 0.07$\pm$0.03 & 0.52$\pm$0.05 & 0.708\\
\hline

$\epsilon_{\text{cut,SED}}$ & TS$_{\text{Cut}}\geq$9.0 & 150 & 1.56$\pm$0.19 & 38.27$\pm$0.57 & $-$0.85$\pm$0.19 & 0.39$\pm$0.05 & 21.19$\pm$1.86 & 0.14$\pm$0.03 & 0.31$\pm$0.06 & 0.737 \\

$\epsilon_{\text{cut,SED}}$ & TS$_{\text{Cut}}\geq$4.0 & 165 & 1.47$\pm$0.18 & 38.58$\pm$0.54 & $-$0.98$\pm$0.18 & 0.41$\pm$0.05 & 20.00$\pm$1.73 & 0.12$\pm$0.03 & 0.35$\pm$0.06 & 0.731 \\

$\textbf{$\epsilon_{\text{cut,SED}}$}$ & $\textbf{TS$_{\text{Cut}}\geq$1.0}$ & $\textbf{175}$ & $\textbf{1.38$\pm$0.17}$ & $\textbf{38.54$\pm$0.53}$ & $\textbf{$-$0.99$\pm$0.18}$ & $\textbf{0.41$\pm$0.05}$ & $\textbf{20.05$\pm$1.70}$ & $\textbf{0.12$\pm$0.03}$ & $\textbf{0.35$\pm$0.06}$ & $\textbf{0.729}$\\

$\epsilon_{\text{cut,SED}}$ & TS$_{\text{Cut}}\geq$0 & 187 & 0.60$\pm$0.14 & 40.20$\pm$0.54 & $-$1.47$\pm$0.18 & 0.54$\pm$0.05 & 15.39$\pm$1.76 & 0.07$\pm$0.03 & 0.50$\pm$0.06 & 0.680 \\
\hline
E$_{\text{Peak}}$ & $-$ & 189 & 0.32$\pm$0.14 & 41.18$\pm$0.48 & $-$2.00$\pm$0.17 & 0.70$\pm$0.05 & 8.95$\pm$1.88 & 0.04$\pm$0.03 & 0.68$\pm$0.05 & 0.659 \\
\hline\hline
\end{tabular}
\label{fp_table}
\end{table*}

\begin{figure*}
\centering
\includegraphics[width=18.0cm]{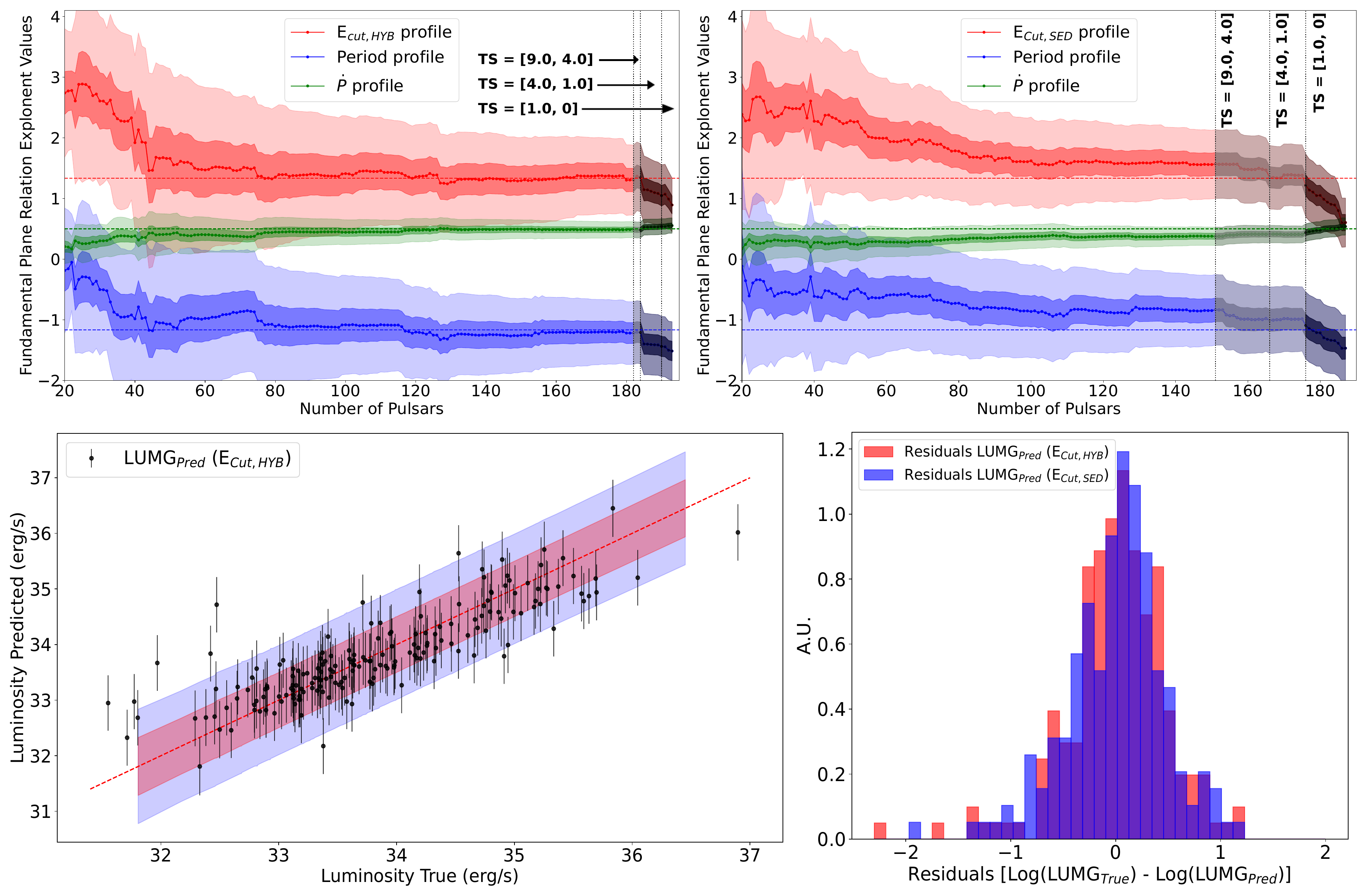}
\caption{
The diagnostic comparison for the performed FP OLS regression analysis. The top left panel displays the profiles of FP exponents ($\epsilon_{\text{cut,HYB}}$, P, $\dot{\text{P}}$) as a function of the increasing number of pulsars included in the analysis, while the top right panel shows these profiles when $\epsilon_{\text{cut,SED}}$ is used instead of $\epsilon_{\text{cut,HYB}}$. The profiles of $\epsilon_{\text{cut}}$, P, $\dot{\text{P}}$ are represented by red, blue, and green colors, respectively, with corresponding 1$\sigma$ (dark-shaded) and 3$\sigma$ (light-shaded) error bands. The expected theoretical values of these parameters are indicated by horizontal dashed lines with the same colors. The light-gray shaded regions highlight the profile areas where pulsars with TS$_{\text{Cut}}<9.0$ are added to the analysis data set, whereas the dark-gray shaded regions indicate areas exhibiting a strong deviation trend. The botton left panel presents a scatter plot comparing true luminosities with those predicted by the FP relation using the $\epsilon_{\text{cut,HYB}}$ parameter with TS$_{\text{Cut}}\geq$4.0, in logarithmic space. The red and blue shaded regions represent 1$\sigma$ and 3$\sigma$ OLS confidence intervals, respectively, with the dashed red line indicating the reference y=x line. The bottom right panel displays the residuals between true and predicted gamma-ray luminosities, with red and blue distributions corresponding to the $\epsilon_{\text{cut,HYB}}$ and $\epsilon_{\text{cut,SED}}$ parameters, respectively.} 
\label{exp_profiles}
\end{figure*}

Table~\ref{fp_table} lists the best-fit exponent values of the FP relation, presenting a comparison between various cut conditions applied on the significance of $\epsilon_\mathrm{cut}$ parameters obtained from different approaches. As is evident from the table, the best-fit values of all the FP relation exponents are compatible with the theoretical expectations within the $\sim$1$\sigma$ level when the $\epsilon_{\text{cut,HYB}}$ spectral cutoff feature is significant (TS$_{\text{Cut}}\geq$9.0) or marginally significant (TS$_{\text{Cut}}\geq$4.0). On the other hand, the best-fit exponent values start to diverge from the theoretical expectations when the pulsars with insignificant cutoff features (TS$_{\text{Cut}}<$4.0) are added into the analysis data set. The profiles of the FP exponents, as a function of increasing number of pulsar sample, are shown in the top panel of Fig.~\ref{exp_profiles}. As an independent cross-check, a similar trend can be observed when using the spectral cutoff energy derived from the pulsar SED ($\epsilon_{\text{cut,SED}}$). In this case, the strong deviation from the theoretically expected values is observed when TS$_{\text{Cut}}<$1.0. This reduction of the TS$_{\text{Cut}}$ threshold, in the case when $\epsilon_{\text{cut,SED}}$ is employed, is due to the fact that the TS$_{\text{Cut}}$ values obtained for $\epsilon_{\text{cut,HYB}}$ are positively biased, especially for the low TS values (see the middle panel of Fig.~\ref{ec_ts_comparison}). 

The investigation of the impact of TS$_{\text{Cut}}$ on the FP relation performed in this subsection reveals an important property of the FP relation. The results presented in Table~\ref{fp_table} and visualized in Fig.~\ref{exp_profiles} prove that the FP relation is valid only when the observed spectral cutoff, $\epsilon_{\text{cut}}$, is significant. Due to the strong dependence of the FP relation on the $\epsilon_{\text{cut}}$ parameter, even the addition of a few pulsars with an insignificant cutoff can cause a significant deviation of estimated FP exponents from the expected theoretical values; consequently, TS$_{\text{Cut}}$ should be taken into account\footnote{Note that the effect of TS$_{\text{Cut}}$ becomes more prominent especially when the OLS approach, which weights all data points equally, is used. An alternative approach can be weighting the data depending on the TS$_{\text{Cut}}$ values.}. An additional test has been performed in order to investigate whether the E$_{\text{Peak}}$ parameter provided in 3PC can be used in the FP relation. The FP analysis using 189 gamma-ray pulsars, with E$_{\text{Peak}}$ values provided in the 3PC, has shown that the best-fit exponent values significantly deviate from the theoretical values at the level of 7.2$\sigma$ for E$_{\text{Peak}}$, 4.9$\sigma$ for P and 4.0$\sigma$ for $\dot{\text{P}}$ (see the last row of Table~\ref{fp_table}), further proving that the E$_{\text{Peak}}$ parameter does not fully represent the highest-energy particles in the pulsar. 

The adjusted R-squared\footnote{Adjusted R-squared is the modified R-squared, which takes into account the impact of the number of predictors in the model, and is given as Adj.~R$^{2}$~=~1–[((1–R$^{2}$)~$\times$~(n–1))/(n–k–1)], where n and k are the size of the data set and the number of independent variables in the model excluding the constant, respectively. Consequently, the use of the Adj.~R$^{2}$ performance metric ensures a fair model comparison by applying a penalty for the inclusion of additional parameters.} values (Adj.~R$^{2}$) given in Table~\ref{fp_table} show how well the pulsar data agree with the FP OLS regression model; consequently, they can provide a performance metric. All the FP models (with $\epsilon_{\text{cut,HYB}}$ and $\epsilon_{\text{cut,SED}}$) give adjusted R$^{2}$ values of $\sim$0.73, indicating that the FP relation provides a reasonable description of the pulsar data and 73$\%$ of the variations seen in the gamma-ray pulsar luminosities can be explained with the FP relation. The bottom left panel of Fig.~\ref{exp_profiles} shows the scatter plot of predicted gamma-ray luminosity values obtained from the FP relation using the $\epsilon_{\text{cut,HYB}}$ TS$_{\text{Cut}}\geq$4.0 data set and true luminosity, together with 68$\%$ (red shaded area) and 95$\%$ (blue shaded area) confidence intervals of the OLS model. The corresponding log-luminosity residual distribution of the prediction is given in the bottom right panel of Fig.~\ref{exp_profiles} in red, together with the residuals obtained using the $\epsilon_{\text{cut,SED}}$ TS$_{\text{Cut}}\geq$4.0 data set (blue) for comparison. As is evident, the predicted luminosities are scattered around the y=x line, and residual distributions of predictions are centered around zero. Clearly, there are some outliers in the data set, indicating that the performance of the FP models can actually be improved when these outliers are carefully identified.

The results presented in this subsection provide a revalidation of the FP relation using the pulsars listed in 3PC, and prove that the FP relation can be used for predicting the luminosities from observational pulsar properties with a reasonable accuracy. In addition, the importance of the cutoff significance is demonstrated, reaching the conclusion that only pulsars exhibiting significant (or marginally significant with TS$_{\text{Cut}}\geq$4.0) spectral cutoffs should be used in the FP relation. Based on the results provided in this subsection, $\epsilon_{\text{cut,HYB}}$ values of pulsars with TS$_{\text{Cut}}\geq$4.0 will be used as a main investigation data set in the following sections.
\subsection{Testing for the emission mechanisms and significance of Fundamental Plane variables}

\begin{table*}
\caption{The results of log-likelihood ratio tests conducted among the variables used in the FP relation. The first column identifies the data set, while the second column lists the variables included in the FP relation. The third column reports the corresponding log-likelihood values of the fits, and the fourth column presents the test statistic improvements between the respective data sets, along with the significance level of statistical improvement indicated in parentheses. The corresponding best-fit exponent values of the FP relations are provided in the fifth column, followed by the adjusted R$^{2}$ values of the fit given in the sixth column. The data set labeled as "A" in the first row, which includes only the $\epsilon_\mathrm{cut}$ variable, provides a base nested model for the comparative tests presented in the table.}
\centering
\small
\renewcommand{\arraystretch}{1.3}
\begin{tabular}{cccccc}
\hline\hline
Data Set & Fit Variables & Log-likelihood & TS Improvement & Best-fit & Adj.~R$^{2}$\\
 &  & log(L) & -2*$\Delta$(log(L)) & Exponent Values & \\
\hline\hline
A&($\epsilon_\mathrm{cut}$) & $-$222.48 & Base model (TS=0) & $\beta$ ($\epsilon_\mathrm{cut}$) = 2.16$\pm$0.27 & 0.263 \\
\hline\hline
B&($\epsilon_\mathrm{cut}$, $\dot{\text{E}}$) & $-$138.67 & 167.62 (12.9$\sigma$) & $\beta$ ($ \epsilon_\mathrm{cut}$) = 1.04$\pm$0.18 & 0.702 \\
& SR regime & & B $-$ A & $\mu$ ($\dot{\text{E}}$) = 0.60$\pm$0.04 & \\ 
\hline
C&($\epsilon_\mathrm{cut}$, $\dot{\text{E}}$, B$_{*}$) & $-$129.69 & 17.96 (4.2$\sigma$) & $\beta$ ($\epsilon_\mathrm{cut}$) = 1.34$\pm$0.19 & 0.728 \\
& CR regime & & C $-$ B & $\mu$ ($\dot{\text{E}}$) = 0.42$\pm$0.05 & \\ 
& & & & $\xi$ (B$_{*}$) = 0.12$\pm$0.03 & \\ 
\hline\hline
D&($\epsilon_\mathrm{cut}$, P) & $-$177.12 & 90.72 (9.5$\sigma$) & $\beta$ ($\epsilon_\mathrm{cut}$) = 2.38$\pm$0.21 & 0.548 \\
& & & D $-$ A & $\gamma$ (P) = 0.62$\pm$0.06 & \\ 
 \hline
E&($\epsilon_\mathrm{cut}$, P, $\dot{\text{P}}$) & $-$129.69 & 94.86 (9.7$\sigma$) & $\beta$ ($\epsilon_\mathrm{cut}$) = 1.34$\pm$0.19 & 0.728 \\
& & & E $-$ D & $\gamma$ (P) = $-$1.21$\pm$0.17 & \\ 
& & & & $\delta$ ($\dot{\text{P}}$) = 0.48$\pm$0.04 & \\ 
\hline\hline
F&($\epsilon_\mathrm{cut}$, P, $\dot{\text{P}}$, SigAvg) & $-$127.58 & 4.22 (2.0$\sigma$) & $\beta$ ($\epsilon_\mathrm{cut}$) = 1.21$\pm$0.20 & 0.733 \\
& & & F $-$ E & $\gamma$ (P) = $-$1.27$\pm$0.13 & \\ 
& & & & $\delta$ ($\dot{\text{P}}$) = 0.49$\pm$0.04 & \\ 
& & & & SigAvg = 0.19$\pm$0.09 & \\ 
\hline
G&($\epsilon_\mathrm{cut}$, P, $\dot{\text{P}}$, TS$_{\text{Cut}}$) & $-$127.81 & 3.76 (1.9$\sigma$) & $\beta$ ($\epsilon_\mathrm{cut}$) = 1.31$\pm$0.19 & 0.732 \\
& & & G $-$ E& $\gamma$ (P) = $-$1.25$\pm$0.17 & \\ 
& & & & $\delta$ ($\dot{\text{P}}$) = 0.49$\pm$0.04 & \\ 
& & & & TS$_{\text{Cut}}$ = 0.10$\pm$0.05 & \\ 
\hline\hline
\end{tabular}
\label{table_signif_test}
\end{table*}

The theoretical FP relations given in Equations~\ref{fp_der} and \ref{fp_obs} are derived under the assumption that the observed emission originates from either the CR regime or the SR regime~\citep{fp2019,fp2022}, thus resulting in different FP exponents depending on the emission regime. Although it was shown in the previous subsection that the estimated FP exponents are compatible with the theoretical values predicted by the CR regime, it is crucial to determine whether emission from the CR regime is significantly favored over the SR regime. The loglinear form of the FP relation given in Eq.~\ref{fp_log_der} allows identification of the preferred gamma-ray emission regime through a likelihood ratio test, since statistically the SR regime relation (L$_{\gamma}\propto\epsilon_\mathrm{cut}^{\beta}$$\dot{\text{E}}^{\mu}$) provides a nested model for the CR regime relation (L$_{\gamma}\propto\epsilon_\mathrm{cut}^{\beta}$B$_{*}^{\xi}$$\dot{\text{E}}^{\mu}$) when the exponent of B$_{*}$, namely $\xi$, is set to zero. Consequently, the likelihood ratio between the CR and SR regimes can be calculated by using a modified version of Eq.~\ref{ts_lambda}, where $\lambda$=0 is replaced with $\xi$=0. Similarly, $\hat{L}$($\xi$) and $\hat{L}$($\xi$=0) represent the maximum likelihoods obtained via the OLS. This approach not only allows for the quantification of the significance of fit improvement for each FP variable but also provides a framework to test for additional relevant variables.

This subsection presents the results of a systematic investigation into the significance of the variables used in the FP relation, using the data set including 184 pulsars exhibiting marginally significant cutoffs ($\epsilon_{\text{cut,HYB}}$, TS$_{\text{Cut}}\geq$4.0; see Table~\ref{fp_table}). Table~\ref{table_signif_test} summarizes the outcomes of likelihood ratio tests, with the base model (denoted as data set~A) including only $\epsilon_\mathrm{cut}$ as a fit variable. The addition of $\dot{\text{E}}$ into the FP relation, denoted as data set~B and corresponding to the SR regime, improves the fit with a significance level of 12.9$\sigma$ compared to data set A. Data set C, representing the CR regime and including an additional variable B$_{*}$ with respect to data set B, is also evaluated. The log-likelihood ratio test between data sets B and C shows an improvement of 4.2$\sigma$, leading to the conclusion that the CR regime is significantly preferred over the SR regime. Consequently, the preference of the gamma-ray emission originating from the CR regime rather than the SR regime, discussed in \cite{fp2022}, has been statistically proven using the pulsars listed in 3PC, providing a significance level on the preference. A similar investigation was performed for the FP relation involving the direct observables ($\epsilon_\mathrm{cut}$, P, $\dot{\text{P}}$). As evident from the table, addition of first P (data set~D), and subsequently $\dot{\text{P}}$ (data set~E) variables significantly improves fit performance at the 9.5$\sigma$ and 9.7$\sigma$ levels, respectively. It is important to note that both ($\epsilon_\mathrm{cut}$,~$\dot{\text{E}}$,~B$_{*}$) and ($\epsilon_\mathrm{cut}$,~P,~$\dot{\text{P}}$) FP relations yield identical log-likelihood values, indicating that they are equivalent in describing the data.

To evaluate potential improvements in model performance, one may consider including additional variables in data set~C (or equivalently data set E). Among these, the average statistical detection significance of pulsars, denoted as 'SigAvg' in the 3PC catalog, and the significance of the spectral cutoff feature (TS$_{\text{Cut}}$) are promising. As demonstrated in Table~\ref{table_signif_test}, the inclusion of SigAvg (data set~F) and TS$_{\text{Cut}}$ (data set~G) yields improvements in model performance at the 2.0$\sigma$ and 1.9$\sigma$ levels, respectively. Although these improvements are not statistically significant, they suggest marginal improvements and could be beneficial when applied to various ML algorithms, such as random forests, to enhance prediction performance. In addition, a model incorporating both TS$_{\text{Cut}}$ and SigAvg, such as ($\epsilon{\text{cut}}$, P, $\dot{\text{P}}$, SigAvg, TS$_{\text{Cut}}$), was tested. However, due to the strong correlation between TS$_{\text{Cut}}$ and SigAvg (Pearson correlation coefficient $r = 0.89$), including both variables did not significantly improve the model (TS improvement $\ll$ 0.1). This strong correlation, which indicates some level of interchangeability between these two variables, arises because higher detection significance generally allows for better characterization of spectral features, such as curvature. Nevertheless, TS$_{\text{Cut}}$ should be preferred, as it provides a more direct measure of spectral curvature and better preserves the theoretical FP exponents, while SigAvg led to deviations from the expected exponent values, suggesting that it introduces unnecessary bias. It is important to point out that both of these variables are not intrinsic pulsar properties, such as $\epsilon_\mathrm{cut}$, P or $\dot{\text{P}}$, but are instead related to the properties of the observed emission, acting as normalization variables for the linear model by accounting for the strength of the observed signal and/or observation time. The likelihood test involving other variables, such as Galactic longitude (G$_{\text{lon}}$) and the characteristic age ($\tau_{c}$) of pulsars, resulted in insignificant improvements of 0.9$\sigma$ and 0.6$\sigma$, respectively.

\subsection{Testing the Fundamental Plane Relation on young and millisecond pulsar populations}

\begin{table*}
\caption{The exponential constants of the Fundamental Plane relation derived using the OLS method for both young and millisecond pulsar data sets. The definitions of the columns are identical to those provided in Table~\ref{fp_table}.}
\centering
\small
\renewcommand{\arraystretch}{1.5}
\begin{tabular}{cc|c|ccc|ccc|c}
\hline\hline
Pulsar & Number of & $\beta$  & $\alpha$ & $\gamma$ & $\delta$ & $\zeta$ & $\xi$ & $\mu$ & Adj.~R$^{2}$\\
Data set & Pulsars & ($\epsilon_\mathrm{cut}$) & (Const.) & (P) & ($\dot{\text{P}}$) & (Const.) & (B$_{*}$) & $(\dot{\text{E}}$) \\
\hline\hline
Young & 71 & 1.53$\pm$0.36 & 37.63$\pm$1.53 & $-$1.17$\pm$0.31 & 0.35$\pm$0.11 & 20.93$\pm$3.21 & -0.06$\pm$0.20 & 0.38$\pm$0.09 & 0.601\\
\hline
Millisecond & 113 & 1.26$\pm$0.21 & 40.03$\pm$1.70 & $-$1.48$\pm$0.28 & 0.55$\pm$0.09 & 14.84$\pm$2.44 & 0.08$\pm$0.16 & 0.51$\pm$0.08 & 0.545 \\

\hline
\end{tabular}
\label{fp_table_ms_nonms}
\end{table*}

\begin{figure*}
\centering
\includegraphics[width=18.0cm]{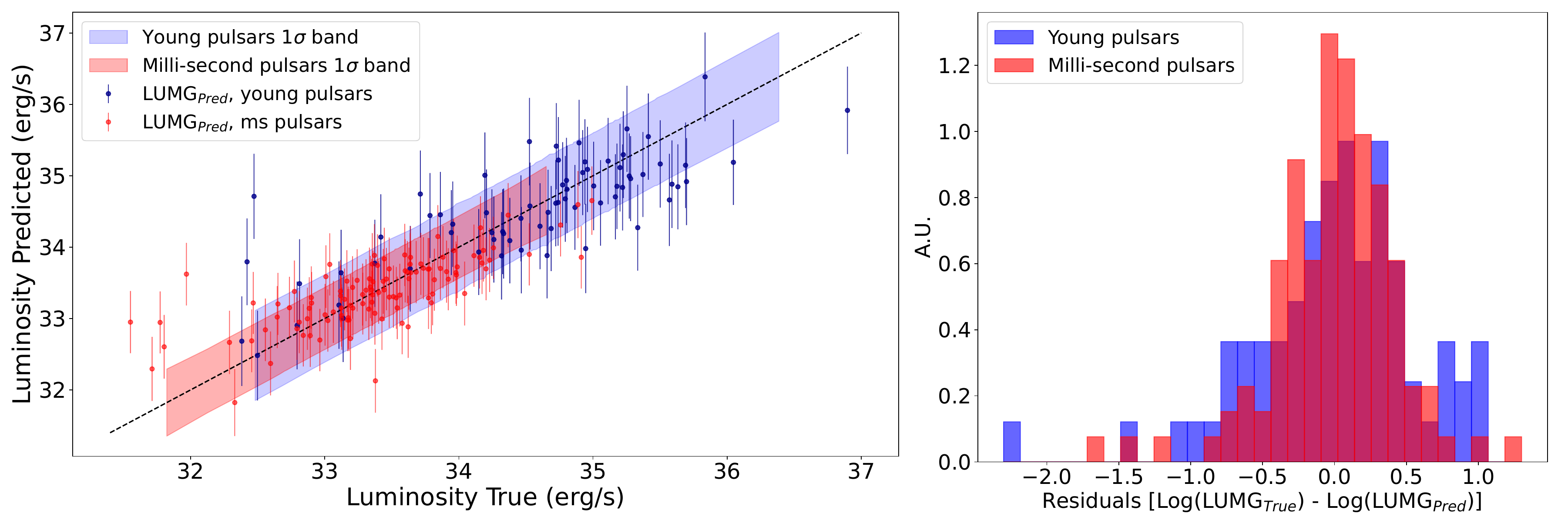}
\caption{The left panel shows a scatter plot comparing the true luminosities with those predicted by the FP relation expressed in logarithmic space. The blue and red markers represent the luminosities of young pulsars and MSPs, respectively, while the shaded regions in corresponding colors represent the 1$\sigma$ confidence intervals derived from OLS regression. The dashed black line denotes the reference y=x line. In the right panel, the residuals between the true and predicted gamma-ray luminosities are shown, with the blue and red distributions representing young pulsars and MSPs, respectively.} 
\label{ms_non_ms_lumg}
\end{figure*}

The data sets used for the validation of FP relations and the investigation of emission mechanisms, as given in Table~\ref{fp_table} and \ref{table_signif_test}, comprise both MSPs and young pulsars within their respective samples. In fact, these two pulsar populations are inherently distinct,  exhibiting different intrinsic properties, such as spin period, period derivative, magnetic field strength at the star surface, and pulsar age. Although the FP relation can be validated when these populations are merged, it is crucial to determine whether it holds true for each pulsar population independently, as a universal FP relation should remain valid within the intrinsic parameter distributions of both MSPs and young pulsars. To address this, the $\epsilon_{\text{cut,HYB}}$ data set with TS$_{\text{Cut}}\geq$4.0 has been subdivided into two distinct data sets comprising only MSPs and young pulsars\footnote{The 'CHARCODE' keyword of 'm' provided in the 3PC data set is used to identify MSPs.}. The exponents of FP relations have been determined for both of these subdata sets, using the OLS method, and are provided in Table~\ref{fp_table_ms_nonms}. The predicted luminosities and the corresponding residuals for the MSPs and young pulsars are shown in the left and right panels of Fig.~\ref{ms_non_ms_lumg}, respectively.

As can be seen from the table, the derived exponents of the FP relations are consistent with theoretical expectations within a 2$\sigma$ uncertainty for both distinct pulsar populations. However, it is crucial to point out that the significantly smaller sample sizes result in much larger errors in the FP exponent parameters, especially when compared to those reported in Table~\ref{fp_table}. The best-fit parameter values for the MSP population exhibit closer alignment with the theoretical FP relations than those of the young pulsar population, likely due to the larger sample size of the MSP population. Additionally, the narrower magnetic field ranges within each population may limit the ability to fully capture underlying dependencies in the FP relation, potentially contributing to the observed slight differences in FP exponents. Consequently, it can be concluded that each individual pulsar population obeys the FP relation within the limits of derived statistical errors. The left panel of Fig.~\ref{ms_non_ms_lumg} illustrates that the low-luminosity part of the combined pulsar data set is predominantly composed of MSPs (red circles), while the high-luminosity part is largely dominated by young pulsars (blue circles), as expected. Furthermore, the right panel of Fig.~\ref{ms_non_ms_lumg} indicates that the luminosity predictions derived from the FP relation for the individual pulsar populations are reasonable and centered around zero, with the MSP population displaying a narrower residual distribution with respect to young pulsars.

\subsection{Investigation of influential observations in the Fundamental Plane Relation}
\label{sec_outliers}

The distribution of residuals derived from the FP relations (see bottom right panels of Fig.~\ref{exp_profiles} and \ref{ms_non_ms_lumg}) reveals the presence of influential outliers within the data set. These outliers potentially degrade the predictive accuracy of the model, thereby increasing the errors associated with the predicted luminosities. Investigation of the common characteristics of these outliers can, in principle, be used for improving the model performance in luminosity prediction, resulting in more precisely constrained estimates of the FP relation exponents. To identify the influential data points, Cook's distance method~\citep{cook1, cook2}, which quantifies the impact\footnote{Cook's distance with higher values indicates a greater influence of specific data points on the model.} of individual observations on the fitted model, has been employed.

\begin{figure*}
\centering
\includegraphics[width=18.0cm]{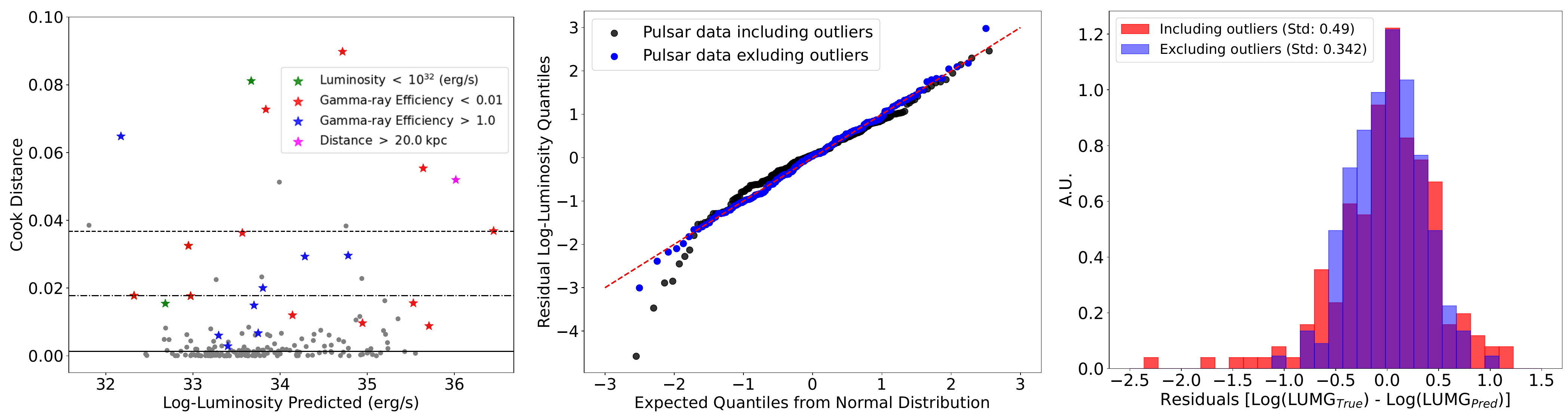}
\caption{Left panel: the Cook distance plot derived from the OLS analysis of 184 pulsars is presented. Pulsars with gamma-ray efficiencies exceeding 1.0 and those below 0.01 are indicated by blue and red stars, respectively. Pulsars exhibiting extremely low luminosities ($<$10$^{32}$ erg/s) are marked by green stars, while the magenta star represents the extragalactic pulsar PSR~J0540$-$6919. The remaining pulsars in the analysis are represented by gray circles. The solid, dashed, and dotted black lines correspond to the median, 90th percentile, and 95th percentile of the Cook distance distribution, respectively. Middle panel: the quantile-quantile plot comparing the data sets with and without outliers, shown by black and blue circles, respectively. The dashed red line represents the reference y=x line. Right panel: the comparison between log-luminosity residual distributions obtained from the data set including outliers (red) and excluding outliers (blue). The standard deviations of the distributions are given in the legend.} 
\label{outliers}
\end{figure*}

The Cook distance plot derived for the OLS regression analysis using the $\epsilon_{\text{cut,HYB}}$ data set with TS$_{\text{Cut}}\geq$4.0 is shown in Fig.~\ref{outliers} (left panel). The plot reveals that the majority of the outliers, particularly those exceeding the 90th percentile of the Cook distance distribution, share common characteristics. Specifically, pulsars exhibiting either exceptionally high ($>$100$\%$) or extremely low ($<$1$\%$) effective gamma-ray efficiencies ($\gamma_{\text{Eff}}$ = L$_{\gamma}$/$\dot{\text{E}}$) are the primary contributors to these influential outliers. Considering that L$_{\gamma}$ is defined as L$_{\gamma}$=4$\pi$f$_{\Omega}$F$_{\text{E}}$D$^{2}$, such outlier effects may arise from either inaccurate distance estimates or the assumption of f$_{\Omega}$=1 used in this analysis based on \cite{fermi3pc,fp2022}. The beaming factor, f$_{\Omega}$, is defined as the ratio of the total beam power averaged over the entire sky to the power in the beam slice that illuminates Earth, averaged in phase, and depends on both the angle $\alpha$ and the inclination $\xi$ between the pulsar's rotation axis and the observer's line of sight (see \cite{Romani_2010} and Eq.~16 in \cite{Abdo_2013_2PC}). Consequently, f$_{\Omega}$ can be less or greater than unity for these pulsars with $\gamma_{\text{Eff}}$ exceeding 100$\%$ or less than 1$\%$, respectively. It is important to note that different emission models predict distinct pulse profile shapes and therefore different values of f$_{\Omega}$. For instance, \cite{petri2011} demonstrated that f$_{\Omega}$ can range between 0.22 and 1.90 within the framework of the striped wind model, whereas \cite{Kalapotharakos_2023} suggested f$_{\Omega}<$1 for the Fermi-LAT pulsar sample. Despite these variations, a realistic range of $\gamma_{\text{Eff}}$, considering the conversion of rotational energy loss into gamma-ray emission, has been estimated to be between 0.01 and unity \citep{Watters_2009}. Pulsars with extremely low luminosities (L$_{\gamma}<10^{32}$ erg/s), which represent the lowest 2nd percentile of the 3PC L$_{\gamma}$ distribution, may also be subject to similar assumptions. Furthermore, the extragalactic gamma-ray pulsar PSR~J0540$-$6919~\citep{lmc_pulsar}, situated in the Tarantula Nebula of the Large Magellanic Cloud with an estimated distance of $\sim$49.7~kpc, is not a member of the Galactic pulsar population. This pulsar is also identified as an outlier in the Cook distance analysis. It is important to note that while the Cook distance plot in Fig. 6 (left panel) provides a visual representation of influential data points, its confidence levels (e.g., 90$\%$ or 95$\%$) are not used as the primary criteria for outlier selection. Instead, outliers are identified based on physically motivated thresholds, specifically, extreme gamma-ray efficiencies ($\gamma_{\text{Eff}}$~$>$~100$\%$ and $<$~1$\%$), very low luminosities (L$_{\gamma}$~$<$~10$^{32}$~erg/s), and the extragalactic nature of PSR~J0540$-$6919. These criteria account for potential biases in distance estimation and beaming factor assumptions, which could systematically affect the FP relation. The Cook distance plot simply illustrates the influence of these physically selected outliers on the regression analysis. Based on the criteria for identifying influential observations outlined above, 23 pulsars have been classified as outliers within the data set. 

The middle panel of Fig.~\ref{outliers} provides a comparison of quantile-quantile plots under two scenarios: one where these outliers remain in the data set (black circles) and another where they are removed (blue circles). As evident from the plot, the residual log-luminosity distribution agrees more closely with the normal distribution when these outliers are excluded. This improvement in distribution is also reflected in the right panel of Fig.~\ref{outliers}, which shows that the accuracy of pulsar luminosity predictions is improved upon the removal of outliers, leading to a smaller standard deviation of residuals tightly centered around the mean, when compared to the data that include outliers. The primary impact of excluding these outliers shows itself in the reduced errors of the estimated FP exponent parameters, thereby resulting in smaller errors on the predicted luminosities. Furthermore, the adjusted R$^{2}$ value of the fit increases from 0.728 when outliers are included to 0.832 when they are removed, highlighting the improved model performance. The resulting FP relations obtained from the OLS analysis of 161 pulsars are derived as
\begin{equation}
\label{fp_outliers}
L_\gamma =
\begin{cases}
  10^{ 40.54 \pm 0.41}\,\epsilon_\mathrm{cut}^{ 1.09 \pm 0.15} \, \text{P}^{ -1.52 \pm 0.14} \, \dot{\text{P}}^{ 0.57 \pm 0.04} & \\
  10^{ 14.22 \pm 1.33}\,\epsilon_\mathrm{cut}^{ 1.09 \pm 0.15} \, \text{B}_{*}^{ 0.10 \pm 0.02} \, \dot{\text{E}}^{ 0.52 \pm 0.04}.  
\end{cases}
\end{equation}

\section{Estimation of pulsar distances using the Fundamental Plane Relation}
\label{sect_4}
The FP relation is primarily used to estimate gamma-ray luminosities of pulsars based on their intrinsic characteristic properties. As was shown and discussed detailed in the previous section, the 3PC pulsar data are in good agreement with the FP relation. One of the most intriguing applications of the FP relation is based on using luminosities to estimate pulsar distances. Such a method becomes particularly useful for the RQ gamma-ray pulsars, whose distances in most cases cannot be measured with the traditional methods. As was mentioned before, there are currently 62 RQ gamma-ray pulsars without any distance estimation in the literature, while their intrinsic properties, such as P, $\dot{\text{P}}$ and $\epsilon_{\text{cut}}$ can be well determined from astronomical observations. Consequently, the luminosity function employing the FP relation can be used to predict unknown distances of RQ $\gamma$-ray pulsars when used together with the traditional luminosity equation as 
\begin{equation}
\label{dist_est1}
L_{\gamma} = 4\pi\text{f}_{\Omega}\text{F}_{\text{E}}\text{D}^{2} = 10^{\alpha}\epsilon_\mathrm{cut}^{\beta}\text{P}^{\gamma}\dot{\text{P}}^{\delta},
\end{equation}
and consequently
\begin{equation}
\label{dist_est2}
\text{D} = \sqrt{ \frac{10^{\alpha}\epsilon_\mathrm{cut}^{\beta}\text{P}^{\gamma}\dot{\text{P}}^{\delta}}{4\pi\text{F}_{\text{E}}}},
\end{equation}
in which the exponent coefficients are obtained from the FP relation. 

Conceptually, it is possible to use the FP relations derived in \cite{fp2022}, shown in Table~\ref{fp_table} or Eq.~\ref{fp_outliers} directly for predicting distances of RQ pulsars from Eq.~\ref{dist_est2}. However, the primary objective of a predictive pulsar distance model should focus on achieving effective generalization to new, unseen data, such as the RQ pulsar sample including 62 pulsars. Therefore, quantification of prediction accuracy of the FP relation using separate training and test data sets becomes crucial for several reasons. For instance, when the same data set is used for both training and testing, the model may learn the specific details of the data set rather than generalizing from it. This can lead to overfitting and introducing overly optimistic performance bias into evaluation metrics, where the model performs extremely good on the training data but poorly on unseen data. Furthermore, the test data set provides an independent evaluation of the model's performance, quantifying how well the model is generalized for unseen data. This approach also provides a background for evaluating different models or fine-tuning of hyperparameters (HPs). Consequently, to ensure that the predictive pulsar distance model’s performance is evaluated properly and generalized reasonably, a traditional approach based on splitting the pulsar data into training and test sets has been used together with cross-validation techniques.

\subsection{Proof of the principle on distance estimation using the Fundamental Plane Relation}
\label{proof_of_principle}

The proof of principle regarding distance estimation using the FP relation should be first verified and quantified on pulsars with known distances, before applying this method on the unseen RQ pulsar data set. For this purpose, the data set of 161 pulsars, excluding outliers as detailed in Sect.~\ref{sec_outliers}, was used. The ‘RepeatedKFold’ function from the {\tt scikit-learn} module~\citep{scikitlearn} was employed to perform a three-fold cross-validation of the loglinear FP relation model. In each pass of the cross-validation process, the data set of 161 pulsars was randomly partitioned into three folds, with each fold (~54 pulsars) taking turns as the test set while the remaining two-thirds were used for model training. This resulted in three distinct train/test splits per pass. To improve the statistical robustness of model evaluation and reduce the influence of any single random partitioning, this three-fold cross-validation was repeated five times per iteration, each time with a newly randomized split of the data. Since each pass produced three train/test configurations, repeating the process five times per iteration resulted in 15 unique train/test realizations per iteration. This ensured that each pulsar appeared in both training and test sets under diverse conditions, minimizing biases introduced by specific partitions and providing a more comprehensive evaluation of model performance. To further improve statistical reliability, 10 independent iterations of this entire process were performed, each with a newly randomized shuffling of the data set. This resulted in a final total of 150 data set realizations. The adjusted R$^{2}$ performance metrics were calculated after each iteration to quantify the accuracy of the models on both training and test data sets. Since this process was repeated 10 times, 150 individual adjusted R$^{2}$ values were obtained for both training and test sets. The distribution of these values provides insight into the stability of model performance across different train/test splits. The choice of five repetitions per iteration and 10 total iterations was made to ensure a sufficiently large number of data set realizations for a statistically robust evaluation, while avoiding unnecessary redundancy. Increasing the number of repetitions beyond this would not significantly alter the distribution of adjusted R$^{2}$ values, indicating that model performance estimates had reached a stable and reliable range. 

\begin{figure}
\centering
\includegraphics[width=9.0cm]{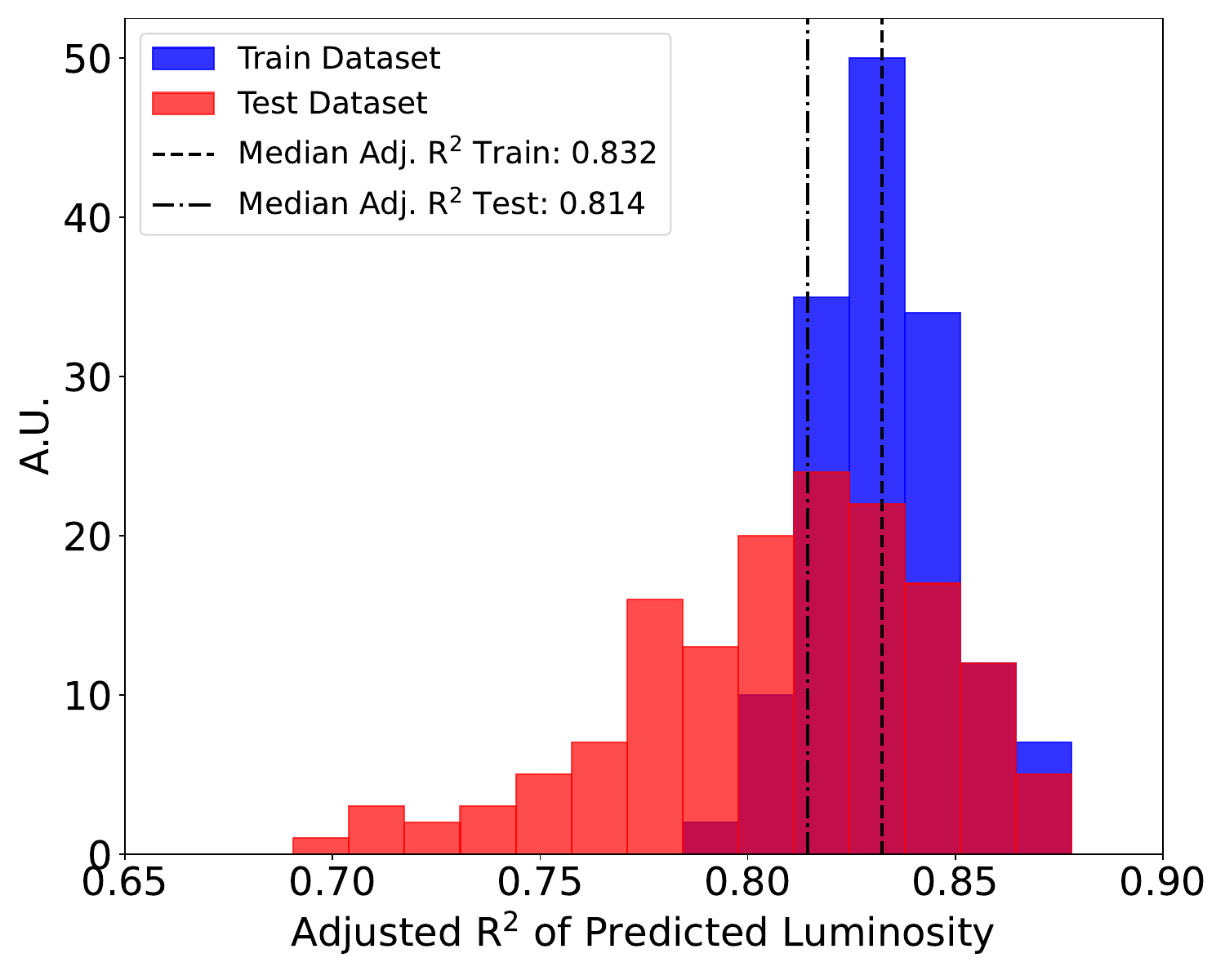}
\caption{The distributions of the adjusted R$^{2}$ values for the predicted gamma-ray luminosities. The blue histogram represents the distribution derived from 150 training data sets, while the red histogram corresponds to the distribution obtained from 150 test data sets. The medians of the training and test distributions are indicated by black dashed and dashed-dotted lines, respectively, with the corresponding values annotated in the figure legend.}
\label{adjr2}
\end{figure}

\begin{figure*}
\centering
\includegraphics[width=18.0cm]{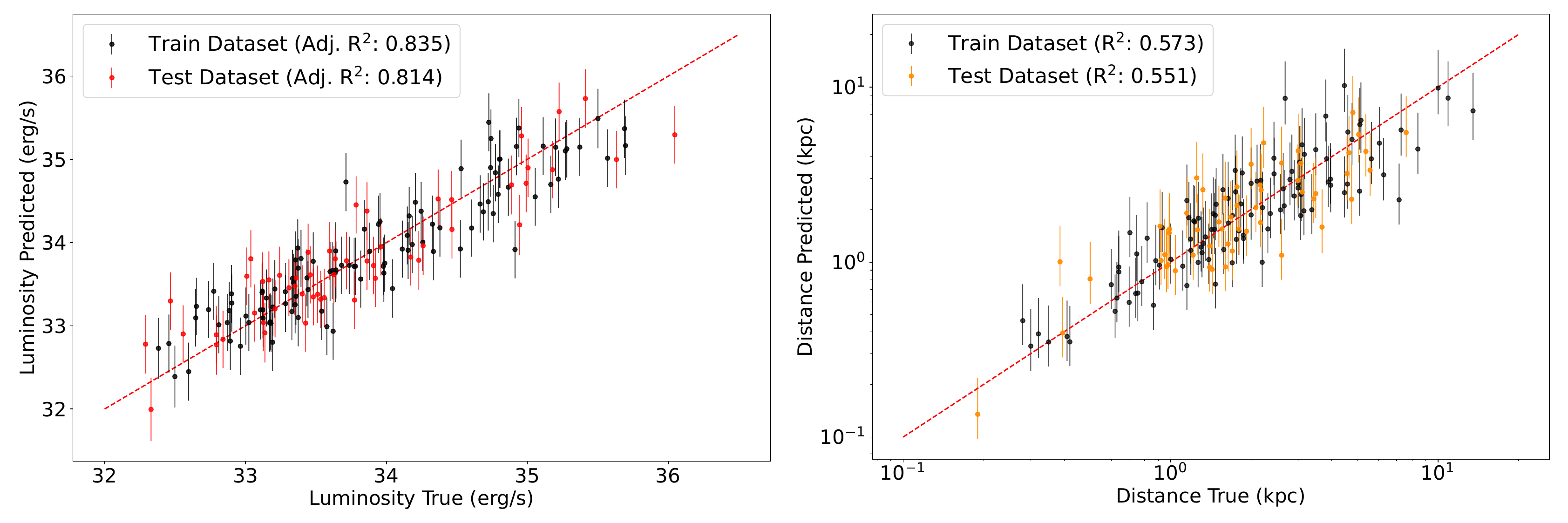}
\caption{Left panel: a scatter plot showing the comparison between the true luminosities and those predicted by the FP relation described in Eq.~\ref{fp_optimized}. The markers in black and red denote the luminosities from the training and test data sets, respectively, with associated 1$\sigma$ statistical prediction errors coming from the OLS method. Right panel: a scatter plot showing the comparison between the true pulsar distances and those derived using the luminosities predicted by the FP relation in Eq.~\ref{fp_optimized} and the subsequent application of Equation Eq.~\ref{dist_est2}. The markers in black and yellow correspond to the pulsar distances from the training and test data sets, respectively, together with 1$\sigma$ statistical asymmetric prediction errors. In both figures, the dashed red line represents the reference y=x line.} 
\label{test_train_lumg_dist}
\end{figure*}

The distribution of adjusted R$^{2}$ performance metrics for the predicted gamma-ray luminosities, derived from 150 train/test samples, is shown in Fig.~\ref{adjr2}. The median adjusted R$^{2}$ values for the training and test data sets are 0.832 and 0.814, respectively. For comparison, when outliers, as discussed in Sect.~\ref{sec_outliers}, are included in the data set, these median values decrease to 0.729 and 0.7, indicating that excluding outliers enhances prediction performance by $\sim$10$\%$ for both the train and test data sets on average. To optimize the model for predicting luminosity, and consequently pulsar distances, the loglinear FP model having adjusted R$^{2}$ values closest to the median of the train/test distributions was selected as the preferred prediction model. This optimized prediction model is given as 
\begin{equation}
\label{fp_optimized}
L_\gamma =
\begin{cases}
  10^{ 40.51 \pm 0.54}\,\epsilon_\mathrm{cut}^{ 1.11 \pm 0.19} \, \text{P}^{ -1.54 \pm 0.18} \, \dot{\text{P}}^{ 0.57 \pm 0.05} & \\
  10^{ 14.13 \pm 1.68}\,\epsilon_\mathrm{cut}^{ 1.11 \pm 0.19} \, \text{B}_{*}^{ 0.09 \pm 0.03} \, \dot{\text{E}}^{ 0.53 \pm 0.05}
\end{cases}
\end{equation}
in both ($\epsilon_\mathrm{cut}$, P, $\dot{\text{P}}$) and ($\epsilon_\mathrm{cut}$, B$_{*}$, $\dot{\text{E}}$) representations of the FP relation. The left panel of Fig.~\ref{test_train_lumg_dist} shows the predicted gamma-ray luminosities using the optimized prediction model given in Eq.~\ref{fp_optimized}, while the right panel shows the corresponding predictions of pulsar distances obtained from predicted gamma-ray luminosities and measured integral Fermi gamma-ray fluxes following Eq.~\ref{dist_est2}, for both train and test data sets. The R$^{2}$ values are used for evaluating pulsar distance predictions and provided in the right panel of Fig.~\ref{test_train_lumg_dist}, 0.573 and 0.551 for the train and test data sets, respectively. The statistical errors on the distance predictions are calculated by propagating the errors coming from the OLS luminosity prediction, which takes into account the full covariance matrix and measured Fermi gamma-ray flux uncertainty. It is important to note that the symmetric errors on the predicted log-luminosity result in asymmetric errors on the predicted distance after error propagation. The upper and lower average relative statistical errors of the distance predictions are at the level of $\sim$61$\%$ and $\sim$29$\%$, respectively, for both training and test data sets. For comparison, the average upper and lower relative distance errors given in the 3PC\footnote{The majority of the statistical errors associated with pulsar distances, 148 out of 204 cases, are reported symmetrically in 3PC. These cases predominantly involve distance estimates derived from the dispersion measure method.} is $\sim$31$\%$ and $\sim$37$\%$, respectively. 

\begin{table}
\caption{Summary of the evaluation of FP exponents for models selected from different percentiles of the adjusted R$^{2}$ distribution across 150 training data sets shown in Fig.~\ref{adjr2}. The first column indicates the percentile at which the model was selected. The second and third columns provide the corresponding adjusted R$^{2}$ values for the training and test data sets, respectively. The fourth, fifth, and sixth columns present the best-fit exponent values of the FP relation, $\epsilon_\mathrm{cut}$, P and $\dot{\text{P}}$, respectively, along with their 1$\sigma$ statistical uncertainties. The median model, highlighted in bold, corresponds to the optimized FP model given in Eq.~\ref{fp_optimized}.}
\centering
\footnotesize
\renewcommand{\arraystretch}{1.1}
\begin{tabular}{c|cc|ccc}
\hline\hline
Percentile & Adj.~R$^{2}$ & Adj.~R$^{2}$  & $\beta$                   & $\gamma$         & $\delta$ \\
Train      & Train        & Test          & ($\epsilon_\mathrm{cut}$) & (P)              & ($\dot{\text{P}}$)\\
\hline\hline
5          & 0.804        & 0.857         & 1.16$\pm$0.17             & $-$1.41$\pm$0.16 & 0.53$\pm$0.04 \\
10         & 0.811        & 0.850         & 1.15$\pm$0.19             & $-$1.51$\pm$0.18 & 0.55$\pm$0.05 \\
32         & 0.825        & 0.835         & 1.14$\pm$0.19             & $-$1.45$\pm$0.16 & 0.55$\pm$0.04 \\
\textbf{50 (median)}& \textbf{0.832}        & \textbf{0.814}         & \textbf{1.11$\pm$0.19}             & \textbf{$-$1.54$\pm$0.18} & \textbf{0.57$\pm$0.05} \\
68         & 0.840        & 0.801         & 1.08$\pm$0.17             & $-$1.51$\pm$0.17 & 0.57$\pm$0.04 \\
90         & 0.855        & 0.767         & 1.19$\pm$0.19             & $-$1.49$\pm$0.17 & 0.57$\pm$0.04 \\
95         & 0.860        & 0.711         & 0.93$\pm$0.17             & $-$1.70$\pm$0.18 & 0.62$\pm$0.05 \\

\hline
\end{tabular}
\label{fp_exponents_deviation_OLS}
\end{table}

To assess the stability of the FP exponents, models were selected at different percentiles of the adjusted R$^{2}$ distribution obtained from 150 train data sets, and the evolution of the exponents was analyzed. The results, summarized in Table~\ref{fp_exponents_deviation_OLS}, indicate a negative correlation between adjusted R$^{2}$ train and R$^{2}$ test values, reflecting differences in the generalization ability of the tested models. Models with the highest adjusted R$^{2}$ train (e.g., 95th percentile, 0.860) exhibit a significantly lower adjusted R$^{2}$ test (0.711), suggesting that the corresponding FP exponents are tuned too closely to the training data, leading to overfitting. On the other hand, models with lower adjusted R$^{2}$ train (e.g., 5th percentile, 0.804) show a higher adjusted R$^{2}$ test, indicating better test performance but potentially underfitting the training data. Across different percentiles, the FP exponents remain largely consistent, with variations mostly within 1$\sigma$ of the median model exponent values. Even at the 95th percentile, the deviations from the median are within 2$\sigma$ and therefore not statistically significant. Given this stability, the optimized FP model given in Eq.~\ref{fp_optimized}, which corresponds to the median model, provides a balanced solution, ensuring reliable exponent estimation while avoiding both overfitting and underfitting issues. Consequently, the optimized FP prediction model presented in Eq.~\ref{fp_optimized} effectively encapsulates the characteristics of the 150 train/test data sets generated through three-fold cross-validation. This model demonstrates that the intrinsic properties of gamma-ray pulsars provided in the FP relation can be used for predicting luminosity and, consequently, obtaining pulsar distances. As shown in both the left and right panels of Fig.~\ref{test_train_lumg_dist}, the predicted luminosity and distance values are distributed around the reference y = x line, indicating a reasonable degree of accuracy. These findings, therefore, provide a proof of principle on distance estimation using the FP relation.

\subsection{Investigation of distance estimation using various machine learning methods}
\label{otherML}

The FP relation has demonstrated its potential for offering reasonable distance estimates for gamma-ray pulsars, as evidenced by the proof of principle presented in the previous subsection, where it was applied to a sample of pulsars with known distances using the OLS method. However, it is crucial to investigate whether incorporating FP variables into other ML techniques, such as random forest regression (RFR; \cite{breiman}) and support vector regression (SVR; \cite{svr_paper}), can improve prediction accuracy. In addition to RFR, two different kernels, linear (SVR-Linear) and radial basis function (SVR-RBF), are tested for the SVR approach. The OLS method minimizes the sum of squared residuals, assuming a strict linear relationship between variables. In contrast, SVR-Linear uses an $\epsilon$-insensitive loss function that allows for errors within a specified margin, therefore improving robustness to outliers, and incorporates a regularization parameter 'C' to balance model complexity and accuracy. Consequently, SVR-Linear adapts better to high-dimensional data and is less sensitive to OLS assumptions, such as normality, homoscedasticity, or independence. The SVR-RBF approach transforms input features into an infinite-dimensional space, capturing nonlinear relationships with the '$\gamma$' parameter controlling the width of the Gaussian function, making it highly flexible and less sensitive to kernel parameter selection. On the other hand, the RFR is an ensemble learning method that constructs multiple decision trees, handling nonlinear relationships and interactions between features without assuming a specific functional form. Due to its inherent averaging and bootstrapping processes, RFR is more robust against overfitting and outliers compared to OLS. 

The primary objective of the study presented in this subsection is to compare distance estimations from various ML techniques to OLS to assess compatibility and also investigate potential improvements. The rationale for the selection of OLS, RFR, and SVR for the analysis of the Fermi 3PC data was motivated by their complementary strengths and suitability for the research objectives. Specifically, OLS was selected as a baseline to explore linear relationships in the data, providing a straightforward benchmark and insight into the simplest linear trends. RFR, as a robust model-independent ensemble method, was chosen for its robustness against overfitting and its capability to model complex, nonlinear relationships. SVR, with its use of kernel functions, was selected for its ability to determine nonlinear patterns, making it particularly suitable for data sets where the relationship between input features and target variables may not follow a simple structure. By comparing the results obtained from these methods, the analysis benefits from a broad range of modeling perspectives, ensuring that the conclusions drawn are not overly reliant on the assumptions or biased by the limitations of a single methodology. Consequently, such an approach enhances the robustness, reliability, and interpretability of the results. 

To test potential improvement in the prediction accuracy, it is crucial to investigate the inclusion of additional pulsar properties, such as the significance of the cutoff feature. The ML algorithms mentioned above offer a direct method for predicting distances using the variables ($\epsilon_\mathrm{cut}$, P, $\dot{\text{P}}$, F$_{E}$), as given in Eq.~\ref{dist_est2}, thus bypassing the intermediate step of predicting gamma-ray luminosity and subsequently converting it to distance, which is done for the proof of principle using the OLS approach in the previous section. Therefore, direct distance estimates using these algorithms can provide an independent cross-check. For a rigorous comparison, the identical 150 train/test data sets employed in Sect.~\ref{proof_of_principle} are used with RFR and SVR algorithms. The comparison of performance metrics is based on the adjusted R$^{2}$ distributions, analogous to the one given in Fig.~\ref{adjr2}, but specifically for distance estimates rather than luminosity. 

\begin{table*}
\caption{A comparison of the results for distance prediction diagnostics using various ML algorithms. The first column specifies the ML algorithm used, while the second column lists the variables employed for training the distance prediction models. The corresponding optimized HP values for each ML algorithm are reported in the third column. The HPs of the RFR ML algorithm include the following: the number of trees in the forest (n$_{\text{est}}$), the minimum number of samples required to split an internal node (spl), the maximum tree depth (D), and the minimum number of samples necessary to constitute a leaf node (S). The HPs of the SVR ML algorithm include the following: 'C' controls the balance between model complexity and accuracy, while '$\epsilon$' defines the margin of tolerance for errors in the $\epsilon$-insensitive loss function. In the SVR-RBF, '$\gamma$' determines the influence of individual data points by controlling the width of the Gaussian kernel, namely the smoothness of the decision boundary. The median and 5th percentile values (representing the 95$\%$ confidence level lower limit) of the adjusted R$^{2}$ distributions derived from the 150 training data sets are shown in the fifth and sixth columns, respectively. Similarly, the seventh and eighth columns report the median and 5th percentile values obtained from the 150 test data sets. It is important to note that the adjusted R$^{2}$ values are computed by comparing the predicted distance values generated by the ML models against the true distance values. The first row, displaying the results from the OLS method, provides a baseline for subsequent comparisons.}
\centering
\small
\renewcommand{\arraystretch}{1.5}
\begin{tabular}{ccc|cc|cc}
\hline\hline
ML & Train & Hyperparameter & Adj.~R$^{2}$ Train & Adj.~R$^{2}$ Train & Adj.~R$^{2}$ Test & Adj.~R$^{2}$ Test\\
Method   & Variables & Values & Median & 5th Percentile & Median & 5th Percentile \\
\hline

\textbf{OLS} & \textbf{($\epsilon_\mathrm{cut}$, P, $\dot{\text{P}}$)} & \textbf{-} & \textbf{0.573} & \textbf{0.455} & \textbf{0.553} & \textbf{0.068} \\

\hline
RFR & ($\epsilon_\mathrm{cut}$, P, $\dot{\text{P}}$, F$_{E}$) & (n$_{\text{est}}$=1200, spl=3, D=15, S=1) & 0.921 & 0.906 & 0.374 & 0.104 
\\
RFR & ($\epsilon_\mathrm{cut}$, P, $\dot{\text{P}}$, F$_{E}$, TS$_{\text{Cutoff}}$) & (n$_{\text{est}}$=1400, spl=3, D=15, S=1) & 0.930  & 0.916 & 0.460 & 0.202 \\
\textbf{RFR} & \textbf{($\epsilon_\mathrm{cut}$, P, $\dot{\text{P}}$, F$_{E}$, TS$_{\text{Cutoff}}$, $\dot{\text{E}}$)} & \textbf{(n$_{\text{est}}$=1500, spl=3, D=20, S=1)} & \textbf{0.942} & \textbf{0.928} & \textbf{0.558} & \textbf{0.352}\\
\hline

SVR-RBF & ($\epsilon_\mathrm{cut}$, P, $\dot{\text{P}}$, F$_{E}$) & (C=5.5, $\epsilon$=0.16, $\gamma$=0.034) & 0.758 & 0.710 & 0.615 & 0.409
\\
SVR-RBF & ($\epsilon_\mathrm{cut}$, P, $\dot{\text{P}}$, F$_{E}$, TS$_{\text{Cutoff}}$) & (C=4.8, $\epsilon$=0.09, $\gamma$=0.03) & 0.779 & 0.730 & 0.637 & 0.415
\\
\textbf{SVR-RBF} & \textbf{($\epsilon_\mathrm{cut}$, P, $\dot{\text{P}}$, F$_{E}$, TS$_{\text{Cutoff}}$, $\dot{\text{E}}$)} & \textbf{(C=1.8, $\epsilon$=0.1, $\gamma$=0.009)} & \textbf{0.759} & \textbf{0.697} & \textbf{0.651} & \textbf{0.500}\\

\hline
SVR-Linear & ($\epsilon_\mathrm{cut}$, P, $\dot{\text{P}}$, F$_{E}$) & (C=1.1, $\epsilon$=0.1) & 0.746 & 0.694 & 0.643 & 0.441
\\
SVR-Linear & ($\epsilon_\mathrm{cut}$, P, $\dot{\text{P}}$, F$_{E}$, TS$_{\text{Cutoff}}$) & (C=100, $\epsilon$=0.1) & 0.753 & 0.697 & 0.658 & 0.473
\\
\textbf{SVR-Linear} & \textbf{($\epsilon_\mathrm{cut}$, P, $\dot{\text{P}}$, F$_{E}$, TS$_{\text{Cutoff}}$, $\dot{\text{E}}$)} & \textbf{(C=0.11, $\epsilon$=0.1)} & \textbf{0.750} & \textbf{0.690} & \textbf{0.662} & \textbf{0.493}\\
\hline

\end{tabular}
\label{mltable}
\end{table*}

\begin{figure}
\centering
\includegraphics[width=9.0cm]{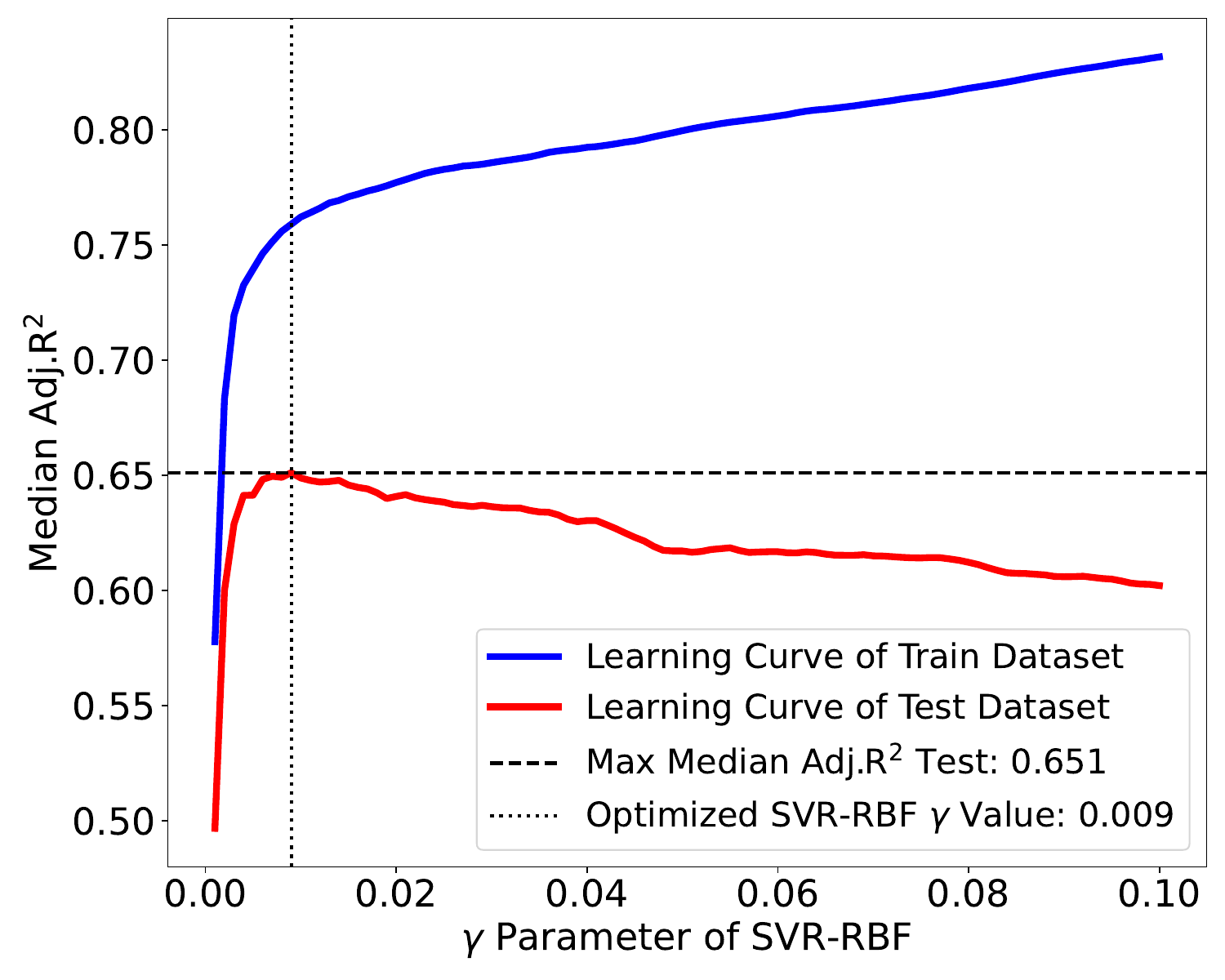}
\caption{The learning curves for the training and test data sets used in the SVR-RBF model, which has been optimized for the $\gamma$ HP value. The y-axis represents the median of the adjusted R$^{2}$ distributions, derived from 150 training data sets (solid blue line) and 150 test data sets (solid red line), for each evaluated $\gamma$ HP value. The horizontal dashed black line marks the highest median adjusted R$^{2}$ value achieved by the testing data sets, while the vertical dotted line denotes the corresponding optimized $\gamma$ HP value. The optimized $\gamma$ HP and the corresponding maximum median adjusted R$^{2}$ values are provided in the legend.}
\label{learning_curve}
\end{figure}

\begin{figure*}
\centering
\includegraphics[width=18.0cm]{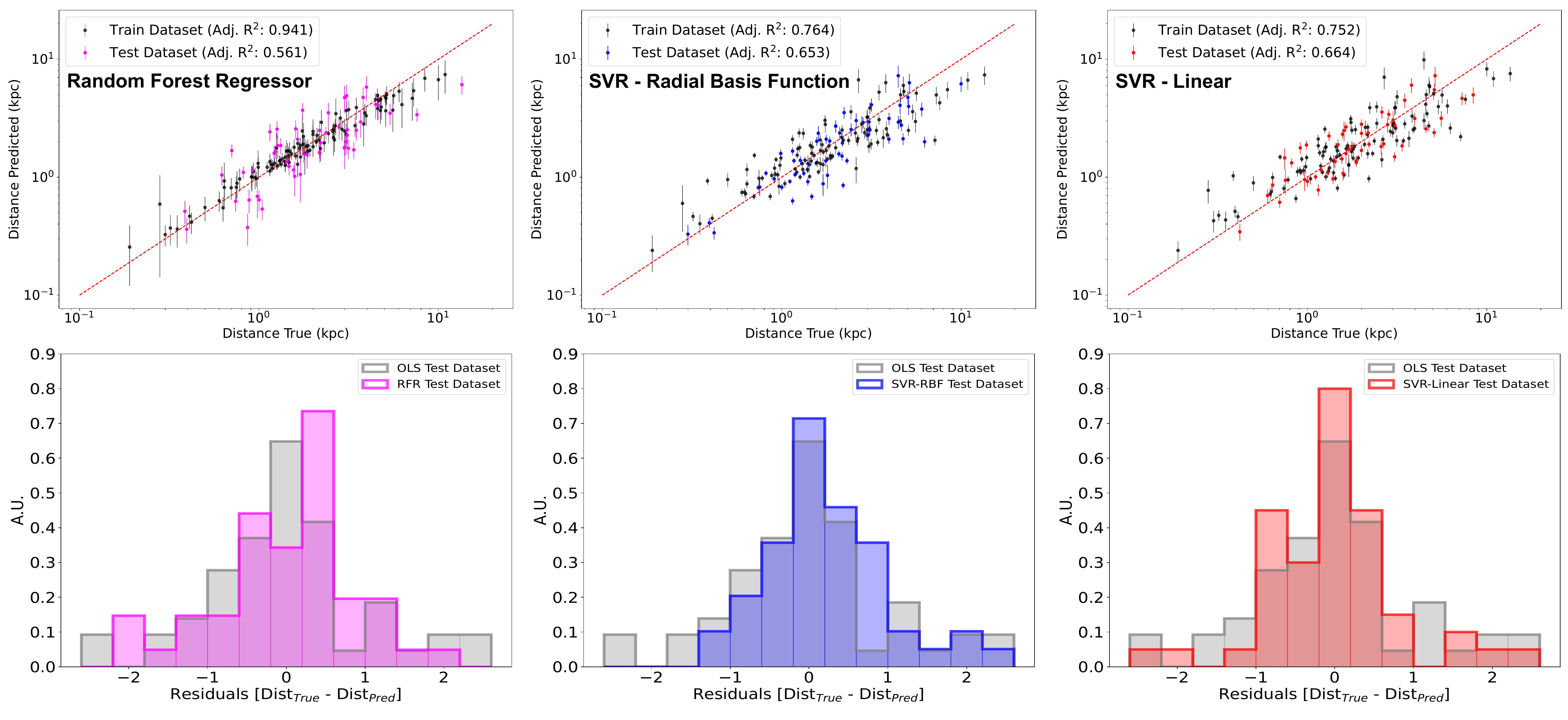}
\caption{Top row: the left, middle, and right panels show the comparison between predicted and true gamma-ray pulsar distances for the train (black circles) and test data sets (magenta, blue, and red circles) using the optimized RFR, SVR-RBF, and SVR-Linear models, respectively, as detailed in Table~\ref{mltable}. The dashed red lines in each plot represent the reference line y=x. The adjusted R$^{2}$ values for both the training and test data sets are provided in the legends of the respective plots. Bottom row: the corresponding residual distributions between the predicted and true pulsar distances for the test data sets are shown. For comparison, the residual distribution from the baseline OLS model is shown as gray histograms, characterized by a median value of 0.01 and 32nd and 68th percentile values of -0.34 and 0.37, respectively. The median values of the residuals for the RFR, SVR-RBF, and SVR-Linear models are 0.23, 0.18, and 0.04, respectively, with corresponding 32nd and 68th percentile ranges of [-0.29, 0.51], [-0.05, 0.51], and [-0.27, 0.31].} 
\label{train_test_other}
\end{figure*}

The comparison of results obtained from different ML approaches is provided in Table~\ref{mltable}. The first row shows the Adj.~R$^{2}$ distribution properties of distance predictions coming from the OLS approach. The median and 5th percentile of the Adj.~R$^{2}$ distributions are provided for both train and test data sets to give insight on the width of distributions. This first row provides a baseline for further comparison. The other variables that can potentially improve prediction accuracy, such as significance of the spectral cutoff (TS$_{\text{Cut}}$), spin-down luminosity ($\dot{\text{E}}$), characteristic pulsar age ($\tau_{\text{c}}$), magnetic field at star surface (B$_{*}$) and at light cylinder (B$_{\text{LC}}$), pulsar location on the Galactic plane (G$_{\text{lon}}$ and G$_{\text{lat}}$), average detection significance (SigAvg), and spectral index of the fitted ECPL model ($\Gamma_{\text{ECPL}}$), are tested in combination with the ($\epsilon_\mathrm{cut}$, P, $\dot{\text{P}}$, F$_{E}$) variables to search for possible improvement. The HPs of each tested ML algorithm, provided in the HP column of Table~\ref{mltable}, are optimized for preventing overfitting of the prediction model by using the learning curves of test and train data sets. An example learning curve obtained for the SVR-RBF model is shown in Fig.~\ref{learning_curve} for the $\gamma$ parameter. As can be seen from the figure, the trend of the train learning curve (blue line) keeps on improving as the $\gamma$ parameter increases, while the test learning curve (red line) peaks at a particular $\gamma$ value and starts decreasing. In this case, the particular $\gamma$ value corresponding to the peak of the median Adj.~R$^{2}$ obtained from test data sets represents the optimized HP value for the SVR-RBF model, keeping the rest of the HP values (C=1.8 and $\epsilon$=0.1 for this particular case) fixed. This procedure is repeated iteratively until all the HPs are individually optimized. With this approach, it is ensured that the model does not suffer from over/underfitting issues. This procedure is applied to all ML models given in Table~\ref{mltable}, and the optimized best HP values are provided. Similar to the approach used in Sect.~\ref{proof_of_principle}, the particular ML model having adjusted R$^{2}$ values closest to the median of the train/test distributions was selected as the preferred distance prediction model.

The table clearly demonstrates that while the RFR approach using the base variables ($\epsilon_\mathrm{cut}$, P, $\dot{\text{P}}$, F$_{\text{E}}$) achieves a very high median adjusted R$^{2}$ value for the training data set, its predictive performance on the test data set is significantly lower compared to the OLS method. However, the subsequent inclusion of the TS$_{\text{Cutoff}}$ and $\dot{\text{E}}$ variables improves the predictive accuracy for both the training and test data sets. This improvement not only allows the RFR approach to reach the predictive power of OLS but also results in a higher 5th percentile value for the test data sets, indicating a narrower distribution. On the other hand, the SVR-RBF and SVR-Linear approaches demonstrate relatively better predictive performance and exhibit narrower distributions of the adjusted R$^{2}$ compared to the OLS and RFR methods, achieving values around $\sim$0.65. This improvement is particularly pronounced with the inclusion of the TS$_{\text{Cutoff}}$ and $\dot{\text{E}}$ variables, contributing to varying degrees of improvement in prediction accuracy. However, the inclusion of further additional variables, namely $\tau_{\text{c}}$, B$_{*}$, B$_{\text{LC}}$, G$_{\text{lon}}$, G$_{\text{lat}}$, SigAvg, and $\Gamma_{\text{ECPL}}$, does not improve prediction performance. Moreover, the observational pulsar properties identified by the 'CHARCODE' keyword in the 3PC catalog, which categorizes pulsars as RL (r), RQ (q), millisecond (m), in a binary system (b), or detected in X-rays (x), were also tested, leading to no pronounced improvement in prediction power.

The top panel of Fig.~\ref{train_test_other} presents scatter plots comparing the predicted and true gamma-ray pulsar distances for both the training and test data sets, derived from three optimized prediction models highlighted in bold and giving the highest adjusted R$^{2}$ in Table~\ref{mltable}: RFR (left), SVR-RBF (middle), and SVR-Linear (right). It is important to point out that the statistical errors of individual pulsar distance estimates, calculated using a bootstrapping approach, are relatively small compared to those obtained with the OLS method across all models. This difference arises because the OLS method incorporates errors from flux measurements (F$_{\text{E}}$) through error propagation, resulting in more conservative and robust distance prediction errors. Nevertheless, the primary objective of this analysis is to compare the distance predictions of pulsars, rather than to focus on their associated errors. The bottom panel of Fig.~\ref{train_test_other} shows the residuals between the predicted and true pulsar distances for the test data set, alongside a comparison with the residuals from the OLS model. It is evident from the plots that, overall, the residual distributions derived from various ML approaches are generally consistent with the OLS distribution, despite some minor differences. Specifically, the residuals from the RFR and SVR-RBF models exhibit a slight skewness, with median values of 0.23 and 0.18, respectively. In contrast, the residual distribution from the SVR-Linear model is more centered around zero, with a median of 0.04, and presents a slightly narrower spread compared to the OLS distribution. It is, theoretically, expected that the linear models can describe the data better owing to the loglinear nature of the theoretical FP relation.

The study presented in this subsection shows that, despite minor differences, all evaluated ML approaches yield distance estimations for gamma-ray pulsars that are comparable to or slightly more accurate than those obtained using the OLS method, while maintaining reasonable accuracy levels. Particularly, even with the baseline FP variables, the SVR approach already shows improved accuracy over the OLS method, while the inclusion of additional variables TS$_{\text{Cutoff}}$ and $\dot{\text{E}}$ further enhances the predictive performance. Consequently, the optimized ML models, highlighted in bold in Table~\ref{mltable}, can potentially be applied for predicting the unknown distances of RQ gamma-ray pulsars.

\section{Prediction of radio-quiet gamma-ray pulsar distances}
\label{pred_results}

\begin{figure*}[ht!]
\centering
\includegraphics[width=18.0cm]{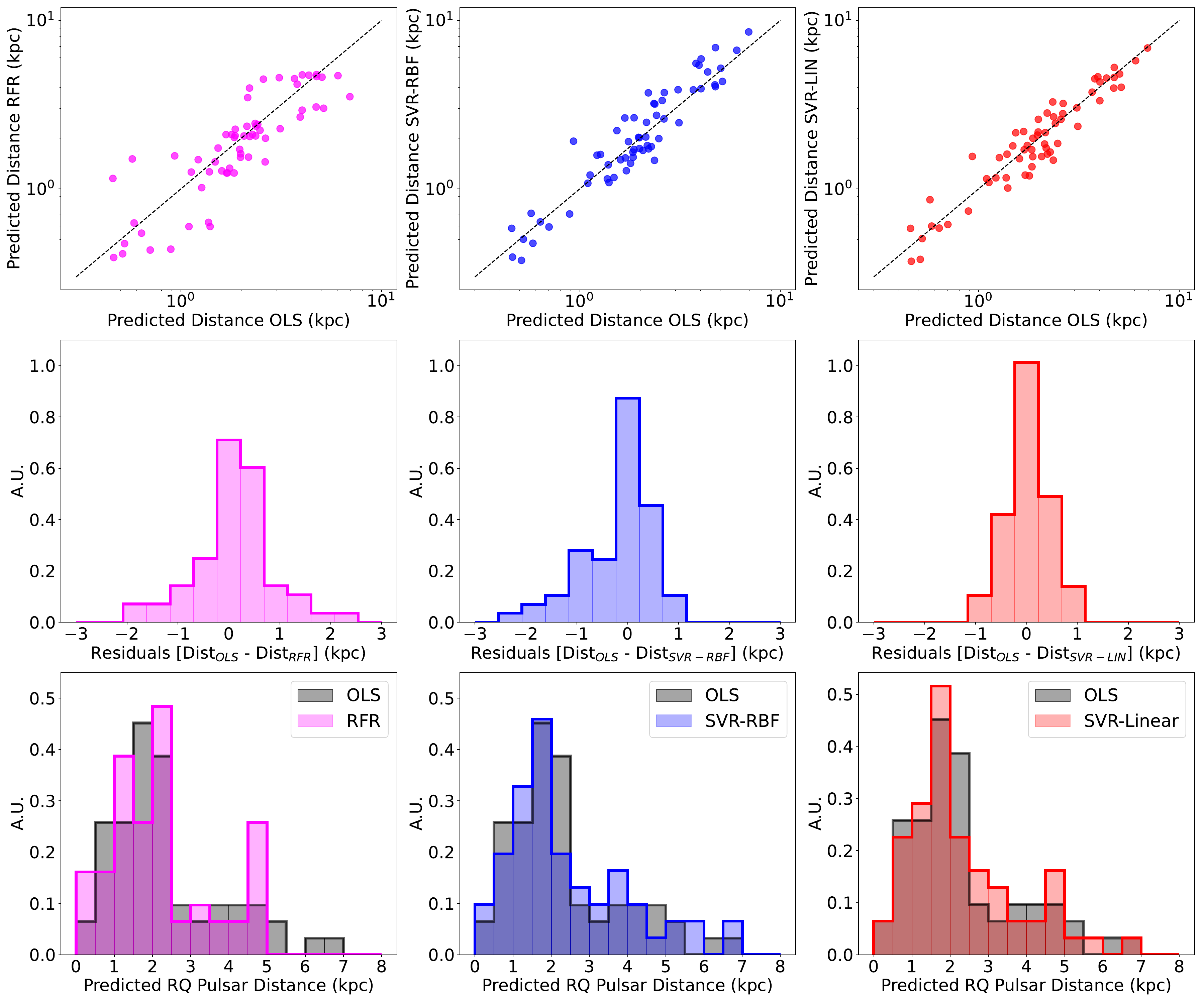}
\caption{Top row: the left, middle, and right panels show the comparison between the predicted RQ gamma-ray pulsar distances using the optimized RFR (magenta), SVR-RBF (blue), and SVR-Linear (red) models, respectively, and the OLS approach. The dashed black lines in each panel represent the reference y=x line. Middle row: the corresponding distance prediction residual distribution between the employed ML models and OLS are shown. Bottom row: the distribution of predicted distances obtained from the optimized RFR (magenta), SVR-RBF (blue), and SVR-Linear (red) models are shown, providing a comparison to the distribution of OLS distance predictions (gray).} 
\label{pred_diag}
\end{figure*}

The proof of principle regarding the potential of the FP relation in estimating the distances to gamma-ray pulsars has been demonstrated through the application of various ML techniques. Given that the distance estimation of RQ gamma-ray pulsars is generally not possible from traditional measurements, this indirect prediction method becomes the most important application of the FP relation. The 3PC contains 62 RQ gamma-ray pulsars without any available distance estimation in the literature, 5 of which are classified as MSPs. The optimized distance prediction models, highlighted in bold in Table~\ref{mltable}, were employed to predict the unknown distances of these 62 RQ pulsars. The results, derived from different ML methods, are provided in Appendix~\ref{B1}. The cross-comparison of distance predictions derived from these ML algorithms, along with associated diagnostics, is summarized in Fig.~\ref{pred_diag}. The top panel illustrates that, overall, the predictions from different ML models are compatible with those from the baseline OLS model, although deviations are observed for the RFR approach, particularly at close and far distances. The middle panel of Fig.~\ref{pred_diag} presents the residual distributions between the OLS and ML models. Evidently, the predictions from the SVR-Linear model demonstrate strong agreement with the OLS results, exhibiting a narrow residual spread ($\sigma_{\text{SVR-LIN}}$=0.431) centered around zero. In contrast, the RFR model shows a broader residual distribution ($\sigma_{\text{RFR}}$=0.839), while the SVR-RBF model displays a slightly skewed residual distribution toward negative values, suggesting an overestimation of distances with respect to the OLS model. The bottom panel of Fig.~\ref{pred_diag} provides the distance distribution of predictions, showing that both the SVR-RBF and SVR-Linear models are generally in good agreement with the OLS model. It is evident, however, that the RFR predictions truncate around $\sim$5 kpc and are unable to predict distances beyond this threshold, whereas the other models can. Although the RFR model can offer reasonable distance predictions, the predictions from the OLS, SVR-RBF, and SVR-Linear models are generally more consistent with each other.

\begin{figure*}
\centering
\includegraphics[width=18.0cm]{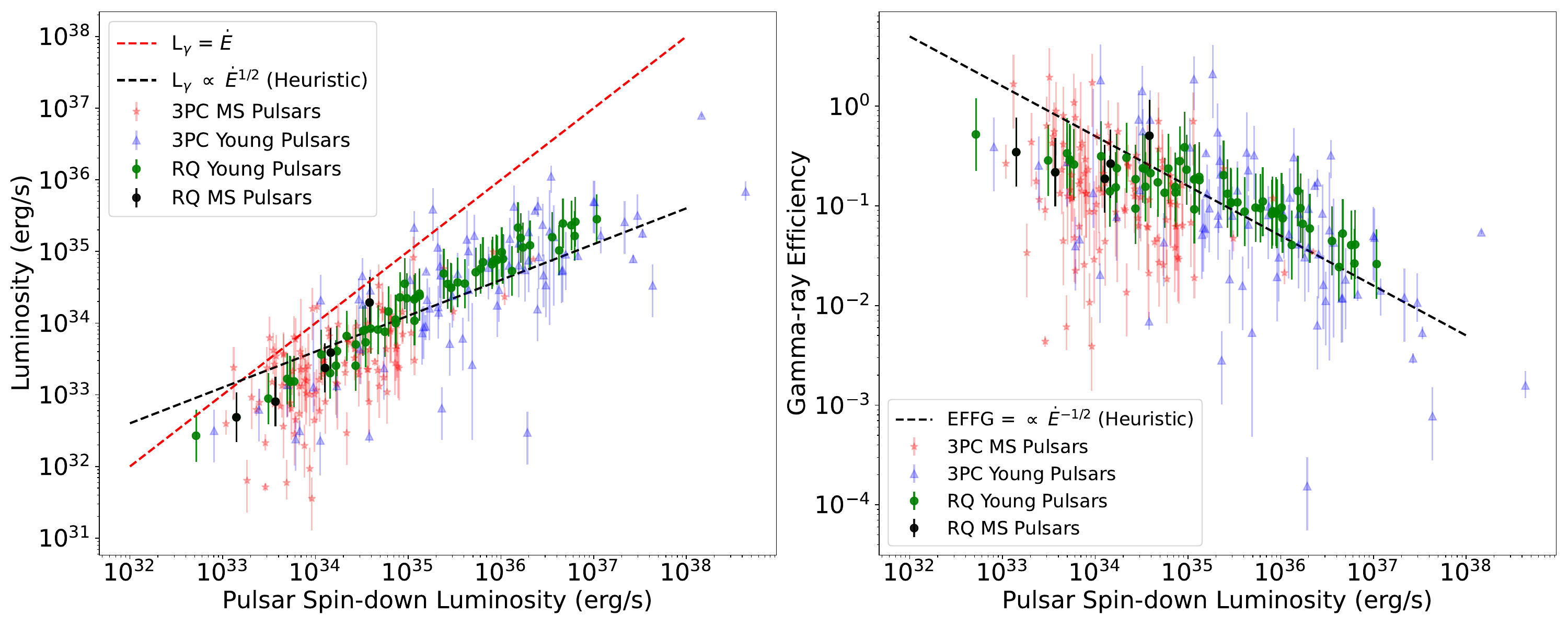}
\caption{Left panel: gamma-ray luminosity vs.~pulsar spin-down power. The blue triangles and red stars mark the young pulsars and MSPs listed in 3PC, respectively. The green and black circles indicate the RQ young pulsars and RQ MSPs with luminosities predicted from the OLS method, respectively. The vertical error bars correspond to statistical luminosity errors taking into account the distance uncertainties. The dashed red line represents 100$\%$ conversion of spin-down power into gamma rays, while the black dashed line indicates the heuristic luminosity, L$^{\text{h}}$~$\propto$~$\dot{\text{E}}^{1/2}$ \citep{heuristic}. Right panel: gamma-ray efficiency, EFFG = L$_{\gamma}$ / $\dot{\text{E}}$, vs.~pulsar spin-down power. The markers and error bars are similar to the ones shown in the left panel. The efficiency corresponding to heuristic luminosity, EFFG$^{\text{h}}$~$\propto$~$\dot{\text{E}}^{-1/2}$ is represented by the black dashed line.}
\label{pred_vs_real}
\end{figure*}

\begin{figure*}
\centering
\includegraphics[width=20.0cm]{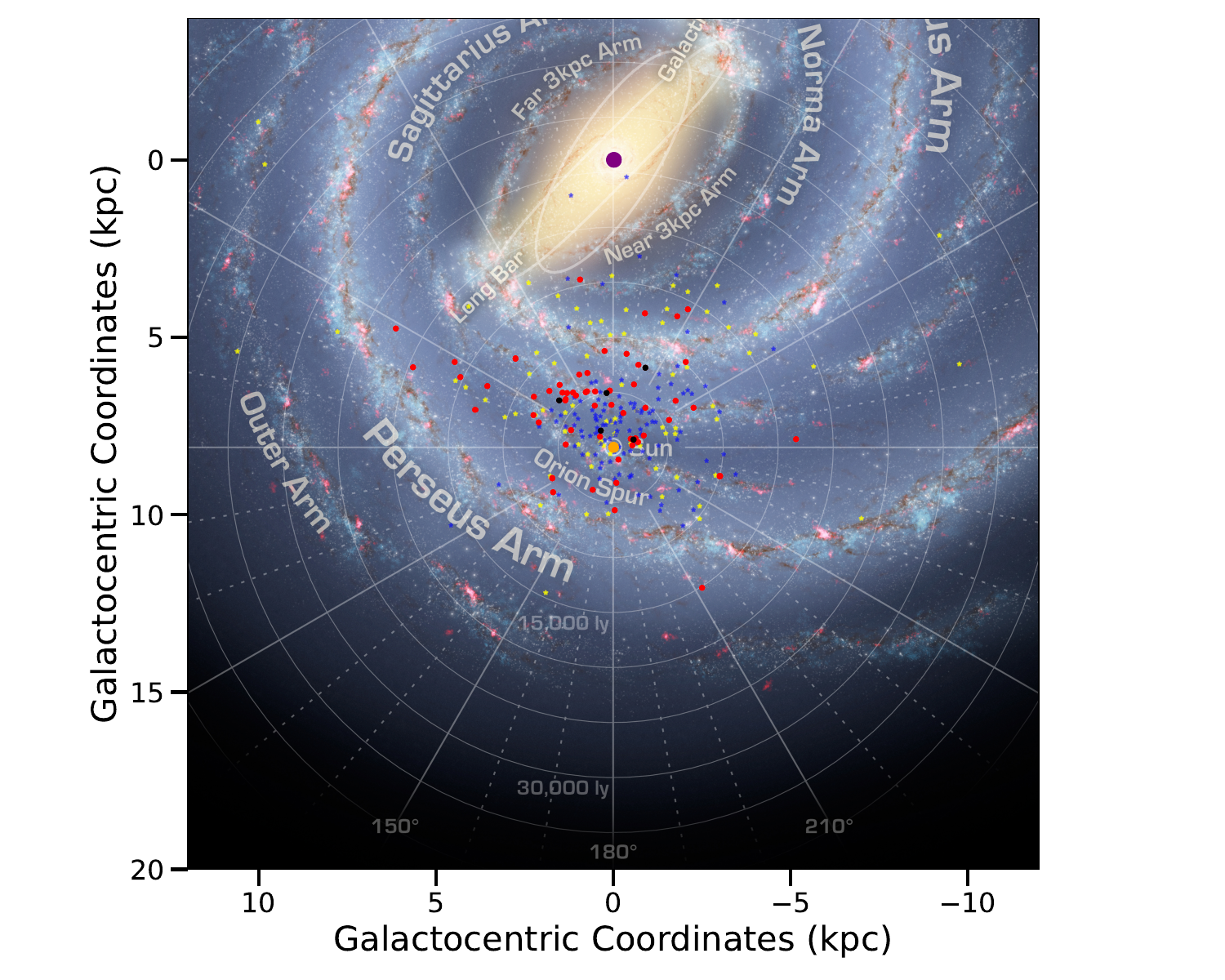}
\caption{The face-on view of the Milky Way indicating the positions of gamma-ray pulsars. The orange circle shows the position of the Sun, while the magenta circle indicates the location of the Galactic center. The yellow and blue stars indicate the location of young pulsars and MSPs in the Galaxy, taking into account the distance information provided in 3PC, respectively. The red and black circles show the position of RQ young pulsars and RQ MSPs with distances predicted from the OLS method, respectively. The plot is produced using the {\tt mw-plot} Python package (see \url{https://milkyway-plot.readthedocs.io/en/stable/index.html}).}
\label{gal_view}
\end{figure*}

Indeed, one of the most informative pulsar characteristic plots, offering valuable insights into the validity and compatibility of RQ pulsar distance predictions with the gamma-ray pulsar populations provided in 3PC, is the scatter plot of gamma-ray luminosity and efficiency as functions of spin-down luminosity. As shown in the left panel of Fig.~\ref{pred_vs_real}, the predicted luminosities of both RQ MSPs and young RQ pulsars are consistent with the gamma-ray pulsar populations provided in 3PC. Especially, the predicted luminosities of RQ MSPs fall well within the 95$\%$ C.L. upper limit of the 3PC MSP population, given as 2.0$\times$10$^{34}$ erg/s. Similarly, the gamma-ray efficiencies of the RQ MSPs dominate the high-efficiency regime (see Fig.~\ref{pred_vs_real}, right panel), with efficiency values ranging from 0.18 to 0.51. Furthermore, predicting the distances of RQ gamma-ray pulsars allows the determination of their spatial distribution within the Milky Way. Figure~\ref{gal_view} presents a face-on view of the Milky Way galaxy, along with the positions of RQ gamma-ray pulsars based on the predicted distances derived using the OLS method. As expected, the predicted distances of all RQ MSPs (indicated by black circles) are concentrated within the solar neighborhood. Considering both the luminosity and distance predictions for RQ MSPs, it is evident that the FP relation has the potential to distinguish between young pulsars and MSPs.

\begin{table*}[ht!]
\caption{Distance prediction of radio-quiet gamma-ray pulsars using different ML approaches. The Char Code column gives the detection codes together with pulsar properties: Q = no radio detection; X = discovered in the X-ray and/or gamma-ray pulsations detected using the X-ray ephemeris; W = black widow, which is a rapidly rotating neutron star in a binary system that emits intense radiation, gradually eroding its low-mass companion star; B = binary system; M = millisecond pulsar. The D3PC column gives the measured distance to the pulsar, while the 3PC Dist.~Method column indicates the measurement method used for obtaining the pulsar distance: P = parallax; K = kinematic (HI absorption); X = X-ray measurement of the hydrogen column density; O = optical methods. The D$_{\text{OLS}}$, D$_{\text{RFR}}$, D$_{\text{SVR-LIN}}$ and D$_{\text{SVR-RBF}}$ columns present the predicted pulsar distances using the OLS, RFR, SVR-Linear, and SVR-RBF methods, respectively.}
\centering
\renewcommand{\arraystretch}{1.3}
\begin{tabular}{cccccccc}
\hline\hline
Pulsar Name & Char Code & 3PC Dist.  & D3PC  & D$_{\text{OLS}}$ & D$_{\text{RFR}}$ & D$_{\text{SVR-LIN}}$ & D$_{\text{SVR-RBF}}$\\
PSR   &           & Method & (kpc) & (kpc)  & (kpc) & (kpc) & (kpc)\\
\hline\hline
J0007+7303   & Q    & K & \textbf{1.4$\pm$0.3}            & \textbf{1.2$^{+0.8}_{-0.3}$}    & 2.6  & 3.7  & 3.9\\
J0633+1746   & XQ   & P & \textbf{0.19$\pm$0.06}          & \textbf{0.13$^{+0.08}_{-0.04}$} & 0.53 & 0.54 & 0.53\\
J1418$-$6058 & Q    & O & \textbf{1.6$\pm$0.7}            & \textbf{2.3$^{+1.4}_{-0.6}$}    & 4.0  & 7.8  & 9.9\\
J1653$-$0158 & MBWQ & P & \textbf{0.74$^{+1.56}_{-0.06}$} & \textbf{0.66$^{+0.40}_{-0.18}$} & 1.66 & 1.63 & 1.27\\
J1809$-$2332 & Q    & K & \textbf{1.7$\pm$1.0}            & \textbf{1.0$^{+0.6}_{-0.3}$}    & 2.7  & 3.1  & 3.4\\
J1836+5925   & Q    & X & \textbf{0.53$\pm$0.27}          & \textbf{0.24$^{+0.15}_{-0.07}$} & 0.54 & 0.71 & 0.56\\
J2021+4026   & Q    & K & \textbf{1.50$\pm$0.45}          & \textbf{0.44$^{+0.27}_{-0.12}$} & 1.83 & 1.54 & 1.52\\
J2030+4415   & Q    & O & \textbf{0.50$^{+0.3}_{-0.2}$}   & \textbf{0.80$^{+0.50}_{-0.22}$} & 1.42 & 1.99 & 1.39\\

\hline\hline
\hline\hline
\end{tabular}
\label{rq_real_dist_test}
\end{table*}

Since the majority of RQ gamma-ray pulsars lack distance measurements, it is challenging to directly evaluate the accuracy of the predictions presented in this section. However, beyond these 62 RQ pulsars, there are nine RQ gamma-ray pulsars, namely PSR~J0007+7303, PSR~J0633+1746, PSR~J1418$-$6058, PSR~J1653$-$0158, PSR~J1809$-$2332, PSR~J1836$-$5925, PSR~J1846$-$0258, PSR~J2021+4026, and PSR~J2030+4415, for which distance measurements from various techniques are available and provided in 3PC. Among these, PSR~J1846$-$0258 lacks an SED model in 3PC, and the fit to the available SED data does not exhibit a significant spectral cutoff. Consequently, it is not possible to apply the FP relation to predict the distance of this pulsar. Nevertheless, the remaining small sample of eight RQ pulsars provides an opportunity to test the accuracy of distance predictions derived from different ML models. The predicted distances for these pulsars, using the ML models employed in this study, are presented in Table~\ref{rq_real_dist_test}. The results show that most of the predictions obtained with the OLS method fall within $\sim$1$\sigma$ of the statistical uncertainties associated with measured distances from various techniques. An exception is PSR~J2021+4026, where the OLS model underestimates the distance. In contrast, other ML methods tend to overestimate the distances for this particular RQ pulsar sample, although the SVR-Linear and SVR-RBF methods generally provide results consistent with each other. Given that the parallax method is the most reliable technique for distance measurement, it can be concluded that, for this small sample of RQ pulsars, the OLS method is the most promising one owing to its simplicity and robustness, providing a further validation that the FP relation can be used for distance prediction of RQ pulsars.

\section{Summary and conclusions}
\label{conc}

In this study, the potential of the FP relation for predicting the distances of RQ gamma-ray pulsars was explored using various ML approaches. A detailed investigation of the FP relation, using the OLS regression and data from 3PC, revealed that the FP exponents are compatible with theoretical expectations at the 1$\sigma$ level, provided that the spectral cutoff feature is significant. However, the inclusion of pulsars with insignificant cutoff features led to a divergence of the exponents from the theoretical values. Consequently, it is concluded that the FP relation is valid only for pulsars exhibiting a significant gamma-ray spectral cutoff in the GeV regime. The FP exponents estimated from the OLS analysis of 184 pulsars from 3PC showing significant cutoffs were found to be ($\epsilon_\mathrm{cut}$,~P,~$\dot{\text{P}}$)~=~(1.34$\pm$0.19, -1.21$\pm$0.17, 0.48$\pm$0.04), or equivalently ($\epsilon_\mathrm{cut}$, B$_{*}$, $\dot{\text{E}}$) = (1.34$\pm$0.19, 0.12$\pm$0.03, 0.42$\pm$0.05). It was also shown that the $\epsilon_\mathrm{cut}$ values obtained from 3PC PLEC4 models and from fitting PLEC models to observational Fermi SED data are interchangeable in the FP relation. This flexibility may be particularly useful for new pulsar detections. The preference for gamma-ray emission originating from the CR regime over the SR regime was evaluated using a log-likelihood ratio test, resulting in a significant preference for the CR regime at the 4.2$\sigma$ level. Additionally, the validity of the FP relation was tested separately for both young pulsar and MSP populations from 3PC. The results indicated that both populations obey the FP relation within 2$\sigma$ of the derived statistical uncertainties. Further analysis of influential observations revealed that most of the outliers in the FP relation correspond to pulsars with either exceptionally high ($>$100$\%$) or extremely low ($<$1$\%$) gamma-ray efficiencies. Based on the adjusted R$^{2}$ performance metric, removing these outlier pulsar data improves the gamma-ray luminosity estimation model by $\sim$10$\%$ compared to the model that includes outliers, leading to more tightly constrained FP relation exponents.

The proof of principle regarding the estimation of gamma-ray pulsar distances using the FP relation was demonstrated with 3PC pulsars that have known distances. It was shown that, using various ML approaches, including the OLS, RFR, SVR-Linear, and SVR-RBF models, it is possible to explain between 55$\%$ and 66$\%$ of variations in the measured distance data. The results also indicated that incorporating additional variables with the FP relation, such as the significance of the spectral cutoff, can enhance predictive model performance. Using these optimized ML models based on pulsars with known distances, predictions for the distances of 62 RQ gamma-ray pulsars were obtained. It is important to note that there are currently no traditional observational distance measurements available for these pulsars in the literature. The predicted distances for both RQ young pulsars and RQ MSPs were found to be consistent with the 3PC populations, in terms of both luminosity and distance. Additionally, the distance predictions for eight unique RQ gamma-ray pulsars, for which traditional distance measurements are available, were cross-checked using the optimized ML predictive models. The majority of the optimized OLS model predictions fell within 1$\sigma$ of the statistical uncertainties associated with the measured distances, while other ML methods generally tended to overestimate the distances.

It is possible to discuss three main factors that can influence the accuracy of the predictions. The first one is the determination of the beaming factor, f$_{\Omega}$, which was assumed to be f$_{\Omega}$=1 throughout this study. As discussed extensively in \cite{fermi3pc, fp2022}, this factor depends heavily on the model and the observer's angle. Furthermore, obtaining the inclination angles ($\alpha$, $\zeta$), which are needed to estimate f$_{\Omega}$, from observations is quite challenging and often unavailable for many pulsars. Although assuming f$_{\Omega}$=1 for all pulsars does not affect the FP relation exponents, but only the scaling, variations in f$_{\Omega}$ contribute to the spread in estimated luminosities. Ideally, applying f$_{\Omega}$ corrections using reliable inclination angle estimates and incorporating this factor as an independent variable in the FP relation could reduce this spread, resulting in more accurate distance predictions. The second factor is the reliable determination of spectral cutoffs and their corresponding significance levels in the CR regime. Since the theoretical FP relation is highly dependent on $\epsilon_\mathrm{cut}$, this factor can substantially impact distance prediction accuracy. In particular, when a second VHE IC component is present, accurately determining $\epsilon_\mathrm{cut}$ becomes more challenging. In such cases, more sophisticated SED modeling, such as joint spectral models that account for both CR and IC components, may be required. Future VHE gamma-ray observatories, such as the Cherenkov Telescope Array (CTA; \cite{cta, HofmannZanin2023}), with their unprecedented sensitivity down to energies as low as $\sim$20~GeV, will be able to effectively resolve these spectral IC components. This will make a more robust determination of the $\epsilon_\mathrm{cut}$ parameter possible, particularly when joint analyses are conducted together with Fermi LAT data. Another important factor are the growing population statistics of known gamma-ray pulsars. Over the past decade, the number of gamma-ray pulsars has increased significantly, from only 7 known pulsars in 2009 to 46 in 1PC, 117 in 2PC, and finally 294 in 3PC. This upward trend is expected to continue, with the forthcoming Fermi Fourth Pulsar Catalog projected to contain at least $\sim$400 gamma-ray pulsars \citep{fermi3pc}. The growing sample size will lead to tighter constraints on the FP exponents and reduce parameter uncertainties, ultimately improving prediction accuracy. Furthermore, applying accurate Shklovskii corrections for all pulsars, along with refining $\dot{\text{E}}$ for individual pulsars based on precise mass and radius measurements, has the potential to further improve the accuracy of distance predictions.

The results presented in this study can be reproduced by reconstructing the optimized prediction models using public statistical packages, such as {\tt scikit-learn} or {\tt statsmodels}. These models can also be applied to estimate distances, especially for newly detected RQ gamma-ray pulsars. For this purpose, the pulsars used in training the optimized models are provided in Table~\ref{knownPulsars} in Appendix~\ref{A1}, where a star symbol ($\star$) marks those used for OLS, a dagger ($\dagger$) marks those used for RFR, a triangle ($\triangle$) marks those used for SVR-Linear, and a section symbol ($\S$) marks those used for SVR-RBF. Based on the findings of this study, the FP relation obtained from the OLS approach, given in Eq.~\ref{fp_optimized}, is recommended for future applications, as it provides a straightforward and effective tool for estimating gamma-ray pulsar distances. The FP exponents in Eq.~\ref{fp_optimized}, in either ($\epsilon_\mathrm{cut}$,~P,~$\dot{\text{P}}$) or ($\epsilon_\mathrm{cut}$, B$_{*}$, $\dot{\text{E}}$) representation, can be used in Eq.~\ref{dist_est2} to directly estimate distances. However, it is important to note that when propagating uncertainties the full covariance matrix from the OLS luminosity prediction model should be taken into account to ensure accurate error estimation. While the OLS method provides a straightforward baseline, ML methods offer additional advantages, particularly in capturing nonlinear dependencies and complex parameter interactions. Tree-based methods, such as RFR, can better identify hidden correlations that may not be well captured by linear models. Similarly, kernel-based approaches like SVR can model nonlinear variations in the data, which may arise owing to complexities in the spectral models, determination of spectral cutoffs, beaming effects, or other astrophysical factors. These advantages are particularly prominent when incorporating additional variables, such as spectral cutoff significance, as demonstrated in this study. While the OLS-based FP relation provides a reliable tool for distance estimation, ML models may be used as complementary approaches, particularly when additional predictive variables become available. Consequently, ML models utilize multidimensional relationships, which can enhance predictive performance and improve generalizability, especially for future data sets. As the number of detected gamma-ray pulsars increases, periodic retraining of the predictive models will be necessary. Expanding the training data set with newly detected pulsars can enhance model generalizability, refine the FP relation, and improve both distance estimates and associated uncertainties. The optimization methods outlined in this study provide a systematic approach for recalibrating both OLS and ML models, ensuring that future predictions remain robust and consistent with the most up-to-date observational data. Additionally, HP tuning, which is essential for ML models, and outlier selection criteria should be revisited as the data set grows, as they play a crucial role in ensuring model robustness and reliability.

This study demonstrates the proof of principle that the distances of RQ pulsars can be predicted with reasonable accuracy by using the FP relation in combination with ML algorithms. Given that traditional distance measurement techniques often fail to estimate distances for RQ gamma-ray pulsars, this study introduces a new methodology that, for the first time, directly uses gamma-ray emission for pulsar distance estimation. The accuracy of these predictions can be further improved with the availability of larger data sets in the future, alongside the refinement of key factors mentioned earlier and the application of more sophisticated ML algorithms.

\begin{appendix}
\section{Properties of 3PC gamma-ray pulsars with known distances}\label{A1}

Table~\ref{knownPulsars} presents the properties of 3PC gamma-ray pulsars with known distances, which were used for model training and testing. Listed parameters include the pulsar period, Shklovskii-corrected period derivative, spin-down luminosity, magnetic ﬁeld strength at the neutron star surface, and gamma-ray luminosity. Spectral cutoff energies derived from the SED data and spectral models are also provided, along with the associated test statistics used for quantifying spectral cutoff signiﬁcance.

\vspace{-10pt} 
\renewcommand{\arraystretch}{1.3}
\begin{center}
\tiny
\begin{longtable}[htbp]{c|ccccc|cccc}
\caption{Properties of 3PC gamma-ray pulsars with known distances used for model training and testing. The LUMG column indicates the pulsar luminosity, taking into account the measured gamma-ray flux and pulsar distance. The P, $\dot{\text{P}}_{\text{Corr}}$, $\dot{\text{E}}_{\text{Corr}}$ and $\dot{\text{B}}_{\text{S,Corr}}$ columns give the period, Shklovskii effect-corrected period derivative, spin-down luminosity, and magnetic field strength at the neutron star surface, respectively. The E$_{\text{Cut,SED}}$, E$_{\text{Cut,B23}}$ and E$_{\text{Cut,BFR}}$ columns report the derived PLEC spectral cutoff values obtained from the 3PC pulsar SED data and synthetic SED data points generated from the PLEC4 b$_{23}$ and b$_{\text{fr}}$ models, respectively. The corresponding test statistic values from the likelihood ratio tests comparing PLEC and PL models are provided in parentheses. The E$_{\text{Cut,HYB}}$ column combines the values from E$_{\text{Cut,B23}}$ and E$_{\text{Cut,BFR}}$, prioritizing the E$_{\text{Cut,BFR}}$ whenever available, and also presents the spectral cutoff values used in the FP relation. Pulsars denoted in bold represent the 184 pulsars used for re-validation of the FP relation in Sect.~\ref{fp_revisited}, exhibiting marginally significant cutoffs with TS$_{\text{Cut}}\geq$4.0. Pulsars marked with an asterisk ($\star$), dagger ($\dagger$), triangle ($\triangle$) and section symbol ($\S$) indicate the pulsars used in the optimized prediction model training for OLS, RFR, SVR-Linear and SVR-RBF approaches, respectively. Please note that identical ML distance prediction models given in Table~\ref{mltable} can be constructed using the relevant training pulsars and the corresponding HP values.} \label{knownPulsars} \\
\hline\hline
Pulsar Name & LUMG & P & $\dot{\text{P}}_{\text{Corr}}$ & $\dot{\text{E}}_{\text{Corr}}$ & $\dot{\text{B}}_{\text{S,Corr}}$ & E$_{\text{Cut,SED}}$ (TS) & E$_{\text{Cut,B23}}$ (TS) & E$_{\text{Cut,BFR}}$ (TS) & E$_{\text{Cut,HYB}}$ (TS) \\
PSR & erg/s & ms & s/s & erg/s & G & GeV & GeV & GeV & GeV \\
\hline
\endfirsthead
\caption[]{\textit{(Continued table from previous page)}} \\
\hline
Pulsar Name & LUMG & P & $\dot{\text{P}}_{\text{Corr}}$ & $\dot{\text{E}}_{\text{Corr}}$ & $\dot{\text{B}}_{\text{S,Corr}}$ & E$_{\text{Cut,SED}}$ (TS) & E$_{\text{Cut,B23}}$ (TS) & E$_{\text{Cut,BFR}}$ (TS) & E$_{\text{Cut,HYB}}$ (TS) \\
PSR & erg/s & ms & s/s & erg/s & G & GeV & GeV & GeV & GeV \\
\hline
\endhead

\hline
\endfoot

\hline
\endlastfoot

\textbf{J0002+6216$^{\star\dagger\S}$}  &  9.01e+33  &  115.36  &  5.97e-15  &  1.53e+35  &  8.40e+11  &  2.25 (29.3) &  2.65 (313.2) &  2.55 (206.0) &  2.55 (206.0) \\ 
\textbf{J0007+7303$^{\dagger\triangle\S}$}  &  1.01e+35  &  315.89  &  3.56e-13  &  4.46e+35  &  1.07e+13  &  4.84 (8439.0) &  5.02 (4924.3) &  4.8 (3457.2) &  4.8 (3457.2) \\ 
\textbf{J0023+0923$^{\dagger\triangle\S}$}  &  3.03e+33  &  3.05  &  8.88e-21  &  1.24e+34  &  1.67e+08  &  1.68 (79.5) &  2.54 (175.6) &  N/A  &  2.54 (175.6) \\ 
\textbf{J0030+0451$^{\star\dagger\triangle\S}$}  &  7.43e+32  &  4.87  &  1.05e-20  &  3.60e+33  &  2.29e+08  &  1.61 (1708.0) &  2.29 (2801.5) &  2.35 (2141.8) &  2.35 (2141.8) \\ 
\textbf{J0034-0534$^{\star\dagger\triangle\S}$}  &  4.38e+33  &  1.88  &  4.47e-21  &  2.66e+34  &  9.28e+07  &  2.48 (264.6) &  3.6 (439.8) &  3.61 (308.0) &  3.61 (308.0) \\ 
\textbf{J0101-6422$^{\star}$}  &  1.56e+33  &  2.57  &  3.44e-21  &  8.01e+33  &  9.52e+07  &  1.66 (293.2) &  2.48 (642.7) &  2.54 (483.7) &  2.54 (483.7) \\ 
\textbf{J0102+4839$^{\star\dagger\triangle}$}  &  9.59e+33  &  2.96  &  1.16e-20  &  1.76e+34  &  1.87e+08  &  3.83 (113.4) &  4.23 (200.7) &  3.85 (155.4) &  3.85 (155.4) \\ 
\textbf{J0106+4855$^{\star\dagger\S}$}  &  2.16e+34  &  83.16  &  4.28e-16  &  2.94e+34  &  1.91e+11  &  2.77 (359.4) &  2.83 (513.9) &  2.99 (314.3) &  2.99 (314.3) \\ 
\textbf{J0205+6449 }  &  7.88e+34  &  65.75  &  1.92e-13  &  2.67e+37  &  3.60e+12  &  5.2 (92.2) &  3.01 (73.8) &  4.12 (55.2) &  4.12 (55.2) \\ 
\textbf{J0218+4232$^{\star\triangle\S}$}  &  5.74e+34  &  2.32  &  7.65e-20  &  2.42e+35  &  4.26e+08  &  3.4 (339.3) &  4.68 (576.0) &  4.56 (428.2) &  4.56 (428.2) \\ 
\textbf{J0248+4230$^{\star\triangle}$}  &  7.95e+32  &  2.6  &  1.68e-20  &  3.78e+34  &  2.12e+08  &  2.1 (5.0) &  1.58 (62.3) &  N/A  &  1.58 (62.3) \\ 
\textbf{J0248+6021$^{\star\dagger\S}$}  &  1.41e+34  &  217.12  &  5.52e-14  &  2.13e+35  &  3.50e+12  &  1.36 (134.4) &  1.72 (598.8) &  1.38 (515.6) &  1.38 (515.6) \\ 
\textbf{J0251+2606$^{\star\dagger\triangle}$}  &  7.99e+32  &  2.54  &  5.29e-21  &  1.28e+34  &  1.17e+08  &  1.94 (43.5) &  2.21 (126.8) &  N/A  &  2.21 (126.8) \\ 
\textbf{J0307+7443$^{\triangle\S}$}  &  2.93e+32  &  3.16  &  1.71e-20  &  2.17e+34  &  2.35e+08  &  1.58 (606.3) &  1.89 (863.4) &  1.72 (593.5) &  1.72 (593.5) \\ 
\textbf{J0312-0921$^{\star}$}  &  4.43e+32  &  3.7  &  1.25e-20  &  9.70e+33  &  2.17e+08  &  1.43 (29.5) &  1.65 (183.1) &  N/A  &  1.65 (183.1) \\ 
\textbf{J0318+0253$^{\star\dagger\triangle}$}  &  1.33e+33  &  5.19  &  1.76e-20  &  4.98e+33  &  3.06e+08  &  2.06 (16.4) &  2.2 (175.4) &  N/A  &  2.2 (175.4) \\ 
\textbf{J0340+4130$^{\dagger\triangle\S}$}  &  5.94e+33  &  3.3  &  5.53e-21  &  6.08e+33  &  1.37e+08  &  3.49 (216.4) &  3.5 (368.0) &  3.34 (243.4) &  3.34 (243.4) \\ 
\textbf{J0418+6635$^{\star\S}$}  &  6.02e+33  &  2.91  &  1.37e-20  &  2.19e+34  &  2.02e+08  &  3.63 (103.6) &  4.23 (125.0) &  4.09 (85.4) &  4.09 (85.4) \\ 
\textbf{J0437-4715 }  &  5.17e+31  &  5.76  &  1.40e-20  &  2.89e+33  &  2.87e+08  &  1.02 (450.4) &  1.38 (1031.6) &  0.7 (1019.6) &  0.7 (1019.6) \\ 
\textbf{J0514-4408$^{\dagger\S}$}  &  6.23e+32  &  320.27  &  2.04e-15  &  2.45e+33  &  8.18e+11  &  0.49 (27.5) &  0.85 (199.7) &  N/A  &  0.85 (199.7) \\ 
\textbf{J0533+6759 }  &  6.38e+33  &  4.39  &  1.26e-20  &  5.90e+33  &  2.38e+08  &  3.45 (137.0) &  3.55 (226.1) &  3.61 (143.7) &  3.61 (143.7) \\ 
\textbf{J0534+2200 }  &  6.85e+35  &  33.65  &  4.20e-13  &  4.35e+38  &  3.81e+12  &  5.34 (534.0) &  8.31 (2238.3) &  8.26 (1217.8) &  8.26 (1217.8) \\ 
\textbf{J0540-6919 }  &  7.89e+36  &  50.65  &  4.79e-13  &  1.46e+38  &  4.98e+12  &  6.59 (28.0) &  5.41 (101.7) &  N/A  &  5.41 (101.7) \\ 
\textbf{J0605+3757 }  &  3.56e+31  &  2.73  &  4.72e-21  &  9.17e+33  &  1.15e+08  &  1.25 (44.5) &  1.55 (185.0) &  N/A  &  1.55 (185.0) \\ 
\textbf{J0610-2100$^{\star\dagger\triangle}$}  &  4.19e+33  &  3.86  &  5.10e-21  &  3.50e+33  &  1.42e+08  &  1.9 (35.0) &  1.99 (249.0) &  N/A  &  1.99 (249.0) \\ 
\textbf{J0613-0200$^{\star\dagger\triangle\S}$}  &  2.78e+33  &  3.06  &  8.88e-21  &  1.22e+34  &  1.67e+08  &  2.55 (572.2) &  3.23 (781.3) &  3.06 (387.2) &  3.06 (387.2) \\ 
\textbf{J0614-3329$^{\star\dagger\triangle\S}$}  &  5.44e+33  &  3.15  &  1.80e-20  &  2.27e+34  &  2.41e+08  &  3.84 (3612.0) &  4.47 (2623.0) &  4.06 (2046.2) &  4.06 (2046.2) \\ 
\textbf{J0621+2514$^{\star\S}$}  &  1.31e+33  &  2.72  &  2.49e-20  &  4.87e+34  &  2.63e+08  &  1.13 (14.4) &  1.47 (36.6) &  N/A  &  1.47 (36.6) \\ 
\textbf{J0631+0646$^{\star\dagger\S}$}  &  4.03e+34  &  110.98  &  3.62e-15  &  1.04e+35  &  6.41e+11  &  2.87 (109.4) &  2.66 (92.1) &  N/A  &  2.66 (92.1) \\ 
\textbf{J0631+1036$^{\star\dagger\triangle}$}  &  1.61e+34  &  287.82  &  1.03e-13  &  1.70e+35  &  5.50e+12  &  3.15 (280.4) &  3.49 (330.7) &  3.39 (198.3) &  3.39 (198.3) \\ 
\textbf{J0633+1746$^{\dagger\triangle\S}$ }  &  1.82e+34  &  237.1  &  1.10e-14  &  3.25e+34  &  1.63e+12  &  2.2 (94921.0) &  2.8 (70631.5) &  2.79 (56579.4) &  2.79 (56579.4) \\ 
\textbf{J0653+4706$^{\dagger\S}$ }  &  1.95e+32  &  4.75  &  2.08e-20  &  7.61e+33  &  3.18e+08  &  0.96 (16.2) &  0.93 (59.2) &  N/A  &  0.93 (59.2) \\ 
\textbf{J0659+1414 }  &  2.64e+32  &  384.92  &  5.50e-14  &  3.80e+34  &  4.65e+12  &  0.54 (78.8) &  0.83 (169.8) &  1.74 (177.5) &  1.74 (177.5) \\ 
\textbf{J0729-1448$^{\star\dagger\triangle\S}$}  &  5.16e+33  &  251.71  &  1.13e-13  &  2.81e+35  &  5.41e+12  &  Inf. (0.0) &  4.43 (10.6) &  N/A  &  4.43 (10.6) \\ 
\textbf{J0740+6620$^{\star\dagger\triangle}$}  &  4.48e+32  &  2.88  &  3.70e-21  &  6.12e+33  &  1.04e+08  &  2.02 (35.1) &  2.91 (72.1) &  N/A  &  2.91 (72.1) \\ 
\textbf{J0742-2822$^{\dagger}$}  &  7.28e+33  &  166.77  &  1.67e-14  &  1.42e+35  &  1.69e+12  &  2.23 (75.8) &  3.26 (151.1) &  N/A  &  3.26 (151.1) \\ 
\textbf{J0751+1807$^{\star\dagger}$}  &  1.48e+33  &  3.48  &  5.93e-21  &  5.56e+33  &  1.45e+08  &  2.0 (170.4) &  1.99 (296.5) &  1.96 (176.2) &  1.96 (176.2) \\ 
\textbf{J0835-4510$^{\star\dagger\triangle\S}$}  &  8.72e+34  &  89.37  &  1.22e-13  &  6.76e+36  &  3.35e+12  &  4.06 (61788.6) &  3.96 (51003.4) &  3.87 (44236.6) &  3.87 (44236.6) \\ 
\textbf{J0908-4913 }  &  2.62e+33  &  106.77  &  1.51e-14  &  4.90e+35  &  1.29e+12  &  1.65 (17.7) &  1.48 (186.5) &  N/A  &  1.48 (186.5) \\ 
J0922+0638  &  3.12e+32  &  430.62  &  1.37e-14  &  6.77e+33  &  2.46e+12  &  13.6 (0.0) &  8.86 (1.6) &  N/A  &  8.86 (1.6) \\ 
\textbf{J0931-1902 }  &  2.39e+33  &  4.64  &  3.32e-21  &  1.31e+33  &  1.26e+08  &  0.83 (18.7) &  0.75 (36.2) &  N/A  &  0.75 (36.2) \\ 
\textbf{J0940-5428 }  &  2.97e+32  &  87.56  &  3.28e-14  &  1.93e+36  &  1.71e+12  &  1.74 (33.2) &  2.5 (213.8) &  N/A  &  2.5 (213.8) \\ 
\textbf{J0952-0607$^{\star\dagger\triangle}$}  &  1.10e+34  &  1.41  &  4.76e-21  &  6.65e+34  &  8.29e+07  &  1.14 (15.5) &  1.47 (63.1) &  N/A  &  1.47 (63.1) \\ 
\textbf{J0955-6150$^{\star\triangle\S}$}  &  4.36e+33  &  1.99  &  1.46e-20  &  7.33e+34  &  1.73e+08  &  2.2 (12.1) &  3.42 (84.8) &  N/A  &  3.42 (84.8) \\ 
\textbf{J1012-4235 }  &  9.32e+31  &  3.1  &  6.55e-21  &  8.68e+33  &  1.44e+08  &  9.52 (0.5) &  5.3 (38.0) &  N/A  &  5.3 (38.0) \\ 
\textbf{J1016-5857$^{\star\dagger\triangle}$}  &  8.38e+34  &  107.4  &  8.04e-14  &  2.56e+36  &  2.97e+12  &  5.52 (78.4) &  4.26 (82.7) &  3.92 (58.4) &  3.92 (58.4) \\ 
\textbf{J1019-5749 }  &  3.88e+35  &  162.51  &  2.01e-14  &  1.85e+35  &  1.83e+12  &  5.29 (17.9) &  5.82 (18.1) &  N/A  &  5.82 (18.1) \\ 
J1023+0038  &  7.25e+33  &  1.69  &  5.23e-21  &  4.28e+34  &  9.51e+07  &  2.87 (135.4) &  N/A  &  N/A  &  N/A  \\ 
\textbf{J1024-0719$^{\star\dagger\triangle\S}$}  &  7.67e+32  &  5.16  &  1.86e-20  &  5.33e+33  &  3.13e+08  &  2.32 (10.4) &  2.57 (102.6) &  N/A  &  2.57 (102.6) \\ 
\textbf{J1028-5819$^{\star\dagger}$}  &  5.94e+34  &  91.4  &  1.42e-14  &  7.34e+35  &  1.15e+12  &  4.72 (679.9) &  4.72 (1478.8) &  3.99 (1129.9) &  3.99 (1129.9) \\ 
\textbf{J1035-6720$^{\star\dagger\triangle\S}$}  &  4.81e+33  &  2.87  &  4.56e-20  &  7.62e+34  &  3.66e+08  &  1.62 (272.1) &  2.21 (694.0) &  2.2 (497.1) &  2.2 (497.1) \\ 
\textbf{J1036-8317$^{\dagger\triangle\S}$}  &  2.29e+33  &  3.41  &  2.90e-20  &  2.88e+34  &  3.18e+08  &  1.33 (61.0) &  2.57 (86.0) &  N/A  &  2.57 (86.0) \\ 
\textbf{J1048+2339$^{\S}$}  &  1.02e+33  &  4.67  &  2.51e-20  &  9.73e+33  &  3.47e+08  &  9.36 (0.8) &  4.5 (48.5) &  N/A  &  4.5 (48.5) \\ 
\textbf{J1048-5832$^{\star\dagger\triangle\S}$}  &  1.86e+35  &  123.71  &  9.55e-14  &  1.99e+36  &  3.48e+12  &  4.13 (1096.8) &  4.33 (1572.8) &  3.91 (1178.8) &  3.91 (1178.8) \\ 
\textbf{J1055-6028 }  &  3.36e+34  &  99.66  &  1.09e-12  &  4.33e+37  &  1.05e+13  &  4.34 (24.9) &  3.9 (40.9) &  N/A  &  3.9 (40.9) \\ 
\textbf{J1057-5226$^{\star\dagger\triangle\S}$}  &  4.34e+33  &  197.11  &  5.84e-15  &  3.01e+34  &  1.09e+12  &  1.25 (11133.7) &  1.63 (13319.6) &  1.51 (12333.6) &  1.51 (12333.6) \\ 
\textbf{J1105-6107 }  &  1.56e+34  &  63.2  &  1.58e-14  &  2.48e+36  &  1.01e+12  &  3.15 (12.2) &  3.59 (59.5) &  N/A  &  3.59 (59.5) \\ 
\textbf{J1112-6103$^{\star\dagger\triangle}$}  &  5.33e+34  &  64.97  &  3.15e-14  &  4.54e+36  &  1.45e+12  &  18.55 (1.0) &  5.78 (13.2) &  N/A  &  5.78 (13.2) \\ 
\textbf{J1119-6127$^{\star\dagger\S}$}  &  3.71e+35  &  409.14  &  4.04e-12  &  2.33e+36  &  4.12e+13  &  11.27 (10.7) &  3.67 (24.3) &  2.49 (33.3) &  2.49 (33.3) \\ 
\textbf{J1124-3653$^{\dagger\triangle\S}$}  &  1.46e+33  &  2.41  &  5.91e-21  &  1.67e+34  &  1.21e+08  &  3.52 (136.4) &  3.43 (260.4) &  3.45 (249.2) &  3.45 (249.2) \\ 
\textbf{J1124-5916$^{\dagger\triangle\S}$}  &  1.68e+35  &  135.54  &  7.51e-13  &  1.19e+37  &  1.02e+13  &  2.27 (279.1) &  3.72 (632.5) &  4.08 (344.2) &  4.08 (344.2) \\ 
\textbf{J1125-5825$^{\star\dagger\triangle\S}$}  &  2.35e+33  &  3.1  &  6.00e-20  &  7.95e+34  &  4.36e+08  &  2.94 (59.2) &  3.28 (40.7) &  N/A  &  3.28 (40.7) \\ 
\textbf{J1125-6014$^{\dagger}$}  &  3.61e+32  &  2.63  &  2.08e-21  &  4.51e+33  &  7.48e+07  &  4.56 (0.5) &  1.73 (17.3) &  N/A  &  1.73 (17.3) \\ 
\textbf{J1137+7528$^{\star\dagger\triangle\S}$}  &  1.06e+33  &  2.51  &  3.20e-21  &  7.96e+33  &  9.07e+07  &  9.4 (0.1) &  1.73 (5.3) &  N/A  &  1.73 (5.3) \\ 
\textbf{J1142+0119$^{\triangle}$}  &  3.63e+33  &  5.08  &  1.50e-20  &  4.52e+33  &  2.79e+08  &  3.39 (67.3) &  3.87 (80.3) &  N/A  &  3.87 (80.3) \\ 
\textbf{J1151-6108$^{\dagger\triangle}$}  &  6.10e+33  &  101.63  &  1.03e-14  &  3.87e+35  &  1.03e+12  &  1.48 (22.4) &  2.45 (94.0) &  N/A  &  2.45 (94.0) \\ 
\textbf{J1207-5050$^{\star\dagger\triangle\S}$}  &  9.20e+32  &  4.84  &  5.88e-21  &  2.05e+33  &  1.71e+08  &  1.41 (22.3) &  1.74 (114.1) &  N/A  &  1.74 (114.1) \\ 
\textbf{J1221-0633$^{\dagger\triangle\S}$}  &  1.09e+33  &  1.93  &  1.09e-20  &  5.92e+34  &  1.46e+08  &  2.46 (44.2) &  3.13 (101.9) &  N/A  &  3.13 (101.9) \\ 
\textbf{J1227-4853$^{\star\triangle\S}$}  &  6.98e+33  &  1.69  &  1.07e-20  &  8.73e+34  &  1.36e+08  &  5.82 (36.3) &  5.67 (70.5) &  5.5 (36.6) &  5.5 (36.6) \\ 
\textbf{J1231-1411$^{\star\dagger}$}  &  2.13e+33  &  3.68  &  7.18e-21  &  5.69e+33  &  1.64e+08  &  1.89 (2996.6) &  2.55 (4329.8) &  2.55 (3067.6) &  2.55 (3067.6) \\ 
\textbf{J1253-5820$^{\dagger\S}$}  &  1.37e+33  &  255.5  &  2.11e-15  &  4.98e+33  &  7.42e+11  &  5.07 (0.0) &  0.82 (35.9) &  N/A  &  0.82 (35.9) \\ 
\textbf{J1301+0833$^{\S}$}  &  2.88e+33  &  1.84  &  6.89e-21  &  4.37e+34  &  1.14e+08  &  1.97 (37.8) &  2.48 (220.5) &  N/A  &  2.48 (220.5) \\ 
\textbf{J1302-3258$^{\triangle\S}$}  &  2.67e+33  &  3.77  &  4.90e-21  &  3.61e+33  &  1.38e+08  &  2.3 (181.6) &  2.34 (297.5) &  2.41 (196.0) &  2.41 (196.0) \\ 
\textbf{J1311-3430$^{\star\S}$}  &  1.81e+34  &  2.56  &  2.05e-20  &  4.83e+34  &  2.32e+08  &  3.92 (605.6) &  5.04 (731.4) &  5.06 (499.0) &  5.06 (499.0) \\ 
\textbf{J1312+0051$^{\star\dagger\triangle}$}  &  3.78e+33  &  4.23  &  8.70e-21  &  4.54e+33  &  1.94e+08  &  1.73 (291.3) &  2.05 (529.3) &  1.93 (422.6) &  1.93 (422.6) \\ 
J1341-6220  &  4.24e+35  &  193.44  &  2.53e-13  &  1.38e+36  &  7.08e+12  &  3538.51 (0.0) &  16.81 (1.7) &  N/A  &  16.81 (1.7) \\ 
\textbf{J1357-6429$^{\star\triangle\S}$}  &  3.39e+34  &  166.19  &  3.54e-13  &  3.05e+36  &  7.77e+12  &  1.31 (47.2) &  1.92 (338.5) &  N/A  &  1.92 (338.5) \\ 
\textbf{J1400-1431 }  &  5.93e+31  &  3.08  &  3.62e-21  &  4.89e+33  &  1.07e+08  &  1.33 (36.7) &  1.98 (188.0) &  N/A  &  1.98 (188.0) \\ 
J1402+1306  &  9.14e+31  &  5.89  &  1.35e-20  &  2.60e+33  &  2.85e+08  &  0.91 (18.2) &  N/A  &  N/A  &  N/A  \\ 
\textbf{J1410-6132$^{\star\triangle\S}$}  &  4.98e+35  &  50.06  &  3.18e-14  &  1.00e+37  &  1.28e+12  &  7.69 (0.2) &  2.24 (27.3) &  N/A  &  2.24 (27.3) \\ 
\textbf{J1418-6058 }  &  9.10e+34  &  110.58  &  1.71e-13  &  4.99e+36  &  4.40e+12  &  4.71 (102.9) &  3.89 (286.8) &  3.6 (290.6) &  3.6 (290.6) \\ 
\textbf{J1420-6048$^{\star\dagger\triangle\S}$}  &  4.90e+35  &  68.21  &  8.24e-14  &  1.03e+37  &  2.40e+12  &  10.41 (1.8) &  3.21 (86.9) &  N/A  &  3.21 (86.9) \\ 
\textbf{J1431-4715$^{\dagger\triangle}$}  &  1.74e+33  &  2.01  &  1.12e-20  &  5.43e+34  &  1.52e+08  &  Inf. (0.0) &  2.16 (56.0) &  N/A  &  2.16 (56.0) \\ 
\textbf{J1446-4701$^{\star\dagger}$}  &  2.26e+33  &  2.19  &  9.63e-21  &  3.62e+34  &  1.47e+08  &  4.0 (36.4) &  3.86 (81.3) &  N/A  &  3.86 (81.3) \\ 
\textbf{J1455-3330 }  &  6.36e+31  &  7.99  &  2.37e-20  &  1.83e+33  &  4.40e+08  &  4.21 (0.4) &  1.45 (16.6) &  N/A  &  1.45 (16.6) \\ 
\textbf{J1509-5850$^{\star\S}$}  &  1.66e+35  &  88.92  &  9.20e-15  &  5.17e+35  &  9.15e+11  &  3.84 (368.1) &  4.12 (531.0) &  4.1 (359.7) &  4.1 (359.7) \\ 
\textbf{J1513-2550$^{\star\dagger\triangle}$}  &  1.43e+34  &  2.12  &  2.09e-20  &  8.68e+34  &  2.13e+08  &  1.71 (33.8) &  3.14 (117.4) &  N/A  &  3.14 (117.4) \\ 
\textbf{J1514-4946$^{\star\dagger\triangle}$}  &  4.13e+33  &  3.59  &  1.71e-20  &  1.46e+34  &  2.51e+08  &  3.58 (684.8) &  4.12 (559.7) &  4.39 (313.8) &  4.39 (313.8) \\ 
J1526-2744  &  5.43e+32  &  2.49  &  3.54e-21  &  9.05e+33  &  9.50e+07  &  1.39 (10.6) &  N/A  &  N/A  &  N/A  \\ 
\textbf{J1531-5610$^{\star\dagger}$}  &  1.76e+34  &  84.21  &  1.38e-14  &  9.12e+35  &  1.09e+12  &  1.23 (22.2) &  1.38 (110.3) &  N/A  &  1.38 (110.3) \\ 
\textbf{J1536-4948$^{\triangle\S}$}  &  9.14e+33  &  3.08  &  2.07e-20  &  2.79e+34  &  2.55e+08  &  5.58 (1043.3) &  6.14 (566.9) &  5.84 (367.9) &  5.84 (367.9) \\ 
\textbf{J1543-5149$^{\triangle\S}$}  &  2.78e+33  &  2.06  &  1.60e-20  &  7.22e+34  &  1.84e+08  &  3.9 (3.6) &  3.3 (94.7) &  N/A  &  3.3 (94.7) \\ 
\textbf{J1544+4937$^{\triangle\S}$}  &  2.53e+33  &  2.16  &  2.14e-21  &  8.40e+33  &  6.88e+07  &  4.77 (10.8) &  3.53 (39.6) &  N/A  &  3.53 (39.6) \\ 
\textbf{J1552+5437$^{\star\dagger\triangle\S}$}  &  2.28e+33  &  2.43  &  2.82e-21  &  7.79e+33  &  8.38e+07  &  3.18 (4.0) &  3.39 (64.4) &  N/A  &  3.39 (64.4) \\ 
\textbf{J1555-2908$^{\star\dagger}$}  &  1.45e+34  &  1.79  &  4.45e-20  &  3.08e+35  &  2.86e+08  &  3.64 (1.3) &  3.97 (25.3) &  N/A  &  3.97 (25.3) \\ 
\textbf{J1600-3053$^{\dagger\triangle\S}$}  &  3.42e+33  &  3.6  &  8.52e-21  &  7.21e+33  &  1.77e+08  &  2.67 (102.0) &  3.08 (118.2) &  N/A  &  3.08 (118.2) \\ 
\textbf{J1614-2230$^{\star\dagger\S}$}  &  1.52e+33  &  3.15  &  3.98e-21  &  5.03e+33  &  1.13e+08  &  1.52 (408.5) &  2.27 (778.2) &  2.08 (607.6) &  2.08 (607.6) \\ 
\textbf{J1614-5048$^{\star\dagger}$}  &  6.40e+34  &  231.89  &  4.92e-13  &  1.56e+36  &  1.08e+13  &  957.46 (0.0) &  3.28 (9.9) &  N/A  &  3.28 (9.9) \\ 
\textbf{J1622-0315$^{\star\triangle\S}$}  &  1.31e+33  &  3.85  &  9.60e-21  &  6.64e+33  &  1.95e+08  &  3.5 (44.3) &  3.75 (79.2) &  N/A  &  3.75 (79.2) \\ 
\textbf{J1625-0021$^{\dagger\triangle\S}$}  &  2.24e+33  &  2.83  &  2.13e-20  &  3.70e+34  &  2.49e+08  &  1.68 (680.2) &  2.12 (918.8) &  1.91 (571.4) &  1.91 (571.4) \\ 
J1627+3219  &  8.71e+33  &  2.18  &  5.48e-21  &  2.08e+34  &  1.11e+08  &  1.37 (26.0) &  N/A  &  N/A  &  N/A  \\ 
\textbf{J1628-3205$^{\star\dagger\S}$}  &  1.92e+33  &  3.21  &  9.98e-21  &  1.19e+34  &  1.81e+08  &  1.74 (26.8) &  2.17 (228.0) &  N/A  &  2.17 (228.0) \\ 
\textbf{J1630+3734$^{\star\dagger\triangle}$}  &  1.01e+33  &  3.32  &  8.04e-21  &  8.67e+33  &  1.65e+08  &  1.66 (30.2) &  1.86 (215.2) &  N/A  &  1.86 (215.2) \\ 
\textbf{J1640+2224$^{\dagger\triangle\S}$}  &  6.21e+32  &  3.16  &  1.84e-21  &  2.31e+33  &  7.73e+07  &  1.48 (7.9) &  2.34 (41.2) &  N/A  &  2.34 (41.2) \\ 
\textbf{J1641+8049$^{\star\triangle}$}  &  2.19e+33  &  2.01  &  9.79e-21  &  4.68e+34  &  1.42e+08  &  1.36 (13.9) &  1.93 (57.2) &  N/A  &  1.93 (57.2) \\ 
\textbf{J1648-4611$^{\star\triangle}$}  &  1.14e+35  &  164.96  &  2.37e-14  &  2.09e+35  &  2.00e+12  &  4.29 (103.0) &  3.82 (69.2) &  3.82 (48.4) &  3.82 (48.4) \\ 
\textbf{J1653-0158$^{\star\triangle\S}$}  &  2.25e+33  &  1.97  &  1.35e-21  &  6.96e+33  &  5.21e+07  &  2.35 (401.3) &  3.09 (711.3) &  3.02 (500.7) &  3.02 (500.7) \\ 
\textbf{J1658-5324$^{\star\triangle\S}$}  &  1.92e+33  &  2.44  &  1.08e-20  &  2.93e+34  &  1.64e+08  &  1.49 (67.1) &  1.72 (531.5) &  1.92 (440.1) &  1.92 (440.1) \\ 
\textbf{J1702-4128$^{\star}$}  &  4.88e+34  &  182.15  &  5.23e-14  &  3.42e+35  &  3.12e+12  &  1.2 (11.5) &  2.0 (45.2) &  N/A  &  2.0 (45.2) \\ 
\textbf{J1705-1906$^{\star\dagger\S}$}  &  2.41e+32  &  298.99  &  4.14e-15  &  6.11e+33  &  1.13e+12  &  1.3 (8.7) &  0.49 (50.5) &  N/A  &  0.49 (50.5) \\ 
\textbf{J1709-4429$^{\dagger\triangle\S}$}  &  1.11e+36  &  102.5  &  9.48e-14  &  3.48e+36  &  3.15e+12  &  4.88 (10165.4) &  4.86 (10007.0) &  4.52 (7613.4) &  4.52 (7613.4) \\ 
\textbf{J1713+0747$^{\star\dagger\triangle\S}$}  &  1.50e+33  &  4.57  &  8.25e-21  &  3.41e+33  &  1.96e+08  &  1.72 (23.0) &  2.49 (153.0) &  N/A  &  2.49 (153.0) \\ 
\textbf{J1718-3825$^{\triangle\S}$}  &  1.51e+35  &  74.67  &  1.32e-14  &  1.25e+36  &  1.00e+12  &  3.69 (131.0) &  3.32 (417.4) &  3.38 (258.8) &  3.38 (258.8) \\ 
\textbf{J1730-2304$^{\star\triangle\S}$}  &  3.95e+32  &  8.12  &  1.47e-20  &  1.09e+33  &  3.50e+08  &  1.74 (5.0) &  1.18 (64.5) &  N/A  &  1.18 (64.5) \\ 
\textbf{J1730-3350$^{\star\dagger\S}$}  &  6.35e+34  &  139.51  &  8.41e-14  &  1.22e+36  &  3.47e+12  &  16.68 (0.9) &  4.02 (39.6) &  N/A  &  4.02 (39.6) \\ 
J1731-4744  &  2.30e+32  &  829.89  &  1.64e-13  &  1.13e+34  &  1.18e+13  &  53.84 (0.0) &  0.27 (0.3) &  N/A  &  0.27 (0.3) \\ 
\textbf{J1732-3131$^{\star\dagger\S}$}  &  8.79e+33  &  196.54  &  2.80e-14  &  1.46e+35  &  2.38e+12  &  1.96 (1647.7) &  2.4 (1858.2) &  2.25 (1366.8) &  2.25 (1366.8) \\ 
\textbf{J1732-5049$^{\star\dagger\S}$}  &  2.37e+33  &  5.31  &  1.14e-20  &  3.00e+33  &  2.49e+08  &  2.58 (5.5) &  2.89 (74.1) &  N/A  &  2.89 (74.1) \\ 
\textbf{J1739-3023$^{\star}$}  &  2.40e+34  &  114.37  &  1.14e-14  &  3.01e+35  &  1.16e+12  &  1.03 (19.8) &  1.54 (74.2) &  N/A  &  1.54 (74.2) \\ 
\textbf{J1740+1000 }  &  6.49e+32  &  154.1  &  2.13e-14  &  2.30e+35  &  1.83e+12  &  0.69 (1.8) &  0.68 (8.4) &  N/A  &  0.68 (8.4) \\ 
\textbf{J1741+1351 }  &  2.03e+33  &  3.75  &  2.80e-20  &  2.10e+34  &  3.28e+08  &  1.49 (11.7) &  2.35 (65.6) &  N/A  &  2.35 (65.6) \\ 
\textbf{J1741-2054$^{\star\dagger\triangle}$}  &  1.28e+33  &  413.7  &  1.70e-14  &  9.48e+33  &  2.68e+12  &  1.01 (1421.4) &  1.1 (3558.1) &  0.97 (3563.4) &  0.97 (3563.4) \\ 
\textbf{J1744-1134$^{\triangle}$}  &  6.92e+32  &  4.07  &  7.11e-21  &  4.16e+33  &  1.72e+08  &  1.14 (325.1) &  1.38 (1050.2) &  1.48 (963.1) &  1.48 (963.1) \\ 
\textbf{J1745+1017$^{\triangle\S}$}  &  1.34e+33  &  2.65  &  2.05e-21  &  4.35e+33  &  7.46e+07  &  2.13 (16.1) &  2.34 (173.0) &  N/A  &  2.34 (173.0) \\ 
\textbf{J1747-2958$^{\dagger\triangle}$}  &  4.30e+35  &  98.83  &  6.11e-14  &  2.50e+36  &  2.49e+12  &  4.12 (110.8) &  3.14 (272.5) &  2.91 (416.8) &  2.91 (416.8) \\ 
\textbf{J1747-4036$^{\star\dagger\triangle\S}$}  &  8.18e+34  &  1.65  &  1.30e-20  &  1.14e+35  &  1.48e+08  &  3.11 (8.3) &  2.89 (166.6) &  N/A  &  2.89 (166.6) \\ 
J1757-2421  &  2.87e+34  &  234.11  &  1.27e-14  &  3.92e+34  &  1.75e+12  &  Inf. (0.0) &  38.42 (0.2) &  N/A  &  38.42 (0.2) \\ 
\textbf{J1801-2451$^{\star\S}$}  &  5.52e+34  &  124.95  &  8.95e-14  &  1.81e+36  &  3.38e+12  &  9.5 (7.6) &  5.6 (35.3) &  N/A  &  5.6 (35.3) \\ 
J1803-6707  &  1.94e+33  &  2.13  &  1.76e-20  &  7.17e+34  &  1.96e+08  &  5.14 (24.7) &  N/A  &  N/A  &  N/A  \\ 
\textbf{J1805+0615$^{\star\dagger\triangle\S}$}  &  9.51e+33  &  2.13  &  1.86e-20  &  7.59e+34  &  2.01e+08  &  2.05 (9.0) &  2.26 (88.7) &  N/A  &  2.26 (88.7) \\ 
\textbf{J1809-2332$^{\star\triangle\S}$}  &  1.47e+35  &  146.79  &  3.44e-14  &  4.29e+35  &  2.27e+12  &  4.02 (2189.3) &  4.04 (2846.7) &  3.64 (2152.3) &  3.64 (2152.3) \\ 
\textbf{J1810+1744$^{\star\dagger\S}$}  &  9.73e+33  &  1.66  &  4.24e-21  &  3.66e+34  &  8.49e+07  &  2.2 (139.4) &  3.55 (340.6) &  3.7 (223.0) &  3.7 (223.0) \\ 
J1811-2405  &  5.55e+33  &  2.66  &  1.29e-20  &  2.71e+34  &  1.87e+08  &  15.66 (0.6) &  9.2 (3.9) &  N/A  &  9.2 (3.9) \\ 
\textbf{J1816+4510$^{\dagger\triangle\S}$}  &  1.49e+34  &  3.19  &  4.37e-20  &  5.31e+34  &  3.78e+08  &  2.06 (152.4) &  3.26 (295.0) &  3.21 (204.8) &  3.21 (204.8) \\ 
\textbf{J1823-3021A$^{\triangle\S}$}  &  9.82e+34  &  5.44  &  3.37e-18  &  8.27e+35  &  4.33e+09  &  3.8 (89.6) &  4.49 (105.7) &  N/A  &  4.49 (105.7) \\ 
\textbf{J1824+1014$^{\triangle}$}  &  6.23e+33  &  4.07  &  5.46e-21  &  3.21e+33  &  1.51e+08  &  3.09 (38.1) &  3.81 (53.0) &  N/A  &  3.81 (53.0) \\ 
J1824-0621  &  1.63e+34  &  3.23  &  9.14e-21  &  1.07e+34  &  1.74e+08  &  4.2 (56.6) &  N/A  &  N/A  &  N/A  \\ 
\textbf{J1824-2452A$^{\S}$}  &  7.73e+34  &  3.05  &  1.61e-18  &  2.24e+36  &  2.25e+09  &  3.38 (76.7) &  3.07 (229.4) &  2.83 (183.4) &  2.83 (183.4) \\ 
\textbf{J1828-1101$^{\star\dagger\triangle\S}$}  &  1.30e+35  &  72.06  &  1.48e-14  &  1.56e+36  &  1.05e+12  &  13.7 (0.2) &  5.44 (18.5) &  N/A  &  5.44 (18.5) \\ 
\textbf{J1831-0952$^{\triangle\S}$}  &  8.82e+34  &  67.27  &  8.31e-15  &  1.08e+36  &  7.57e+11  &  1.22 (0.0) &  0.93 (154.1) &  N/A  &  0.93 (154.1) \\ 
\textbf{J1833-1034}  &  1.79e+35  &  61.9  &  2.02e-13  &  3.36e+37  &  3.58e+12  &  8.48 (24.5) &  4.65 (14.4) &  5.21 (70.6) &  5.21 (70.6) \\ 
\textbf{J1833-3840$^{\dagger\triangle}$}  &  7.29e+33  &  1.87  &  1.77e-20  &  1.08e+35  &  1.84e+08  &  2.89 (1.8) &  2.2 (42.6) &  N/A  &  2.2 (42.6) \\ 
J1835-3259B  &  4.50e+34  &  1.83  &  4.39e-20  &  2.83e+35  &  2.87e+08  &  4.72 (1.3) &  N/A  &  N/A  &  N/A  \\ 
\textbf{J1836+5925 }  &  2.08e+34  &  173.26  &  1.50e-15  &  1.14e+34  &  5.16e+11  &  1.84 (23172.8) &  2.45 (29292.5) &  2.46 (24519.2) &  2.46 (24519.2) \\ 
J1837-0604  &  7.52e+34  &  96.31  &  4.49e-14  &  1.98e+36  &  2.10e+12  &  50.64 (0.1) &  8.66 (3.3) &  N/A  &  8.66 (3.3) \\ 
\textbf{J1843-1113$^{\triangle\S}$}  &  4.33e+33  &  1.85  &  9.34e-21  &  5.82e+34  &  1.33e+08  &  4.71 (3.0) &  3.19 (124.0) &  N/A  &  3.19 (124.0) \\ 
J1846-0258  &  5.55e+34  &  326.57  &  7.11e-12  &  8.06e+36  &  4.88e+13  &  10.75 (0.0) &  N/A  &  N/A  &  N/A  \\ 
J1852-1310  &  5.82e+32  &  4.31  &  9.27e-21  &  4.57e+33  &  2.02e+08  &  31.6 (0.0) &  N/A  &  N/A  &  N/A  \\ 
J1853-0004  &  1.68e+34  &  101.44  &  5.57e-15  &  2.11e+35  &  7.61e+11  &  2.09 (0.0) &  1.19 (1.6) &  N/A  &  1.19 (1.6) \\ 
\textbf{J1855-1436 }  &  1.59e+34  &  3.59  &  1.09e-20  &  9.29e+33  &  2.00e+08  &  4.87 (9.8) &  5.78 (12.0) &  N/A  &  5.78 (12.0) \\ 
\textbf{J1857+0143$^{\dagger\triangle}$}  &  2.92e+34  &  139.77  &  3.10e-14  &  4.49e+35  &  2.11e+12  &  68.65 (0.0) &  1.17 (7.9) &  N/A  &  1.17 (7.9) \\ 
J1857+0943  &  1.27e+33  &  5.36  &  1.73e-20  &  4.44e+33  &  3.08e+08  &  Inf. (0.0) &  N/A  &  N/A  &  N/A  \\ 
\textbf{J1858-2216$^{\triangle\S}$}  &  1.16e+33  &  2.38  &  3.50e-21  &  1.03e+34  &  9.24e+07  &  1.37 (140.1) &  1.94 (295.9) &  1.94 (215.9) &  1.94 (215.9) \\ 
\textbf{J1901-0125$^{\star\triangle\S}$}  &  1.30e+34  &  2.79  &  3.58e-20  &  6.48e+34  &  3.20e+08  &  4.18 (40.6) &  3.6 (124.8) &  N/A  &  3.6 (124.8) \\ 
\textbf{J1902-5105$^{\star\dagger\triangle\S}$}  &  7.62e+33  &  1.74  &  8.58e-21  &  6.43e+34  &  1.24e+08  &  3.22 (130.8) &  3.97 (302.5) &  3.69 (248.6) &  3.69 (248.6) \\ 
\textbf{J1903-7051$^{\star\dagger\triangle\S}$}  &  5.46e+32  &  3.6  &  7.92e-21  &  6.70e+33  &  1.71e+08  &  2.49 (97.3) &  2.47 (135.8) &  N/A  &  2.47 (135.8) \\ 
\textbf{J1907+0602$^{\star\dagger\S}$}  &  2.36e+35  &  106.64  &  8.65e-14  &  2.81e+36  &  3.07e+12  &  4.13 (1181.5) &  4.01 (1268.8) &  3.68 (907.0) &  3.68 (907.0) \\ 
\textbf{J1908+2105$^{\dagger}$}  &  3.95e+33  &  2.56  &  1.31e-20  &  3.09e+34  &  1.86e+08  &  5.01 (8.0) &  5.12 (16.1) &  N/A  &  5.12 (16.1) \\ 
\textbf{J1913+0904$^{\star\triangle\S}$}  &  2.12e+34  &  163.25  &  1.76e-14  &  1.60e+35  &  1.72e+12  &  1.83 (8.8) &  2.21 (57.6) &  N/A  &  2.21 (57.6) \\ 
J1913+1011  &  4.66e+34  &  35.91  &  3.38e-15  &  2.88e+36  &  3.53e+11  &  301.82 (0.1) &  35.55 (0.2) &  N/A  &  35.55 (0.2) \\ 
\textbf{J1921+0137$^{\star\dagger\S}$}  &  3.35e+34  &  2.5  &  1.90e-20  &  4.81e+34  &  2.21e+08  &  4.93 (21.8) &  4.3 (65.8) &  N/A  &  4.3 (65.8) \\ 
\textbf{J1921+1929$^{\star\triangle\S}$}  &  2.48e+33  &  2.65  &  3.65e-20  &  7.74e+34  &  3.15e+08  &  1284.59 (0.0) &  2.62 (9.9) &  N/A  &  2.62 (9.9) \\ 
\textbf{J1925+1720$^{\dagger\triangle}$}  &  2.90e+34  &  75.66  &  1.05e-14  &  9.54e+35  &  9.01e+11  &  2.2 (2.5) &  1.83 (26.4) &  N/A  &  1.83 (26.4) \\ 
\textbf{J1932+2220$^{\star\dagger\triangle\S}$}  &  7.34e+34  &  144.46  &  5.70e-14  &  7.46e+35  &  2.90e+12  &  5.29 (0.3) &  2.56 (18.1) &  N/A  &  2.56 (18.1) \\ 
\textbf{J1935+2025$^{\star\triangle\S}$}  &  5.49e+34  &  80.0  &  6.04e-14  &  4.64e+36  &  2.23e+12  &  1.7 (11.2) &  1.78 (112.7) &  N/A  &  1.78 (112.7) \\ 
\textbf{J1939+2134$^{\dagger\triangle}$}  &  2.33e+34  &  1.56  &  1.05e-19  &  1.10e+36  &  4.10e+08  &  2.88 (5.5) &  3.21 (76.9) &  N/A  &  3.21 (76.9) \\ 
\textbf{J1946-5403$^{\star\dagger}$}  &  1.56e+33  &  2.71  &  2.69e-21  &  5.33e+33  &  8.64e+07  &  1.45 (101.3) &  1.82 (447.7) &  1.29 (405.2) &  1.29 (405.2) \\ 
\textbf{J1952+3252$^{\star\triangle\S}$}  &  1.59e+35  &  39.53  &  5.83e-15  &  3.72e+36  &  4.86e+11  &  3.36 (1284.3) &  3.75 (2061.6) &  3.71 (1393.4) &  3.71 (1393.4) \\ 
\textbf{J1959+2048$^{\star\triangle}$}  &  3.02e+33  &  1.61  &  1.24e-20  &  1.17e+35  &  1.43e+08  &  1.77 (66.5) &  2.05 (369.4) &  2.14 (255.9) &  2.14 (255.9) \\ 
\textbf{J2006+0148$^{\star\dagger\triangle}$}  &  2.15e+33  &  2.16  &  3.29e-21  &  1.28e+34  &  8.53e+07  &  4.27 (15.9) &  4.16 (20.2) &  N/A  &  4.16 (20.2) \\ 
\textbf{J2006+3102$^{\star\dagger\triangle}$}  &  4.63e+34  &  163.7  &  2.49e-14  &  2.24e+35  &  2.04e+12  &  3.24 (28.4) &  3.08 (56.9) &  N/A  &  3.08 (56.9) \\ 
\textbf{J2017+0603$^{\dagger\triangle}$}  &  8.34e+33  &  2.9  &  8.52e-21  &  1.38e+34  &  1.59e+08  &  3.53 (479.5) &  3.5 (555.7) &  3.85 (347.9) &  3.85 (347.9) \\ 
\textbf{J2017-1614$^{\star\dagger\triangle\S}$}  &  1.61e+33  &  2.31  &  2.41e-21  &  7.67e+33  &  7.54e+07  &  3.38 (49.3) &  4.13 (65.6) &  N/A  &  4.13 (65.6) \\ 
\textbf{J2021+3651$^{\star\dagger\triangle}$}  &  1.91e+35  &  103.75  &  9.48e-14  &  3.35e+36  &  3.17e+12  &  3.14 (3079.2) &  3.49 (4479.0) &  3.24 (3851.9) &  3.24 (3851.9) \\ 
\textbf{J2021+4026 }  &  2.16e+35  &  265.33  &  5.48e-14  &  1.16e+35  &  3.86e+12  &  2.27 (2783.8) &  2.57 (6106.0) &  2.69 (5037.0) &  2.69 (5037.0) \\ 
\textbf{J2022+3842$^{\star\dagger\triangle}$}  &  3.17e+35  &  48.58  &  8.62e-14  &  2.97e+37  &  2.07e+12  &  6.07 (1.7) &  2.52 (104.9) &  N/A  &  2.52 (104.9) \\ 
\textbf{J2030+3641 }  &  4.55e+34  &  200.13  &  6.51e-15  &  3.20e+34  &  1.16e+12  &  1.64 (426.3) &  1.97 (759.2) &  1.97 (553.1) &  1.97 (553.1) \\ 
\textbf{J2030+4415$^{\dagger\triangle\S}$}  &  1.32e+33  &  227.07  &  5.05e-15  &  1.70e+34  &  1.08e+12  &  1.21 (255.1) &  1.56 (930.0) &  1.6 (701.7) &  1.6 (701.7) \\ 
\textbf{J2032+4127$^{\star\dagger\triangle}$}  &  5.28e+34  &  143.25  &  1.16e-14  &  1.56e+35  &  1.31e+12  &  4.14 (1022.6) &  4.3 (561.0) &  4.01 (433.8) &  4.01 (433.8) \\ 
\textbf{J2039-3616$^{\star\dagger\S}$}  &  1.33e+33  &  3.27  &  6.34e-21  &  7.16e+33  &  1.46e+08  &  1.53 (12.5) &  2.41 (76.0) &  N/A  &  2.41 (76.0) \\ 
\textbf{J2039-5617$^{\dagger\triangle\S}$}  &  5.21e+33  &  2.65  &  1.20e-20  &  2.54e+34  &  1.80e+08  &  3.63 (176.5) &  4.9 (196.8) &  4.41 (124.2) &  4.41 (124.2) \\ 
\textbf{J2042+0246$^{\star\dagger\triangle\S}$}  &  2.86e+32  &  4.53  &  1.41e-20  &  5.97e+33  &  2.55e+08  &  1.0 (58.3) &  1.08 (159.1) &  N/A  &  1.08 (159.1) \\ 
\textbf{J2043+1711$^{\star\dagger\S}$}  &  6.57e+33  &  2.38  &  4.80e-21  &  1.41e+34  &  1.08e+08  &  3.63 (299.1) &  3.88 (559.0) &  3.86 (369.0) &  3.86 (369.0) \\ 
\textbf{J2043+2740$^{\star\dagger\S}$}  &  2.37e+33  &  96.13  &  1.24e-15  &  5.50e+34  &  3.49e+11  &  1.05 (88.2) &  1.4 (237.5) &  N/A  &  1.4 (237.5) \\ 
\textbf{J2047+1053$^{\star\S}$}  &  3.98e+33  &  4.29  &  1.37e-20  &  6.83e+33  &  2.45e+08  &  5.53 (12.2) &  6.15 (14.6) &  N/A  &  6.15 (14.6) \\ 
\textbf{J2051-0827$^{\star\triangle\S}$}  &  6.46e+32  &  4.51  &  1.28e-20  &  5.50e+33  &  2.43e+08  &  1.55 (8.8) &  1.81 (48.6) &  N/A  &  1.81 (48.6) \\ 
\textbf{J2052+1219$^{\dagger\S}$}  &  1.70e+34  &  1.99  &  2.09e-21  &  1.05e+34  &  6.52e+07  &  6.29 (5.4) &  7.4 (12.2) &  N/A  &  7.4 (12.2) \\ 
\textbf{J2115+5448$^{\triangle\S}$}  &  8.10e+33  &  2.61  &  2.93e-20  &  6.50e+34  &  2.80e+08  &  3.05 (66.6) &  2.41 (39.2) &  N/A  &  2.41 (39.2) \\ 
J2116+1345  &  2.59e+33  &  2.22  &  2.64e-21  &  9.55e+33  &  7.75e+07  &  4.53 (8.3) &  N/A  &  N/A  &  N/A  \\ 
\textbf{J2124-3358$^{\star\dagger\triangle\S}$}  &  7.80e+32  &  4.93  &  7.57e-21  &  2.49e+33  &  1.95e+08  &  1.59 (1569.3) &  1.98 (2005.1) &  1.79 (1423.3) &  1.79 (1423.3) \\ 
\textbf{J2129-0429$^{\star\dagger\triangle\S}$}  &  2.73e+33  &  7.61  &  2.30e-19  &  2.06e+34  &  1.34e+09  &  1.72 (73.6) &  2.72 (169.4) &  N/A  &  2.72 (169.4) \\ 
J2205+6012  &  5.60e+33  &  2.41  &  1.99e-20  &  5.62e+34  &  2.22e+08  &  92.17 (0.0) &  0.88 (2.7) &  N/A  &  0.88 (2.7) \\ 
\textbf{J2208+4056$^{\star\dagger}$}  &  3.15e+32  &  636.96  &  5.28e-15  &  8.07e+32  &  1.86e+12  &  0.34 (8.4) &  0.61 (36.2) &  N/A  &  0.61 (36.2) \\ 
\textbf{J2214+3000$^{\star\dagger\S}$}  &  1.41e+33  &  3.12  &  1.33e-20  &  1.72e+34  &  2.06e+08  &  1.69 (303.8) &  2.18 (1653.0) &  2.07 (1186.7) &  2.07 (1186.7) \\ 
\textbf{J2215+5135$^{\star\dagger\S}$}  &  1.53e+34  &  2.61  &  2.38e-20  &  5.27e+34  &  2.52e+08  &  3.51 (205.5) &  4.78 (199.4) &  4.55 (128.2) &  4.55 (128.2) \\ 
\textbf{J2229+6114$^{\dagger\S}$}  &  2.59e+35  &  51.65  &  7.53e-14  &  2.16e+37  &  2.00e+12  &  5.01 (1432.4) &  5.09 (2220.6) &  4.84 (1729.3) &  4.84 (1729.3) \\ 
\textbf{J2234+0944$^{\star\triangle\S}$}  &  5.95e+32  &  3.63  &  1.34e-20  &  1.11e+34  &  2.23e+08  &  3.45 (63.3) &  3.29 (177.8) &  3.01 (129.6) &  3.01 (129.6) \\ 
\textbf{J2240+5832$^{\star\dagger\S}$}  &  6.22e+34  &  139.94  &  1.53e-14  &  2.20e+35  &  1.48e+12  &  3.74 (21.2) &  4.06 (38.4) &  N/A  &  4.06 (38.4) \\ 
\textbf{J2241-5236$^{\triangle\S}$}  &  3.25e+33  &  2.19  &  6.99e-21  &  2.63e+34  &  1.25e+08  &  1.83 (540.7) &  2.24 (730.2) &  1.93 (723.9) &  1.93 (723.9) \\ 
\textbf{J2256-1024$^{\dagger\triangle}$}  &  4.25e+33  &  2.29  &  1.16e-20  &  3.83e+34  &  1.65e+08  &  2.9 (19.2) &  3.21 (170.9) &  2.58 (150.3) &  2.58 (150.3) \\ 
\textbf{J2302+4442$^{\star\triangle\S}$}  &  3.47e+33  &  5.19  &  1.34e-20  &  3.77e+33  &  2.66e+08  &  2.62 (924.8) &  2.97 (1393.0) &  3.02 (861.0) &  3.02 (861.0) \\ 
\textbf{J2310-0555$^{\S}$}  &  1.62e+33  &  2.61  &  4.96e-21  &  1.10e+34  &  1.15e+08  &  1.54 (66.1) &  2.05 (176.0) &  N/A  &  2.05 (176.0) \\ 
\textbf{J2317+1439$^{\dagger\triangle}$}  &  2.13e+32  &  3.45  &  3.00e-21  &  2.89e+33  &  1.03e+08  &  174.01 (0.0) &  0.32 (11.5) &  N/A  &  0.32 (11.5) \\ 
\textbf{J2339-0533$^{\star\dagger\S}$}  &  5.90e+33  &  2.88  &  1.36e-20  &  2.25e+34  &  2.01e+08  &  3.51 (688.4) &  3.95 (655.9) &  4.03 (384.6) &  4.03 (384.6) \\ 
\end{longtable}
\end{center}

\newpage
\section{Properties of Radio Quiet Gamma-ray Pulsars}\label{B1}

Table~\ref{unknownPulsars} provides the properties of radio-quiet gamma-ray pulsars from the 3PC catalog, along with distance predictions obtained from different optimized machine learning models. The table includes the measured gamma-ray ﬂux and the predicted distances from OLS, RFR, SVR-RBF, and SVR-Linear models. For OLS predictions, the corresponding statistical uncertainties are also provided.

\renewcommand{\arraystretch}{1.3}
\begin{center}
\small
\begin{longtable}{c|ccccc|cccc}
\caption{
Properties of radio-quiet gamma-ray pulsars from the 3PC, along with distance predictions obtained using various ML approaches. The column definitions are identical with those provided in Table~\ref{knownPulsars}. The G100 column lists the measured integral gamma-ray flux above 100~MeV. The D$_{\text{OLS}}$, D$_{\text{RFR}}$, D$_{\text{SVR-RBF}}$ and D$_{\text{SVR-LIN}}$ columns display the predicted distances derived from optimized OLS, RFR, SVR-RBF, and SVR-Linear models, as described in Sect.~\ref{sect_4}. For the OLS method, the 1$\sigma$ upper and lower bounds of the distance prediction errors are included, while only the best-predicted distances are presented for the other ML models. Conservative average relative statistical errors of approximately $\sim$61$\%$ (upper) and $\sim$29$\%$ (lower), obtained from the OLS approach, may be assumed for the RFR, SVR-RBF, and SVR-Linear predictions.} 
\label{unknownPulsars} \\
\hline\hline
Pulsar Name & E$_{\text{Cut,HYB}}$ (TS) & P & $\dot{\text{P}}_{\text{Corr}}$ & $\dot{\text{E}}_{\text{Corr}}$ & G100 & D$_{\text{OLS}}$ & D$_{\text{RFR}}$ & D$_{\text{SVR-RBF}}$ & D$_{\text{SVR-LIN}}$ \\
PSR & GeV & ms & s/s & erg/s & erg/cm$^{2}$/s & kpc & kpc & kpc & kpc \\
\hline
\endfirsthead
\caption[]{\textit{(Continued table from previous page)}} \\
\hline
Pulsar Name (PSR) & E$_{\text{Cut}}$ (TS) & P & $\dot{\text{P}}_{\text{Corr}}$ & $\dot{\text{E}}_{\text{Corr}}$ & G100 & D$_{\text{OLS}}$ & D$_{\text{RFR}}$ & D$_{\text{SVR-RBF}}$ & D$_{\text{SVR-LIN}}$ \\
PSR & GeV & ms & s/s & erg/s & erg/cm$^{2}$/s & kpc & kpc & kpc & kpc \\
\hline
\endhead

\hline
\endfoot

\hline
\endlastfoot

J0357+3205 & 1.21 (2191.0) & 444.11 & 1.31$\times$10$^{-14}$ & 5.90$\times$10$^{33}$ & (6.01$\pm$0.18)$\times$10$^{-11}$ & \textbf{0.46$^{+0.29}_{-0.13}$} & 0.39 & 0.39 & 0.37 \\
J0359+5414 & 2.38 (194.7)  & 79.43  & 1.67$\times$10$^{-14}$ & 1.32$\times$10$^{36}$ & (1.98$\pm$0.13)$\times$10$^{-11}$ & \textbf{4.74$^{+2.88}_{-1.31}$} & 4.77 & 4.04 & 4.58 \\
J0554+3107 & 1.59 (469.6)  & 464.96 & 1.43$\times$10$^{-13}$ & 5.60$\times$10$^{34}$ & (1.87$\pm$0.09)$\times$10$^{-11}$ & \textbf{1.84$^{+1.14}_{-0.51}$} & 1.24 & 1.65 & 1.71 \\
J0622+3749 & 0.91 (550.4)  & 333.21 & 2.54$\times$10$^{-14}$ & 2.71$\times$10$^{34}$ & (1.77$\pm$0.09)$\times$10$^{-11}$ & \textbf{1.10$^{+0.69}_{-0.31}$} & 0.60 & 1.08 & 1.15 \\
J0633+0632 & 3.05 (941.4)  & 297.41 & 7.96$\times$10$^{-14}$ & 1.19$\times$10$^{35}$ & (9.56$\pm$0.24)$\times$10$^{-11}$ & \textbf{1.40$^{+0.86}_{-0.39}$} & 0.60 & 1.09 & 1.01 \\
J0734-1559 & 3.66 (425.2)  & 155.14 & 1.25$\times$10$^{-14}$ & 1.32$\times$10$^{35}$ & (4.58$\pm$0.14)$\times$10$^{-11}$ & \textbf{2.17$^{+1.33}_{-0.60}$} & 1.54 & 1.81 & 1.75 \\
J0744-2525 & 2.38 (148.2)  & 92.0   & 9.31$\times$10$^{-16}$ & 4.72$\times$10$^{34}$ & (1.73$\pm$0.12)$\times$10$^{-11}$ & \textbf{1.98$^{+1.20}_{-0.55}$} & 1.54 & 2.01 & 2.17 \\
J0802-5613 & 1.36 (223.3)  & 273.97 & 2.78$\times$10$^{-15}$ & 5.33$\times$10$^{33}$ & (6.59$\pm$0.60)$\times$10$^{-12}$ & \textbf{1.38$^{+0.87}_{-0.39}$} & 1.26 & 1.39 & 1.61 \\
J1023-5746 & 3.42 (257.0)  & 111.5  & 3.80$\times$10$^{-13}$ & 1.08$\times$10$^{37}$ & (1.46$\pm$0.12)$\times$10$^{-10}$ & \textbf{4.02$^{+2.48}_{-1.12}$} & 2.93 & 3.93 & 3.33 \\
J1044-5737 & 3.43 (686.1)  & 139.03 & 5.46$\times$10$^{-14}$ & 8.02$\times$10$^{35}$ & (1.14$\pm$0.04)$\times$10$^{-10}$ & \textbf{2.20$^{+1.33}_{-0.60}$} & 2.04 & 1.73 & 1.61 \\
J1057-5851 & 1.06 (198.6)  & 620.37 & 1.01$\times$10$^{-13}$ & 1.66$\times$10$^{34}$ & (1.32$\pm$0.17)$\times$10$^{-11}$ & \textbf{1.27$^{+0.81}_{-0.36}$} & 1.02 & 1.60 & 1.53 \\
J1105-6037 & 3.18 (135.7)  & 194.94 & 2.18$\times$10$^{-14}$ & 1.16$\times$10$^{35}$ & (3.07$\pm$0.35)$\times$10$^{-11}$ & \textbf{2.41$^{+1.48}_{-0.68}$} & 2.41 & 2.73 & 2.44 \\
J1111-6039 & 4.21 (93.1)   & 106.69 & 1.95$\times$10$^{-13}$ & 6.35$\times$10$^{36}$ & (5.91$\pm$0.34)$\times$10$^{-11}$ & \textbf{6.06$^{+3.72}_{-1.68}$} & 4.70 & 6.64 & 5.76 \\
J1135-6055 & 3.77 (235.6)  & 114.5  & 7.82$\times$10$^{-14}$ & 2.06$\times$10$^{36}$ & (4.58$\pm$0.22)$\times$10$^{-11}$ & \textbf{4.71$^{+2.87}_{-1.30}$} & 3.06 & 4.14 & 3.96 \\
J1139-6247 & 6.12 (28.7)   & 120.48 & 4.06$\times$10$^{-15}$ & 9.17$\times$10$^{34}$ & (1.92$\pm$0.21)$\times$10$^{-11}$ & \textbf{3.93$^{+2.49}_{-1.12}$} & 2.67 & 5.42 & 4.62 \\
J1203-6242 & 3.96 (113.8)  & 100.6  & 4.45$\times$10$^{-14}$ & 1.73$\times$10$^{36}$ & (3.74$\pm$0.36)$\times$10$^{-11}$ & \textbf{5.04$^{+3.07}_{-1.40}$} & 4.60 & 5.19 & 4.81 \\
J1208-6238 & 5.94 (47.1)   & 440.68 & 3.31$\times$10$^{-12}$ & 1.53$\times$10$^{36}$ & (3.73$\pm$0.40)$\times$10$^{-11}$ & \textbf{6.95$^{+4.39}_{-1.97}$} & 3.52 & 8.54 & 6.85 \\
J1231-5113 & 1.0 (295.2)   & 206.61 & 1.15$\times$10$^{-16}$ & 5.16$\times$10$^{32}$ & (1.07$\pm$0.07)$\times$10$^{-11}$ & \textbf{0.46$^{+0.30}_{-0.13}$} & 1.15 & 0.58 & 0.58 \\
J1231-6511 & 1.95 (174.8)  & 247.52 & 2.82$\times$10$^{-14}$ & 7.34$\times$10$^{34}$ & (1.21$\pm$0.14)$\times$10$^{-11}$ & \textbf{2.63$^{+1.61}_{-0.74}$} & 1.44 & 2.60 & 2.79 \\
J1335-5656 & 3.81 (59.1)   & 3.24   & 1.26$\times$10$^{-20}$ & 1.46$\times$10$^{34}$ & (8.22$\pm$0.85)$\times$10$^{-12}$ & \textbf{1.98$^{+1.20}_{-0.55}$} & 1.61 & 1.73 & 2.58 \\
J1350-6225 & 3.49 (70.5)   & 138.16 & 8.88$\times$10$^{-15}$ & 1.33$\times$10$^{35}$ & (3.60$\pm$0.41)$\times$10$^{-11}$ & \textbf{2.36$^{+1.45}_{-0.66}$} & 2.36 & 3.19 & 2.67 \\
J1358-6025 & 3.83 (102.5)  & 60.53  & 3.01$\times$10$^{-15}$ & 5.35$\times$10$^{35}$ & (3.17$\pm$0.36)$\times$10$^{-11}$ & \textbf{3.68$^{+2.23}_{-1.03}$} & 4.51 & 3.87 & 3.75\\
J1413-6205 & 3.81 (640.6)  & 109.74 & 2.74$\times$10$^{-14}$ & 8.18$\times$10$^{35}$ & (1.78$\pm$0.06)$\times$10$^{-10}$ & \textbf{1.84$^{+1.12}_{-0.51}$} & 2.02 & 1.54 & 1.35\\
J1422-6138 & 3.33 (96.1)   & 340.98 & 9.68$\times$10$^{-14}$ & 9.64$\times$10$^{34}$ & (5.35$\pm$0.25)$\times$10$^{-11}$ & \textbf{1.86$^{+1.16}_{-0.52}$} & 2.26 & 2.64 & 1.99\\
J1429-5911 & 3.59 (582.3)  & 115.84 & 2.39$\times$10$^{-14}$ & 6.07$\times$10$^{35}$ & (1.13$\pm$0.04)$\times$10$^{-10}$ & \textbf{2.06$^{+1.25}_{-0.57}$} & 2.06 & 1.69 & 1.55 \\
J1447-5757 & 1.77 (214.8)  & 158.73 & 1.18$\times$10$^{-14}$ & 1.17$\times$10$^{35}$ & (2.34$\pm$0.26)$\times$10$^{-11}$ & \textbf{1.96$^{+1.19}_{-0.55}$} & 1.71 & 2.02 & 2.08 \\
J1459-6053 & 3.81 (639.4)  & 103.15 & 2.53$\times$10$^{-14}$ & 9.09$\times$10$^{35}$ & (1.23$\pm$0.04)$\times$10$^{-10}$ & \textbf{2.27$^{+1.38}_{-0.62}$} & 2.10 & 1.78 & 1.65\\
J1522-5735 & 2.79 (443.1)  & 204.28 & 6.25$\times$10$^{-14}$ & 2.89$\times$10$^{35}$ & (7.59$\pm$0.40)$\times$10$^{-11}$ & \textbf{1.86$^{+1.13}_{-0.51}$} & 2.07 & 1.71 & 1.55\\
J1528-5838 & 2.39 (185.6)  & 355.69 & 2.48$\times$10$^{-14}$ & 2.17$\times$10$^{34}$ & (1.81$\pm$0.16)$\times$10$^{-11}$ & \textbf{1.75$^{+1.10}_{-0.49}$} & 1.33 & 1.91 & 1.81\\
J1615-5137 & 2.30 (55.5)   & 179.28 & 1.06$\times$10$^{-14}$ & 7.28$\times$10$^{34}$ & (3.34$\pm$0.30)$\times$10$^{-11}$ & \textbf{1.68$^{+1.02}_{-0.47}$} & 2.10 & 2.63 & 2.19\\
J1620-4927 & 4.15 (171.7)  & 171.94 & 1.05$\times$10$^{-14}$ & 8.15$\times$10$^{34}$ & (1.29$\pm$0.05)$\times$10$^{-10}$ & \textbf{1.22$^{+0.76}_{-0.34}$} & 1.49 & 1.58 & 1.16\\
J1623-5005 & 3.71 (34.2)   & 85.07  & 4.16$\times$10$^{-15}$ & 2.67$\times$10$^{35}$ & (6.11$\pm$0.50)$\times$10$^{-11}$ & \textbf{2.20$^{+1.33}_{-0.61}$} & 3.96 & 3.71 & 2.82\\
J1624-4041 & 2.45 (318.1)  & 167.86 & 4.73$\times$10$^{-15}$ & 3.94$\times$10$^{34}$ & (2.74$\pm$0.15)$\times$10$^{-11}$ & \textbf{1.60$^{+0.98}_{-0.44}$} & 1.28 & 1.49 & 1.51 \\
J1641-5317 & 1.86 (301.2)  & 175.11 & 3.70$\times$10$^{-15}$ & 2.72$\times$10$^{34}$ & (1.48$\pm$0.12)$\times$10$^{-11}$ & \textbf{1.69$^{+1.04}_{-0.47}$} & 1.24 & 1.53 & 1.70\\
J1649-3012 & 2.60 (138.7)  & 3.42   & 1.29$\times$10$^{-20}$ & 1.27$\times$10$^{34}$ & (9.04$\pm$0.84)$\times$10$^{-12}$ & \textbf{1.48$^{+0.89}_{-0.41}$} & 1.44 & 1.17 & 1.80\\
J1650-4601 & 2.96 (157.1)  & 127.12 & 1.51$\times$10$^{-14}$ & 2.91$\times$10$^{35}$ & (5.61$\pm$0.59)$\times$10$^{-11}$ & \textbf{2.15$^{+1.30}_{-0.60}$} & 3.48 & 2.48 & 2.16\\
J1714-3830 & 4.88 (38.7)   & 84.13  & 7.03$\times$10$^{-14}$ & 4.66$\times$10$^{36}$ & (9.12$\pm$0.92)$\times$10$^{-11}$ & \textbf{4.74$^{+2.92}_{-1.33}$} & 4.64 & 6.89 & 5.25\\
J1736-3422 & 2.85 (54.0)   & 347.22 & 6.51$\times$10$^{-14}$ & 6.14$\times$10$^{34}$ & (1.75$\pm$0.27)$\times$10$^{-11}$ & \textbf{2.64$^{+1.65}_{-0.76}$} & 2.00 & 3.72 & 3.21\\
J1742-3321 & 2.02 (50.1)   & 143.27 & 1.27$\times$10$^{-15}$ & 1.71$\times$10$^{34}$ & (1.45$\pm$0.26)$\times$10$^{-11}$ & \textbf{1.53$^{+0.95}_{-0.44}$} & 1.75 & 2.22 & 2.15\\
J1744-7619 & 1.77 (675.6)  & 4.69   & 9.67$\times$10$^{-21}$ & 3.71$\times$10$^{34}$ & (1.97$\pm$0.62)$\times$10$^{-11}$ & \textbf{0.58$^{+0.35}_{-0.16}$} & 0.63 & 0.47 & 0.60\\
J1746-3239 & 2.46 (296.4)  & 199.54 & 6.56$\times$10$^{-15}$ & 3.26$\times$10$^{34}$ & (5.13$\pm$0.29)$\times$10$^{-11}$ & \textbf{1.13$^{+0.69}_{-0.31}$} & 1.26 & 1.21 & 1.09\\
J1803-2149 & 4.27 (97.4)   & 106.33 & 1.95$\times$10$^{-14}$ & 6.40$\times$10$^{35}$ & (8.92$\pm$0.69)$\times$10$^{-11}$ & \textbf{2.57$^{+1.57}_{-0.71}$} & 4.48 & 3.34 & 2.59\\
J1813-1246 & 3.19 (1636.8) & 48.07  & 1.76$\times$10$^{-14}$ & 6.24$\times$10$^{36}$ & (2.48$\pm$0.06)$\times$10$^{-10}$ & \textbf{2.36$^{+1.45}_{-0.65}$} & 2.06 & 1.47 & 1.48\\
J1817-1742 & 4.81 (32.7)   & 149.74 & 2.06$\times$10$^{-14}$ & 2.42$\times$10$^{35}$ & (2.85$\pm$0.45)$\times$10$^{-11}$ & \textbf{3.79$^{+2.36}_{-1.09}$} & 4.18 & 5.55 & 4.50\\
J1826-1256 & 3.61 (1020.5) & 110.24 & 1.21$\times$10$^{-13}$ & 3.57$\times$10$^{36}$ & (4.13$\pm$1.29)$\times$10$^{-11}$ & \textbf{1.79$^{+1.09}_{-0.49}$} & 2.10 & 1.42 & 1.19\\
J1827-0849 & 10.46 (12.4)  & 2.24   & 1.10$\times$10$^{-20}$ & 3.84$\times$10$^{34}$ & (2.95$\pm$0.36)$\times$10$^{-11}$ & \textbf{2.34$^{+1.51}_{-0.67}$} & 2.44 & 3.21 & 3.28\\
J1827-1446 & 0.96 (42.2)   & 499.19 & 4.53$\times$10$^{-14}$ & 1.44$\times$10$^{34}$ & (1.94$\pm$0.18)$\times$10$^{-11}$ & \textbf{0.93$^{+0.59}_{-0.26}$} & 1.57 & 1.92 & 1.56\\
J1838-0537 & 3.81 (57.5)   & 145.75 & 4.51$\times$10$^{-13}$ & 5.75$\times$10$^{36}$ & (1.20$\pm$0.13)$\times$10$^{-10}$ & \textbf{4.02$^{+2.47}_{-1.13}$} & 4.74 & 5.90 & 4.31\\
J1844-0346 & 2.24 (121.3)  & 112.85 & 1.55$\times$10$^{-13}$ & 4.25$\times$10$^{36}$ & (4.59$\pm$0.53)$\times$10$^{-11}$ & \textbf{4.34$^{+2.68}_{-1.22}$} & 4.73 & 4.94 & 4.54\\
J1846+0919 & 2.42 (585.5)  & 225.55 & 9.93$\times$10$^{-15}$ & 3.42$\times$10$^{34}$ & (3.57$\pm$0.12)$\times$10$^{-11}$ & \textbf{1.37$^{+0.84}_{-0.38}$} & 0.63 & 1.14 & 1.16\\
J1906+0722 & 4.47 (98.1)   & 111.53 & 3.59$\times$10$^{-14}$ & 1.02$\times$10$^{36}$ & (8.59$\pm$0.80)$\times$10$^{-11}$ & \textbf{3.09$^{+1.89}_{-0.86}$} & 4.57 & 3.87 & 3.03\\
J1932+1916 & 2.61 (369.6)  & 208.22 & 9.32$\times$10$^{-14}$ & 4.07$\times$10$^{35}$ & (6.54$\pm$0.45)$\times$10$^{-11}$ & \textbf{2.13$^{+1.29}_{-0.59}$} & 2.35 & 2.03 & 1.84\\
J1954+2836 & 3.71 (544.7)  & 92.71  & 2.12$\times$10$^{-14}$ & 1.05$\times$10$^{36}$ & (1.07$\pm$0.03)$\times$10$^{-10}$ & \textbf{2.48$^{+1.50}_{-0.68}$} & 2.23 & 1.99 & 1.86\\
J1957+5033 & 1.32 (1187.2) & 374.81 & 7.08$\times$10$^{-15}$ & 5.31$\times$10$^{33}$ & (2.62$\pm$0.08)$\times$10$^{-11}$ & \textbf{0.70$^{+0.45}_{-0.20}$} & 0.43 & 0.59 & 0.61\\
J1958+2846 & 2.81 (1040.5) & 290.4  & 2.12$\times$10$^{-13}$ & 3.41$\times$10$^{35}$ & (1.06$\pm$0.03)$\times$10$^{-10}$ & \textbf{1.71$^{+1.04}_{-0.47}$} & 1.24 & 1.28 & 1.21\\
J2017+3625 & 2.17 (564.1)  & 166.75 & 1.36$\times$10$^{-15}$ & 1.16$\times$10$^{34}$ & (7.52$\pm$0.46)$\times$10$^{-11}$ & \textbf{0.64$^{+0.39}_{-0.18}$} & 0.55 & 0.64 & 0.58\\
J2028+3332 & 1.74 (1068.3) & 176.71 & 4.86$\times$10$^{-15}$ & 3.48$\times$10$^{34}$ & (5.71$\pm$0.19)$\times$10$^{-11}$ & \textbf{0.89$^{+0.54}_{-0.25}$} & 0.44 & 0.71 & 0.74\\
J2034+3632 & 1.92 (86.6)   & 3.65   & 1.73$\times$10$^{-21}$ & 1.41$\times$10$^{33}$ & (1.24$\pm$0.78)$\times$10$^{-11}$ & \textbf{0.57$^{+0.35}_{-0.16}$} & 1.50 & 0.72 & 0.86\\
J2055+2539 & 1.51 (2702.3) & 319.56 & 4.10$\times$10$^{-15}$ & 4.96$\times$10$^{33}$ & (5.32$\pm$0.10)$\times$10$^{-11}$ & \textbf{0.51$^{+0.32}_{-0.14}$} & 0.41 & 0.38 & 0.38\\
J2111+4606 & 5.00 (246.2)  & 157.84 & 1.63$\times$10$^{-13}$ & 1.63$\times$10$^{36}$ & (4.89$\pm$0.15)$\times$10$^{-11}$ & \textbf{5.14$^{+3.16}_{-1.42}$} & 3.01 & 4.34 & 4.01\\
J2139+4716 & 1.10 (957.7)  & 282.85 & 1.78$\times$10$^{-15}$ & 3.11$\times$10$^{33}$ & (2.70$\pm$0.10)$\times$10$^{-11}$ & \textbf{0.52$^{+0.33}_{-0.15}$} & 0.47 & 0.50 & 0.51\\
J2238+5903 & 3.66 (489.2)  & 162.74 & 9.69$\times$10$^{-14}$ & 8.87$\times$10$^{35}$ & (6.64$\pm$0.24)$\times$10$^{-11}$ & \textbf{3.12$^{+1.90}_{-0.86}$} & 2.27 & 2.47 & 2.35\\
\end{longtable}
\end{center}

\end{appendix}

\begin{acknowledgements}

\modulolinenumbers[8]

E.O.A. gratefully acknowledge financial support from the TÜBİTAK Research Institute for Fundamental Sciences. I extend my sincere gratitude to Gerrit Spengler for his invaluable active contributions and constructive feedback, which were instrumental to the development of this work. I also wish to express my sincere appreciation to Elena Amato for her feedback, as well as to Ertuğrul Karamanlı for his comprehensive analysis cross-checks, which collectively enhanced the scientific rigor and overall quality of this paper. The Jupyter notebook used for the OLS-based predictions, together with associated data ﬁles, is available at \href{https://doi.org/10.5281/zenodo 15293226}{Zenodo doi:10.5281/zenodo.15293226} to support transparency and reproducibility. Additional materials may be included in future updates to the repository.

\end{acknowledgements}

\bibliographystyle{aasjournal}
\bibliography{pulsar}



\end{document}